\documentclass{emulateapj}
\usepackage{natbib}
\newcommand{\lunits}{\ensuremath{\textrm{erg~s}^{-1}}}

\newcommand{\hubbleunits}{\textrm{km~s}\ensuremath{^{-1}}~\textrm{Mpc}\ensuremath{^{-1}}}
\newcommand{\ha}{\textrm{H}\ensuremath{\alpha}}
\newcommand{\hb}{\textrm{H}\ensuremath{\beta}}
\newcommand{\hg}{\textrm{H}\ensuremath{\gamma}}

\newcommand{\nii}{[\textrm{N}~\textsc{ii}]}
\newcommand{\oii}{[\textrm{O}~\textsc{ii}]}
\newcommand{\oiii}{[\textrm{O}~\textsc{iii}]}
\newcommand{\hii}{\textrm{H}~\textsc{ii}}
\newcommand{\halam}{\textrm{H}\ensuremath{\alpha~\lambda6563}}
\newcommand{\hblam}{\textrm{H}\ensuremath{\beta~\lambda4861}}
\newcommand{\hglam}{\textrm{H}\ensuremath{\gamma~\lambda4340}}
\newcommand{\oilam}{[\textrm{O}~\textsc{i}]~\ensuremath{\lambda6300}}
\newcommand{\oiilam}{\oii~\ensuremath{\lambda3727}}
\newcommand{\oiiilam}{[\textrm{O}~\textsc{iii}]~\ensuremath{\lambda5007}}
\newcommand{\niilam}{[\textrm{N}~\textsc{ii}]~\ensuremath{\lambda6584}} 
\newcommand{\oiidoublet}{[\textrm{O}~\textsc{ii}]~\ensuremath{\lambda\lambda3726,3729}}
\newcommand{\oiiidoublet}{[\textrm{O}~\textsc{iii}]~\ensuremath{\lambda\lambda4959,5007}}
\newcommand{\siidoublet}{[\textrm{S}~\textsc{ii}]~\ensuremath{\lambda\lambda6716,6731}}

\newcommand{\neonlam}{[\textrm{Ne}~\textsc{iii}]~\ensuremath{\lambda3869}} 

\newcommand{\ewhanii}{\textrm{EW}(\ha+\nii)}
\newcommand{\ewhb}{\textrm{EW}(\textrm{H}\ensuremath{\beta})}
\newcommand{\oiiha}{\mbox{[\textrm{O}~\textsc{ii}]/\textrm{H}\ensuremath{\alpha}}}
\newcommand{\niiha}{\nii/\ha}
\newcommand{\oiiihb}{\mbox{[\textrm{O}~\textsc{iii}]/\textrm{H}\ensuremath{\beta}}}
\newcommand{\oiiioii}{\oiii/\oii}
\newcommand{\oiiiha}{\mbox{[\textrm{O}~\textsc{iii}]/\textrm{H}\ensuremath{\alpha}}}
\newcommand{\hahb}{\ha/\hb}
\newcommand{\hbhg}{\hb/\hg}
\newcommand{\hbha}{\hb/\ha}
\newcommand{\dnbreak}{D\ensuremath{_{n}(4000)}}
\newcommand{\ebv}{E(\bv)}
\newcommand{\ehbha}{\mbox{\ensuremath{E}(\textrm{H}\ensuremath{\beta}-\textrm{H}\ensuremath{\alpha})}}
\newcommand{\ehghb}{E(\textrm{H}\ensuremath{\gamma}-\textrm{H}\ensuremath{\beta})}

\newcommand{\rv}{\ensuremath{R_V}}
\newcommand{\aha}{\ensuremath{A(\ha)}}
\newcommand{\mb}{\ensuremath{M_B}}
\newcommand{\lir}{\ensuremath{L({\rm IR})}}
\newcommand{\lfir}{\ensuremath{L({\rm FIR})}}
\newcommand{\lb}{\ensuremath{L(B)}}
\newcommand{\lu}{\ensuremath{L(U)}}
\newcommand{\lbsun}{\lb\ensuremath{_{\sun}}}
\newcommand{\lblbsun}{\lb/\lbsun}
\newcommand{\lulha}{\lu/\lha}
\newcommand{\msun}{\ensuremath{\mathcal{M}_{\sun}}}
\newcommand{\lsun}{\ensuremath{L_{\sun}}}
\newcommand{\zsun}{\textrm{Z}\ensuremath{_{\sun}}}
\newcommand{\pagel}{\ensuremath{R_{23}}}
\newcommand{\lha}{\mbox{\ensuremath{L}(\textrm{H}\ensuremath{\alpha})}}
\newcommand{\loii}{\mbox{{\ensuremath{L}([\textrm{O}~\textsc{ii}])}}}

\newcommand{\loiii}{\mbox{{\ensuremath{L}([\textrm{O}~\textsc{iii}])}}}
\newcommand{\loiiilam}{\mbox{{\ensuremath{L}([\textrm{O}~\textsc{iii}]~\ensuremath{\lambda5007})}}}
\newcommand{\lhb}{\ensuremath{L(\textrm{H}\beta)}}
\newcommand{\lhbobs}{\ensuremath{L(\textrm{H}\beta)_{\rm obs}}}
\newcommand{\luobs}{\ensuremath{L(U)_{\rm obs}}}
\newcommand{\loiiobs}{\loii\ensuremath{_{\rm obs}}}
\newcommand{\loiiiobs}{\loiii\ensuremath{_{\rm obs}}}
\newcommand{\logoh}{\ensuremath{12+\log\,(\textrm{O}/\textrm{H})}}
\newcommand{\zzsun}{\textrm{Z}/\textrm{Z}\ensuremath{_{\sun}}}
\newcommand{\logsunoh}{\ensuremath{12+\log\,(\textrm{O}/\textrm{H})_{\sun}}}

\newcommand{\onpettini}{\ensuremath{(\oiii/\hb})/(\nii/\ha)}
\newcommand{\logu}{\ensuremath{\log\,U}}
\newcommand{\sfrunits}{\ensuremath{\mathcal{M}_{\sun}~\textrm{yr}^{-1}}}
\newcommand{\sfr}{\ensuremath{\psi}}
\newcommand{\sfroii}{\sfr(\oii)}
\newcommand{\sfrha}{\sfr(\ha)}
\newcommand{\sfrhb}{\sfr(\hb)}
\newcommand{\sfrir}{\sfr(\textrm{IR})}
\newcommand{\sfruband}{\sfr(U)}

\newenvironment{inlinefigure}{
\def\@captype{figure}
\noindent\begin{minipage}{0.999\linewidth}\begin{center}}
{\end{center}\end{minipage}\smallskip}

\slugcomment{ApJ, accepted}
\shortauthors{Moustakas, Kennicutt, \& Tremonti}
\shorttitle{Optical SFR Indicators}

\begin{document}

\title{Optical Star-Formation Rate Indicators}

\author{John Moustakas\altaffilmark{1}, Robert~C. Kennicutt,
  Jr.\altaffilmark{1,2}, \& Christy~A. Tremonti\altaffilmark{1}} 
\altaffiltext{1}{Steward Observatory, University of Arizona, 933 N
  Cherry Ave., Tucson, AZ 85721, USA}
\altaffiltext{2}{Institute of Astronomy, University of Cambridge,
  Madingley Road, Cambridge CB3 0HA, United Kingdom} 


\setcounter{footnote}{2}

\begin{abstract}
Using integrated optical spectrophotometry for $412$ star-forming
galaxies at $z\sim0$, and fiber-aperture spectrophotometry for
$120,846$ SDSS galaxies at $z\sim0.1$, we investigate the \halam,
\hblam, \oiilam, and \oiiilam{} nebular emission lines and the
$U$-band luminosity as quantitative star-formation rate (SFR)
indicators.  We demonstrate that the extinction-corrected \halam{}
luminosity is a reliable SFR tracer even in highly obscured
star-forming galaxies.  We find that variations in dust reddening
dominate the systematic uncertainty in SFRs derived from the observed
\hb, \oii, and $U$-band luminosities, producing a factor of $\sim1.7$,
$\sim2.5$, and $\sim2.1$ scatter in the mean transformations,
respectively.  We show that \oii{} depends weakly on variations in
oxygen abundance over a wide range in metallicity,
$\logoh=8.15-8.7$~dex ($\zzsun=0.28-1.0$), and that in this
metallicity interval galaxies occupy a narrow range in ionization
parameter ($-3.8\lesssim \logu \lesssim-2.9$~dex).  We show that the
scatter in \oiiilam{} as a SFR indicator is a factor of $3-4$ due to
its sensitivity to metal abundance and ionization.  We develop
empirical SFR calibrations for \hb{} and \oii{} parameterized in terms
of the $B$-band luminosity, which remove the systematic effects of
reddening and metallicity, and reduce the SFR scatter to $\pm40\%$ and
$\pm90\%$, respectively, although individual galaxies may deviate
substantially from the median relations.  Finally, we compare the
$z\sim0$ relations between blue luminosity and reddening, ionization,
and \oiiha{} ratio against measurements at $z\sim1$ and find broad
agreement.  We emphasize, however, that optical emission-line
measurements including \ha{} for larger samples of intermediate- and
high-redshift galaxies are needed to test the applicability of our
locally derived SFR calibrations to distant galaxies.
\end{abstract}



\section{INTRODUCTION}\label{sec:intro}

Over the past twenty years many techniques have been devised to
estimate the global star-formation rates (SFRs) of galaxies.  The
first quantitative analysis of SFRs using the \halam{} nebular
recombination line was undertaken by \citet{kenn83}.  Today, with the
advent of large, multi-wavelength surveys of galaxies, virtually every
part of the electromagnetic spectrum has been explored as a means of
deriving SFRs \citep[e.g.,][]{kenn83, buat89, condon92, david92}.  Two
indicators directly associated with massive star formation are the
ultraviolet (UV; $\lambda\lambda1200-2500$~\AA) and nebular
recombination line luminosities.  These techniques were particularly
important in establishing the order-of-magnitude rise in the SFR
density of the universe from the present day to $z=1$ and beyond
\citep{lilly96, madau96, cowie97, glazebrook99, steidel99, hopkins04}.
Despite considerable progress in measuring SFRs of distant galaxies,
however, their accuracy remains limited by a wide range of systematic
uncertainties \citep{cram98, glazebrook99, bell01a, sullivan01,
hopkins01a, hopkins03, charlot02, rosa-gonzalez02, kewley02b,
kewley04, hirashita03, bell03}.  Well-calibrated SFR diagnostics with
well-understood systematic uncertainties are needed to improve
constraints on the evolution of the cosmic SFR, and to investigate the
physical processes responsible for this evolution.  In this paper we
carry out a detailed empirical analysis of rest-frame optical SFR
indicators.

One of the most well-understood SFR indicators is \halam.  The \ha{}
luminosity is directly proportional to the hydrogen-ionizing radiation
from massive ($\gtrsim10$~\msun) stars, and therefore provides a
near-instantaneous ($\lesssim10$~Myr) measure of the SFR with minimal
dependence on the physical conditions of the ionized gas
\citep{kenn98}.  Unfortunately, \ha{} is only observable from the
ground at $z\lesssim0.4$ in the optical, and at $0.7\lesssim
z\lesssim2.5$ through the near-infrared atmospheric windows
(Fig.~\ref{fig:zlambda}).  Given the observational difficulties of
observing \ha{} at high redshift, therefore, the \oiilam{} nebular
emission line has been suggested as an alternative SFR indicator
\citep{gallagher89, kenn92b, guzman97, barbaro97, jansen01,
salamanca02, rosa-gonzalez02, hopkins03, kewley04, mouhcine05}.  Its
intrinsic strength and blue rest-frame wavelength allow it to be
measured even in low signal-to-noise (S/N) spectra at $z\lesssim1.5$
in the optical, and at $2\lesssim z\lesssim5.2$ in the near-infrared
(Fig.~\ref{fig:zlambda}).  However, the \oii{} luminosity depends
explicitly on the chemical abundance and excitation state of the
ionized gas, and suffers a larger amount of dust extinction than \ha.
Therefore, unlike the Balmer recombination lines, \oii{} is not
directly proportional to the SFR and must be calibrated either
empirically \citep{kenn92b, rosa-gonzalez02, kewley04}, or
theoretically \citep{barbaro97, charlot01}.

\citet{gallagher89} presented the first quantitative analysis of
\oii{} as a SFR indicator using \oii{} and \hb{} measurements of a
sample of $75$ blue-irregular galaxies.  \citet{kenn92b} improved upon
this study by obtaining integrated optical ($3650-7150$~\AA) spectra
of $55$ galaxies spanning the Hubble sequence, including a handful of
peculiar objects, and derived the first empirical calibration of
\oii{} relative to \ha.  Previously, the equivalent width of \oii{}
was used only indirectly to infer ongoing star formation
\citep[e.g.,][]{broadhurst88, colless90, dressler84, couch87}.
Subsequently, \citet{jansen01} and \citet{kewley04} used the Nearby
Field Galaxy Survey \citep[NFGS;][]{jansen00a, jansen00b}, an imaging
and spectrophotometric survey of a representative sample of $\sim200$
nearby galaxies, to quantify how variations in reddening and chemical
abundance affect the observed \oii{} luminosity, and to improve the
\oii{} SFR calibration.  \citet{jansen01} showed that the observed
\oiiha{} ratio varies by a factor of $\sim7$ near $\mb^{*}$,
predominantly due to variations in dust reddening, and that the
\oiiha{} ratio is inversely proportional to galaxy luminosity
\citep[see also][]{carter01, salamanca02}.  Comparing the NFGS
observations to photoionization models, \citet{kewley04} concluded
that the \oiiha{} ratio is also strongly dependent on the heavy
element abundance and varies weakly with the strength of the ionizing
radiation field.

However, all these studies have been based on relatively small samples
of optically selected normal galaxies exhibiting modest
current-to-past-averaged star-formation rates, normal morphologies,
and typical or lower-than-average infrared luminosities.  From models
of hierarchical galaxy formation \citep[e.g.,][]{somerville99}, we
anticipate a higher incidence of extreme or bursty star formation at
high redshift due to a greater frequency of mergers/interactions and
larger reservoirs of neutral gas \citep[e.g.,][]{hammer05}.
Therefore, SFR calibrations based on local samples of normal galaxies
may not apply at high redshift.

Large spectrophotometric surveys of the nearby universe such as the
Sloan Digital Sky Survey \citep[SDSS;][]{york00} provide the
opportunity to study the emission-line properties of galaxies with
unprecedented statistical precision \citep[e.g.,][]{brinchmann04,
tremonti04}.  \citet{hopkins03} present a thorough analysis of
multi-wavelength SFR diagnostics in the SDSS.  However, the fraction
of light subtended by the $3\arcsec$ fiber-optic aperture utilized by
the SDSS depends on the distance and intrinsic properties (size,
bulge-to-disk ratio, surface-brightness distribution, etc.) of each
individual galaxy.  For example, in the star-forming galaxy sample
studied by \citet{tremonti04} the median light fraction is only
$\sim25\%$.  Furthermore, the SDSS's strict magnitude-limited
selection ($r<17.7$~mag) targets primarily the most luminous
present-day galaxies, which are unlikely to be representative of
high-redshift star-forming galaxies.  Therefore, empirical SFR
calibrations based on SDSS observations should not be applied blindly
to distant samples.

As part of a larger effort to characterize the physical properties of
star-forming galaxies, we have obtained high S/N optical
spectrophotometry ($3600-6900$~\AA{} at $\sim8$~\AA{} FWHM resolution)
for a diverse sample of $417$ nearby galaxies \citep[hereafter
MK05]{moustakas05a}.  This survey targets objects that represent a
small fraction of the galaxy population today, but which may be more
typical of high-redshift samples, including starbursts,
interacting/merging systems, and dusty, infrared-luminous galaxies;
the sample also includes a large number of normal star-forming
galaxies.  We utilize the long-slit drift-scanning technique developed
by \citet{kenn92a} to obtain spatially integrated spectra at
intermediate spectral resolution, which makes our observations
well-matched to traditional long-slit spectroscopy of distant
galaxies.  We supplement our new observations with the NFGS to
increase the number of normal galaxies, and to identify any selection
biases in our diverse sample of galaxies.  In addition, we compare our
analysis of optical SFR diagnostics against a sample of $\sim120,000$
star-forming galaxies from the SDSS in order to assess the effects of
statistical incompleteness and aperture bias in our integrated galaxy
sample and in the SDSS, respectively.

We use these data to study the \oiilam, \hblam, and \oiiilam{} nebular
emission lines as quantitative SFR diagnostics by comparing them
against SFRs derived from the extinction-corrected \ha{} luminosity.
We also investigate the $U$-band luminosity as a SFR indicator in an
effort to determine whether far-optical broadband photometry offers
comparable precision to empirically calibrated emission-line
diagnostics for deriving SFRs of distant galaxies.  In
\S\ref{sec:data} we present our integrated and SDSS samples.  In
\S\ref{sec:analysis} we explore empirical correlations between optical
SFR diagnostics and global properties such as luminosity, dust
extinction, and chemical abundance in order to understand the dominant
sources of uncertainty that limit the application of these
diagnostics.  In \S\ref{sec:results} we derive new empirical SFR
calibrations.  Finally, \S\ref{sec:applications} we discuss the
applicability of our new calibrations to intermediate-redshift galaxy
samples, and present our conclusions in \S\ref{sec:conclusions}.  To
compute distances and absolute magnitudes (always on the Vega system)
we adopt $\Omega_{0}+\Omega_{\Lambda}=1$, $\Omega_{0}=0.3$, and
$H_{0}=70$~\hubbleunits{} \citep{spergel03, freedman01}.  Following
convention, we give both emission- and absorption-line equivalent
widths as positive numbers in the rest frame-of-reference.  Finally,
we adopt $\logsunoh=8.7$~dex as the solar oxygen abundance
\citep{prieto01,holweger01}.

\section{THE DATA}\label{sec:data}

\subsection{The Integrated Galaxy Sample}\label{sec:sample}   

Our integrated galaxy sample consists of our own spectroscopic
observations (MK05) and the NFGS \citep{jansen00a, jansen00b}.  We
briefly summarize the relevant details of each survey, and refer the
reader to the original papers for more information regarding the
sample selection and data reductions.  The MK05 survey provides high
S/N optical ($3600-6900$~\AA) spectrophotometry at $\sim8$~\AA{} FWHM
resolution for $417$ nearby galaxies.  Based on a variety of internal
and external comparisons MK05 find that the relative
spectrophotometric precision of the data is $\sim4\%$.  The survey
roughly divides into four major subsamples: (1) $\sim125$ galaxies
selected from the First Byurakan Survey of UV-excess galaxies
\citep{markarian89}; (2) $\sim100$ infrared-luminous galaxies selected
from the IRAS Warm and Bright Galaxy Surveys \citep{kim95, veilleux95,
veilleux99}; (3) $\sim35$ morphologically selected interacting/merging
systems drawn from the Ph.~D. thesis sample of \citet{turner98}; and
(4) $\sim130$ normal galaxies selected from a volume-limited \ha- and
UV-imaging survey of the $11$~Mpc local volume (R.~C. Kennicutt
et~al., 2006, in preparation), and the Ursa Major cluster
\citep{tully96}.  These samples are intentionally chosen to span the
diverse range of galaxies in the local universe with active star
formation, both to improve our understanding of galactic star
formation and to serve as a more comprehensive reference sample for
lookback studies.  To provide a more representative, complementary
sample of nearby galaxies we turn to the NFGS.  The NFGS is a
photometric and spectroscopic survey of $196$ galaxies selected to
reproduce the $B$-band luminosity function \citep{jansen00a}.  The
wavelength coverage ($3600-7100$~\AA), spectral resolution
($\sim6$~\AA{} FWHM), relative spectrophotometric precision
($\sim6\%$), and S/N of the NFGS are well-matched to our own
observations.  Excluding several broad-line AGN and one BL~Lac, the
combined integrated galaxy sample consists of spectrophotometry for
$589$ galaxies.

To measure fluxes and equivalent widths of the nebular emission lines
we utilize {\sc ispec1d}, a spectral synthesis fitting code described
in detail in MK05.  Using {\sc ispec1d} we find the non-negative
linear combination of population synthesis models \citep{bruzual03}
that optimally reproduces the observed stellar continuum.  Subtracting
the model continuum from the data results in a pure nebular
emission-line spectrum self-consistently corrected for underlying
stellar absorption.  We measure the emission-line fluxes and
equivalent widths of the strong nebular lines, tabulated in MK05,
using simultaneous, multi-Gaussian profile fitting with physically
motivated constraints on the intrinsic velocity widths and redshifts
of the Balmer and forbidden lines.  For consistency we re-measure the
emission-line fluxes in the NFGS spectra using {\sc ispec1d}.  We find
our measurements broadly consistent with \citet{jansen00b}, except at
low equivalent widths where we argue that our technique is more
reliable.

To define a sample of star-forming galaxies and to ensure that we can
measure reliably the nebular reddening (\S\ref{sec:reddening}), we
impose a $3\sigma$ S/N cut on the \ha{} and \hb{} emission lines,
which removes $28$ objects from the integrated sample.  We verify that
most of these objects are early-type (E/S0) galaxies with effectively
zero star formation.  However, three galaxies (NGC~1266, UGC~05101,
and CGCG~049-057) fail our S/N criterion on \hb{} because they are
dusty, infrared-luminous galaxies.  All three objects have
well-detected \ha{} emission, and in only one do we marginally detect
\oii.  Since these galaxies comprise $<1\%$ of the full integrated
sample, and just $4\%$ of objects with $\lir>10^{11}~\lsun$
(\S\ref{sec:ha_sfr}), we do not expect our conclusions to be biased
with respect to highly obscured galaxies.  However, these types of
objects emphasize the need for either \ha{} or infrared observations
to ensure a complete census of star formation in galaxies.

We use traditional emission-line diagnostic diagrams to differentiate
galaxies with nebular emission powered by star formation versus
galaxies with an admixture of star formation and AGN activity
\citep{baldwin81, veilleux87, ho97, kewley01b}.  In
Figure~\ref{fig:bpt} we plot the observed \oiii/\hb{} line-ratio as a
function of \nii/\ha{} for the MK05 data (squares) and the NFGS
(triangles).  For comparison with the integrated line-ratios, we
overplot the emission-line sequence traced by individual \hii{}
regions (small points) in spiral galaxies \citep{mccall85, zaritsky94,
vanzee98} and dwarf galaxies \citep{izotov94, izotov97, izotov98}.
The solid curve in Figure~\ref{fig:bpt} empirically segregates normal
star-forming galaxies from AGN based on an analysis of $\sim10^{5}$
SDSS galaxies by \citet{kauffmann03c}.  The dashed curve defines the
theoretical boundary between AGN and star-forming galaxies presented
by \citet{kewley01a}.  We conservatively adopt the
\citet{kauffmann03c} curve to remove objects with AGN activity,
although in \S\ref{sec:results} we explore the effect of including AGN
on our results.  Finally, we classify objects without $1\sigma$
\oiiilam{} or \niilam{} detections as star-forming galaxies using
either the condition $\log\,(\niiha)<-0.4$ \citep{tremonti04}, or
using the \siidoublet/\ha{} and \oilam/\ha{} versus \oiiilam/\hb{}
diagnostic diagrams and the theoretical boundaries defined by
\citet{kewley01a}.  After rejecting another $9$ objects that cannot be
classified using any of the above methods, our final integrated sample
consists of $412$ star-forming galaxies.

Figure~\ref{fig:bpt} shows that the sequences formed by \hii{} regions
and star-forming galaxies overlap across the full range of
emission-line ratios in the \niiha{} versus \oiiihb{} plane
\citep{kenn92b, lehnert94, kobulnicky99a, charlot01}.  Therefore, to
first order, techniques developed to analyze the physical properties
of individual \hii{} regions may be used to interpret the integrated
spectral properties of galaxies \citep[e.g.,][]{kobulnicky99a,
stasinska01, pilyugin04b, moustakas05c}.

In Figure~\ref{fig:properties} we compare the spectrophotometric
properties of the MK05 and NFGS samples.  We plot the distributions of
the combined sample as dashed, unshaded histograms.  We collect
$B$-band photometry for both samples, listed in order of preference,
from \citet{devac91}, LEDA\footnote{\url{http://leda.univ-lyon1.fr}}
\citep{prugniel98}, or by synthesizing photometry directly from the
spectra, as described in MK05.  These magnitudes have been corrected
for foreground Galactic extinction \citep[$\rv=3.1$;][]{odonnell94,
schlegel98}, but not for nebular emission lines or inclination effects
since these corrections are challenging or impossible to make at high
redshift.  Distances for the MK05 sample are based on either primary
or secondary measurements when available, or the \citet{mould00}
multi-attractor linear infall model.  Distances for the NFGS sample
are based exclusively on the infall model.  We assign a fixed $15\%$
uncertainty to objects without a published distance error.

In Figure~\ref{fig:properties}\emph{a} we plot the $B$-band luminosity
distributions for the MK05 and NFGS samples, using $M_{\sun,
B}=+5.42$~mag to convert between \lblbsun{} and \mb.  Despite the
significantly different selection criteria, we find the luminosity
distributions of the two samples qualitatively similar.  The MK05
survey includes a larger number of faint dwarf galaxies
($\mb\gtrsim-16$~mag) and galaxies more luminous than
$\mb\simeq-20$~mag relative to the NFGS.  The combined distribution,
however, is fairly uniform between $\mb\simeq-16$ and $-22$~mag, and
spans the range between $\sim-12$~mag and $\sim-22.5$~mag.  In
Figure~\ref{fig:properties}\emph{b} we plot the distribution of dust
extinction at \ha, $\aha=2.52\,\ebv$, as determined from the Balmer
decrement (\S\ref{sec:reddening}).  The median (mean) extinction for
the combined sample is $0.51$~mag ($0.59\pm0.50$~mag), ranging from
zero to $2.62$~mag in IC~0750, a highly reddened Sab galaxy in the
MK05 sample.  Figure~\ref{fig:properties}\emph{c} characterizes the
metallicity distribution of these samples, using the methodology in
\S\ref{sec:oii_metallicity} to estimate the gas-phase abundance.  The
median (mean) metallicity is $8.54$~dex ($8.50\pm0.21$~dex), and spans
the range $7.77<\logoh<8.84$~dex ($0.12\lesssim\zzsun\lesssim1.4$).
The metallicity distribution of the MK05 sample extends to lower
values than the NFGS due to the inclusion of a larger number of
low-luminosity galaxies.  Finally, in
Figure~\ref{fig:properties}\emph{d} we plot the distribution of
observed \oiiilam/\oiilam{} flux ratios for both samples.  The
\oiiioii{} ratio characterizes the hardness of the ionizing radiation
field, or the ionization parameter of the photoionized gas
\citep{shields90, kewley02a}.  In the MK05 sample the median (mean)
logarithmic \oiiioii{} ratio is $-0.33$~dex ($-0.27\pm0.32$~dex), but
spans a factor of $\sim65$ in excitation, whereas the NFGS
distribution peaks in a narrow range around $-0.45\pm0.20$~dex.
Although in the local universe low-mass galaxies undergoing a strong
burst of star formation typically exhibit the highest \oiiioii{}
ratios, stronger ionizing radiation fields may be much more common in
high-redshift, massive galaxies \citep[e.g.,][]{pettini01}.

In summary, we find that by combining our survey with the NFGS we
achieve wide coverage of the physical parameter space spanned by the
$z=0$ population of star-forming galaxies, from normal galaxies that
dominate the mass density, to dwarfs and optical/infrared starbursts
that likely dominate at high redshift \citep[e.g.,][]{flores99,
hammer05, bell05}.  In \S\ref{sec:applications} we show that the
diversity of this sample allows us to construct SFR diagnostics that
may be applied at high redshift.

\subsection{The SDSS Sample}\label{sec:sdsssample}

To complement our sample of galaxies with integrated spectra we culled
the SDSS fourth data release \citep[DR4;][]{adelman05} to define a
complete sample of nearby star-forming galaxies.  We use DR4 galaxies
in the SDSS Main Galaxy Sample \citep{strauss02}, which have Petrosian
$r$ magnitudes between $14.5<r<17.77$~mag and $r$-band Petrosian
half-light surface brightnesses $\mu_{50}\leq24.5$~mag~arcsec$^{-2}$
\citep[corrected for foreground Galactic extinction;][]{schlegel98}.
We only include galaxies having $z>0.033$ to ensure that \oiidoublet{}
lies within the spectral range of the SDSS spectrograph.  In addition,
we elect to remove galaxies where less than $10\%$ of the $r$-band
light falls in the fiber in order to remove cases of extreme aperture
bias, while preserving the approximate magnitude-limited nature of the
sample \citep[see also][]{tremonti04}.  The $10\%$ cut on light
fraction removes less than $4\%$ of the sample, whereas a cut at
$30\%$ removes over half of the sample and introduces biases which are
difficult to quantify.  The above requirements result in a parent
sample of $360,902$ SDSS galaxies.

Emission-line fluxes and equivalent widths for these galaxies have
been measured\footnote{\url{http://www.mpa-garching.mpg.de/SDSS}}
using a customized continuum fitting code based on the
\citet{bruzual03} populations synthesis models and described by
\citet{tremonti04} (\S\ref{sec:sample}).  We have compared the results
of this code and {\sc ispec1d} and we find excellent agreement among
the resulting emission-line flux and equivalent width measurements.
To define a sample of star-forming galaxies we impose $3\sigma$
detections of the \ha{} and \hb{} emission lines, which eliminates
$\sim50\%$ of the parent sample (all early-type), leaving $173,540$
emission-line galaxies.  Finally, we remove objects contaminated by
AGN activity using the methodology described in \S\ref{sec:sample},
resulting in a sample of $120,846$ star-forming galaxies.  The mean
$r$-band light fraction for this sample is $25\pm9\%$.

In Figure~\ref{fig:sdssproperties} we compare the distributions of
spectrophotometric properties for the SDSS and integrated galaxy
samples.  To facilitate a direct comparison we plot the distributions
normalized to the total number of galaxies in each sample.  For the
SDSS we compute $M_{B, {\rm Vega}}$ using the Petrosian $g_{\rm
AB}$-band magnitude and $(g-r)_{\rm AB}$ color, and the following
relation (M.~R.~Blanton et~al. 2006, in preparation):

\begin{equation}
B_{\rm Vega} = g_{\rm AB} + 0.3915\,(g-r)_{\rm AB} + 0.087,
\label{eq:MB}
\end{equation}

\noindent where the $g$- and $r$-band magnitudes have been corrected
for foreground Galactic extinction \citep{schlegel98} and
k-corrected\footnote{Using {\tt k-correct} version 4.1.3, which is
available at \url{http://cosmo.nyu.edu/blanton/kcorrect}.} to $z=0$
\citep{blanton03}.  The standard deviation of the color term in
equation~(\ref{eq:MB}) is $0.15$~mag, which we add in quadrature to
the measured photometric uncertainties.

Figure~\ref{fig:sdssproperties}\emph{a} shows that the SDSS galaxies
provide complete coverage of the bright end of the $B$-band luminosity
distribution, centered on $-20.1$~mag.  By comparison, the integrated
sample spans a much broader range in absolute magnitude, uniformly
ranging from the most luminous objects in the SDSS to
$\mb\simeq-16$~mag, including a handful of $\mb>-16$~mag dwarf
galaxies.  Throughout our analysis we will attribute many of the
differences in physical properties between the integrated and SDSS
samples to the different luminosity distributions, although aperture
bias in the SDSS observations are also important, as we discuss below.

Figure~\ref{fig:sdssproperties}\emph{b} compares the distributions of
\aha{} in the SDSS and integrated galaxy samples.  The median (mean)
extinction in the SDSS is $0.83$~mag ($0.85\pm0.41$~mag), whereas in
the integrated sample \aha{} peaks at zero and decreases almost
monotonically to $\aha\simeq2.6$~mag.  We attribute the observed
differences in these distributions to two effects.  First, in the
local universe there exists a correlation between luminosity and dust
extinction, whereby luminous galaxies contain more dust, on average,
than less luminous galaxies
\citep[\S\ref{sec:calibrations};][]{buat96, wang96, tully98,
adelberger00, jansen01, stasinska04}.  Consequently, \aha{} in the
SDSS peaks at higher values both because the mean luminosity of the
SDSS sample is larger, and because the SDSS misses less-extincted,
lower-luminosity galaxies.  In addition, aperture bias in the SDSS may
be important.  If the centers of spiral galaxies have higher optical
depths to dust compared to the optical depth averaged over the whole
galaxy \citep[e.g.,][]{valentijn94, jansen94}, then the mean
extinction in a fiber-optic survey of star-forming galaxies will be
higher, on average, than the mean extinction of an integrated
spectroscopic survey.  \citet{kewley05}, however, find that extinction
does not depend on enclosed light fraction based on a comparison of
integrated and nuclear spectra in the NFGS.

In Figure~\ref{fig:sdssproperties}\emph{c} we compare the metallicity
distributions of the two samples, adopting the empirical abundance
calibrations given in \S\ref{sec:oii_metallicity}.  Relative to the
integrated galaxy sample, the distribution of oxygen abundances in the
SDSS peaks strongly at higher abundance.  The median metallicity in
the SDSS is $\logoh\simeq8.70$~dex ($\zzsun\simeq1$), compared to
$\sim8.54$~dex ($\zzsun\simeq0.7$) in the integrated sample.  We
attribute these differences to a combination of two effects: the
luminosity-metallicity correlation and aperture effects.  Locally, and
at high redshift, luminous star-forming galaxies obey a
luminosity-metallicity correlation, whereby luminous galaxies are more
metal-rich than low-luminosity galaxies \citep[J.~Moustakas
et~al. 2006, in preparation;][]{skillman89a, zaritsky94, richer95,
garnett02, melbourne02, tremonti04, kobulnicky03b, kobulnicky04}.  The
median \mb{} magnitude difference of the two samples is $-1.33$~mag.
Adopting the slope of the $B$-band luminosity-metallicity correlation
found by \citet{tremonti04}, $-0.185$~dex~mag$^{-1}$, we predict a
median metallicity difference of $+0.25$~dex.  This value has the
correct sign, and, given all the uncertainties, is roughly consistent
with the measured median metallicity difference of $+0.16$~dex.
Aperture effects may also drive the SDSS metallicities to higher
values, since the centers of spiral galaxies are typically more
metal-rich than their outskirts \citep[e.g.,][]{mccall85, oey93,
zaritsky94, kenn96, vanzee98, pilyugin04a}.  \citet{tremonti04}
estimate that aperture bias in the SDSS may lead to an overestimate of
the globally averaged metallicity by at least $+0.1$~dex \citep[see
also][]{kewley05}.

Finally, in Figure~\ref{fig:sdssproperties}\emph{d} we compare the
distributions of the observed \oiiilam/\oiilam{} ratios in the SDSS
and integrated galaxy samples.  The median (mean) ionization parameter
of the SDSS galaxies is $-0.54$~dex ($-0.52\pm0.15$~dex), lower on
average and more narrowly peaked than the distribution of ratios in
the integrated galaxy sample.  The observed differences are consistent
with the higher mean luminosity and metallicity of the SDSS galaxies
\citep{dopita00, kewley01b}, and with the inclusion of a larger number
of extreme starbursts with hard ionizing radiation fields in the
integrated galaxy sample.

To summarize, we find significant differences among the SDSS and
integrated samples, which is not surprising given the different
selection criteria.  Although the integrated sample contains $<1\%$
the number of objects in the SDSS sample, it intentionally spans a
broader range of physical properties such as luminosity, metallicity,
dust extinction, and ionization.  By comparison, the strength of the
SDSS sample is its statistical completeness of the bright end of the
luminosity function, although aperture bias must be carefully treated.
In the following analysis of optical SFR diagnostics we discuss all
these effects in detail.

\section{ANALYSIS}\label{sec:analysis}

\subsection{Nebular Reddening}\label{sec:reddening}

To quantify the amount of dust reddening we compute the Balmer
decrement, \hahb, where all the Balmer emission lines have been
corrected for underlying stellar absorption as discussed in
\S\ref{sec:data}.  We define the color excess due to dust reddening,
\ehbha{}, using the relation

\begin{equation}
\ehbha \equiv -2.5\,\log\,\left[\frac{ (\hahb)_{\rm int} }{
    (\hahb)_{\rm obs} }\right], 
\label{eq:ehbha}
\end{equation}

\noindent where $(\hahb)_{\rm obs}$ is the observed decrement and
$(\hahb)_{\rm int}$ is the intrinsic Balmer decrement.  We assume the
case~B recombination value $(\hahb)_{\rm int} = 2.86$, which is
appropriate for an individual \hii{} region at a typical electron
temperature and density \citep{storey95}.  To account for the small
variation in $(\hahb)_{\rm int}$ with electron temperature we
propagate a $5\%$ systematic uncertainty in $(\hahb)_{\rm int}$ into
the total error in \ehbha.  To relate \ehbha{} to the broad-band color
excess, \ebv, we introduce an attenuation curve,
$k(\lambda)\equiv\,A(\lambda)/\ebv$, to obtain the expression

\begin{equation}
\ebv \equiv \frac{\ehbha}{k(\hb)-k(\ha)},
\label{eq:ebv}
\end{equation}

\noindent where $k(\hb)$ and $k(\ha)$ are the values of $k(\lambda)$
at $4861$~\AA{} and $6563$~\AA, respectively
\citep[e.g.,][]{calzetti01}.  

In order to de-redden our integrated emission-line fluxes we must
assume a functional form for $k(\lambda)$.  The most common practice
is to neglect radiative transfer effects due to variations in geometry
\citep[which, in general, is not a good assumption:][]{witt92,
witt00}, and adopt a Milky Way or LMC/SMC extinction law.
Alternatively, \citet{charlot00} advocate a power-law attenuation
curve, $k(\lambda)\propto\lambda^{-0.7}$, based on a multi-wavelength
analysis of nearby starburst galaxies and simple radiation transfer
arguments.  Because optical extinction and attenuation curves are
similar (unlike in the ultraviolet), we opt for the simplest solution
and adopt the \citet{odonnell94} Milky Way extinction curve.
Equation~(\ref{eq:ebv}) then becomes $\ebv=0.874\,\ehbha$.

In \S\ref{sec:results} we also consider using the \hbhg{} Balmer
decrement to account for dust extinction, since \hglam{} is
observationally accessible across the same range of redshifts as the
emission-line SFR diagnostics considered in this paper
(Fig.\ref{fig:zlambda}).  However, obtaining a reliable measurement of
\hg{} is challenging because of its intrinsic weakness and the
combined effects of stellar absorption and dust extinction.  In the
following, we adopt an intrinsic \hbhg{} ratio of $2.14$ assuming an
electron temperature of $10,000$~K; this ratio varies by just $\pm3\%$
across a broad range of nebular conditions \citep{storey95}.  Using
the \citet{odonnell94} Milky Way extinction curve, the analogous
relations to equations~(\ref{eq:ehbha}) and (\ref{eq:ebv}) for the
\hbhg{} ratio become $\ehghb=-2.5\,\log\,(2.14/\hbhg)$ and
$\ebv=2.05\,\ehghb$, respectively.

In Figure~\ref{fig:ebv_compare} we compare the reddenings determined
using \hahb{} and \hbhg{} for the MK05 and NFGS samples (squares and
triangles with error bars, respectively), and the SDSS (small points
without error bars).  The \emph{solid} line is the line-of-equality
for the two measurements.  For this comparison we apply a $7\sigma$
S/N cut on \hg, and we require that $\ewhb>10$~\AA{} in emission.  We
impose this minimum equivalent width because the \hg{} emission-line
measurement is very sensitive to the quality of the continuum
subtraction.  For the SDSS galaxies we find no median systematic
offset and a dispersion of $0.15$~mag.  The \ebv{} measurements based
on \hbhg{} for the MK05 and NFGS samples are displaced systematically
toward higher reddening by $0.08$ and $0.15$~mag, respectively, and
the residual scatter is significant: $0.22$ and $0.38$~mag,
respectively.  We attribute the larger discrepancy between \ebv{}
using \hbhg{} and \ebv{} using \hahb{} in the integrated samples to
their lower spectral resolution ($\sim8$ and $\sim6$~\AA{} FWHM,
respectively), compared to the SDSS ($\sim4$~\AA).  \citet{liang04b}
discuss in detail the biases inherent to low spectral resolution
spectroscopy.  Thus, while \hbhg{} offers a direct measurement of the
reddening in the absence of \ha, the errors are formidable
($\gtrsim50\%$), even in the best case scenario when detailed
continuum subtraction can be used and the line S/N and equivalent
width are moderately high.  Fortunately, at high redshift galaxies are
more gas rich and line equivalent widths are generally larger
\citep[e.g.,][]{kobulnicky03b}.  Thus, using \hbhg{} as a reddening
diagnostic in higher-redshift samples may be more reliable
\citep[e.g.,][]{flores04}.


\subsection{\halam{} Star-Formation Rates}\label{sec:ha_sfr}

The \halam{} nebular emission line is one of the primary diagnostics
used to estimate the SFRs of galaxies in the local universe
\citep{kenn83, kenn92b, gallego95, kenn98, nakamura04, brinchmann04}.
The largest uncertainties affecting \ha-based SFRs are due to dust
absorption of Lyman-continuum photons within individual \hii{}
regions, dust attenuation in the general interstellar medium of the
galaxy, and uncertainties in the shape of the initial mass function
\citep{kenn98}.  Throughout this paper we adopt the theoretical
calibration between the \ha{} luminosity, \lha, and the SFR, \sfr{},
given by \citet{kenn98}:

\begin{equation}
\sfrha = 7.9\times10^{-42}\ \frac{\lha}{\lunits}\ \ \ \sfrunits,
\label{eq:ha_sfr}
\end{equation}

\noindent where \lha{} has been corrected for underlying stellar
absorption and interstellar dust attenuation \citep[see
also][]{kenn94}.  This transformation has been computed for solar
metallicity, the \citet{salpeter55} IMF with lower- and upper-mass
cutoffs of $0.1$ and $100$~\msun, respectively.
Equation~(\ref{eq:ha_sfr}) also assumes that no Lyman-continuum
photons are absorbed by dust, a point that we return to below.

In this paper we adopt the extinction-corrected \ha{} luminosity as
our fiducial SFR tracer.  However, it is instructive to assess the
absolute accuracy of \ha{} as a SFR indicator by comparing it against
other independent indicators.  Several authors have carried out this
exercise using the ultraviolet, infrared, and radio luminosity
\citep{buat96, cram98, glazebrook99, bell01a, sullivan01, hopkins01a,
kewley02b, buat02, hopkins03, hirashita03, bell03}.  These studies
reveal the numerous challenges of deriving absolute SFRs because each
indicator is coupled to the true SFR by different physical processes,
and they each suffer a variety of systematic uncertainties.
Nevertheless, we choose to compare \sfrha{} to the SFR derived from
the bolometric infrared luminosity, $\lir\equiv L(8-1000~\micron)$.
The infrared luminosity measures the amount of stellar radiation
absorbed and re-emitted by dust grains.  Assuming solar metallicity,
that the dust re-radiates $100\%$ of the bolometric luminosity, and
that star formation has been continuous for the past $10-100$~Myr,
\citet{kenn98} provides the following transformation given the same
IMF used in equation~(\ref{eq:ha_sfr}):

\begin{equation}
\sfrir = 4.5\times10^{-44}\ \frac{\lir}{\lunits}\ \ \ \sfrunits.
\label{eq:ir_sfr}
\end{equation}

\noindent The coefficient in equation~(\ref{eq:ir_sfr}) will be
different in galaxies with significant dust heating from old stellar
populations, or lower overall dust content \citep{kenn98, bell03,
hirashita03}.  

We estimate \lir{} for our integrated galaxy sample using the
following procedure.  Fluxes from IRAS at $12$, $25$, $60$ and
$100$~\micron{} for our survey have been tabulated by MK05.  Ranked in
order of preference, these fluxes are from the large optical galaxy
catalog \citep{rice88}, the IRAS Bright Galaxy Survey
\citep{soifer89}, or the Faint Source Catalog \citep{moshir90}.  IRAS
fluxes for the NFGS have been gathered from the same references.  If
no detection is reported at $12$ or $25$~\micron, we use the empirical
ratios (correct to $\pm30\%$) given in \citet{bell03}:
$S_{\nu}(12~\micron) / S_{\nu}(100~\micron) = 0.0326$ and
$S_{\nu}(25~\micron) / S_{\nu}(60~\micron) = 0.131$.  We confirm the
validity of these ratios for the subset of our sample with detections
in all four IRAS bands.  To extrapolate the infrared spectral energy
distribution beyond $100$~\micron, we construct a modified blackbody
with dust emissivity proportional to $\lambda^{-1}$ \citep{gordon00,
bell03}.  We determine the temperature and normalization of the
modified blackbody using the $S_{\nu}(60~\micron) /
S_{\nu}(100~\micron)$ flux ratio for each object.  Finally, we
integrate numerically between $8-1000$~\micron{} to obtain \lir.  On
average, our \lir{} estimates are a factor of $1.84\pm0.22$ (ranging
from factors of $1.28-3.19$) larger than the far-infrared luminosity,
$\lfir\equiv L(40-120~\micron)$ \citep{helou88}.  We assume a
systematic uncertainty of $15\%$ in \lir{} \citep{bell03}, which we
add in quadrature to the reported IRAS flux uncertainties.

To verify our technique for estimating \lir, which uses flux
measurements in all four IRAS bands, we compare our results against
the methods developed by \citet{dale01}, \citet{dale02}, and
\citet{sanders96}.  \citet{dale01} present a semi-empirical technique
for estimating $L(3-1100~\micron)$ by arguing that the infrared
spectral energy distributions of dusty galaxies can be parameterized
in terms of a single parameter, the $S_{\nu}(60~\micron) /
S_{\nu}(100~\micron)$ ratio, which characterizes the relative level of
star formation activity in galaxies.  Subsequently, \citet{dale02} use
new submillimeter observations to show that their original model
over-predicts the amount of cold dust emission in quiescent galaxies,
and they present an updated semi-empirical relation for deriving
$L(3-1100~\micron)$ based on IRAS fluxes at $25$, $60$, and
$100$~\micron.  Finally, \citet{sanders96} present an empirical
relation for deriving $L(8-1000~\micron)$ based on all four IRAS
bands.  However, because this relation is optimized for luminous- and
ultra-luminous infrared galaxies, it tends to over-predict \lir{} in
galaxies with a larger amount of cold dust \citep{dale01}.  With these
caveats in mind, we find that our \lir{} values are $19\%$, $16\%$,
and $7\%$ smaller, respectively, than the three calibrations described
above.  We attribute part of the difference with the \citet{dale01}
and \citet{dale02} estimates to the different definitions of \lir.
For example, as much as $\sim10\%$ of the total infrared energy budget
in quiescent systems is emitted at $3-8$~\micron{} \citep[their
Table~2]{dale02}.  Because we adopt a systematic uncertainty of $15\%$
in our estimate of \lir, we conclude, therefore, that our results are
consistent with these other methods. 


In Figure~\ref{fig:ha_sfr} we plot the \sfrha/\sfrir{} ratio against
\lb{} for our integrated sample, using the observed \ha{} luminosity
in panel (\emph{a}) and the extinction-corrected \ha{} luminosity in
panel (\emph{b}).  The \emph{solid} line indicates equality of the two
SFR measurements.  Here and throughout much of our analysis we adopt
the $B$-band luminosity as our preferred independent variable.  As we
discuss in \S\ref{sec:calibrations}, the two dominant sources of
scatter in optical SFR diagnostics, dust extinction and metallicity,
correlate strongly with \lb.  Furthermore, \lb{} serves as an
observationally convenient surrogate for stellar mass that is
available for both our local samples and most high-redshift samples.
In panel (\emph{b}) we also plot the distribution of
$S_{\nu}(60~\micron) / S_{\nu}(100~\micron)$ ratios for the MK05 and
NFGS samples, illustrating the higher average level of star formation
activity in the MK05 sample.


To first order we find that \sfrha{} based on the observed \ha{}
luminosity is systematically offset from \sfrir{} by a median (mean)
amount $-0.27$~dex ($-0.26$~dex), and that the scatter in the ratio is
$0.37$~dex, or a factor of $\sim2.3$ (Fig.~\ref{fig:ha_sfr}\emph{a}).
To second order, we observe a strong systematic dependence on \lb{} in
the sense that \sfrha/\sfrir{} decreases by a factor of $\sim100$ over
a factor of $\sim3000$ increase in \lb.  In
Figure~\ref{fig:ha_sfr}\emph{b} we show the effect of correcting \ha{}
for extinction using the observed Balmer decrement.  The median (mean)
systematic difference reduces to $0.00$~dex ($0.02$~dex), and the
scatter decreases to $0.22$~dex.  Moreover, extinction-correcting
\ha{} removes much of the second-order trend seen in panel
(\emph{a}). 

Consequently, we find that correcting \ha{} for extinction using a
simple Milky Way extinction curve and the observed Balmer decrement
gives \ha{} SFRs that are consistent with \sfrir{} with a precision of
$\pm70\%$ and no systematic offset, even in many of the most
dust-obscured galaxies in our sample \citep[see also][]{kewley02b}.
At face value, this agreement may seem surprising, given the simple
assumptions built into equations~(\ref{eq:ha_sfr}) and
(\ref{eq:ir_sfr}).  For example, our \ha{} SFR calibration assumes
that none of the Lyman-continuum radiation from massive stars is
absorbed by dust.  Including this effect, which may be a significant
correction \citep[e.g.,][]{inoue01}, would increase the inferred SFR
at a given \ha{} luminosity.  However, the agreement between \sfrir{}
and the extinction-corrected \sfrha{} suggests that Lyman-continuum
extinction is not significant for our sample.  We note, however, four
infrared-luminous galaxies in the MK05 sample (CGCG239-011~W,
IRAS~17208-0014, NGC~3628, and UGC~09618~N) which fall significantly
below the $\sfrha=\sfrir$ line in Figure~\ref{fig:ha_sfr}\emph{b}.
Assuming \sfrir{} represents the actual SFR in these objects, the
extinction-corrected \sfrha{} would underestimate the SFR by up to an
order-of-magnitude.  In addition, \sfrir{} is susceptible to several
simplifying assumptions which we have neglected in our comparison.
The most important assumption may be that $100\%$ of the bolometric
luminosity associated with the current star formation episode is
absorbed by dust and re-emitted into the infrared.  In fact, at low
luminosity ($\mb\gtrsim-19$~mag), we find that $\sfrha/\sfrir>1$ for
our sample in a systematic sense.  If the extinction-corrected
\sfrha{} reflects the actual SFR in these objects, then \sfrir{} would
under-estimate the SFR by factors of $1.5-5$.

Clearly, all these effects warrant a more in-depth analysis.  However,
our emphasis in this paper is to explore rest-frame optical
emission-line diagnostics.  Consequently, in the remainder of this
paper we adopt the \ha{} luminosity, suitably corrected for stellar
absorption and extinction using the observed Balmer decrement, as our
fiducial SFR tracer, and we assume equation~(\ref{eq:ha_sfr}) is the
appropriate transformation from luminosity to SFR.  







\subsection{\hblam{} Star-Formation Rates}\label{sec:hb_sfr}

Above $z\sim0.4$, \ha{} becomes inaccessible to ground-based optical
spectrographs.  In the search for reliable, self-consistent SFR
measurements at all redshifts, the higher-order Balmer lines such as
\hb{} offer a promising alternative.  The advantages and disadvantages
of using \hb{} to measure the SFR were originally discussed by
\citet{kenn92b}, who discouraged its use due to the difficulty of
accounting for underlying stellar absorption from moderate-resolution
spectroscopy of galaxies with $\ewhanii\lesssim50$~\AA.  However,
population synthesis modeling of the stellar continuum ensures
accurate removal of the absorption underlying the nebular emission
lines \citep[\S\ref{sec:data}; MK05;][]{tremonti04}.  

As a SFR diagnostic, \hb, like all the Balmer lines, inherits the same
strengths and weaknesses of \ha: it is equally sensitive to variations
in the IMF and to absorption of Lyman-continuum photons by dust within
star-forming regions.  In addition, \hb{} suffers more interstellar
dust attenuation and is fractionally more sensitive to underlying
stellar absorption.  For example, assuming that \ha{} experiences one
magnitude of extinction and that stellar absorption is a $20\%$
correction, the observed \hb{} line will be $0.18$ times the strength
of \ha.  Despite these uncertainties, when available, \hb{} may be a
superior SFR diagnostic than the more commonly used \oiilam{} nebular
emission line (\S\ref{sec:oii_sfr}).

In Figure~\ref{fig:lb_hbha} we explore the systematic effects of
stellar absorption and dust reddening on \hb{} as a precision SFR
indicator.  This comparison includes all objects having
$\ewhb>5$~\AA{} in emission.  Below this limiting equivalent width,
\hb{} measurements are very uncertain, even when using population
synthesis to subtract the stellar continuum.  For reference, in the
MK05 and NFGS samples the mean stellar absorption underlying the \hb{}
emission line is $4.4\pm0.6$~\AA{} and $3.9\pm0.5$~\AA, respectively.

Figure~\ref{fig:lb_hbha}\emph{a} plots the observed \hbha{} ratio
without corrections for dust reddening or stellar absorption at \hb.
The \emph{dashed} line in panels (\emph{a}) and (\emph{b}) indicates
the intrinsic Balmer decrement, $\log\,(\hbha)_{\rm int}=-0.46$~dex
(see \S\ref{sec:reddening}).  By definition, galaxies whose \ha{} and
\hb{} fluxes have been corrected for reddening using the \hahb{} ratio
would lie along this line, and \hb{} would mirror \ha{} as a star
formation tracer.  We find that the observed \hbha{} ratio varies
systematically with \lb{} due to the combined effects of stellar
absorption and reddening, and that the median (mean) ratio is
$-0.84$~dex ($-0.86\pm0.22$~dex).  In Figure~\ref{fig:lb_hbha}\emph{b}
we show the effect of correcting \hb{} for underlying stellar
absorption.  The data shift toward the intrinsic ratio in a
luminosity-dependent way because, locally, luminous/massive galaxies
have, on average, lower emission-line equivalent widths.  The
dependence on \lb{} becomes less pronounced, the median (mean) ratio
increases to $-0.53$~dex ($-0.54$~dex), and the scatter reduces to
just $0.07$~dex.  The residual dependence of $(\hbha)_{\rm obs}$ on
\lb{} reflects the luminosity-dust correlation, which we discuss in
more detail in \S\ref{sec:calibrations}.  The low scatter in this
figure is due to the \hahb{} ratio's sensitivity to variations in
reddening.

Finally, in Figure~\ref{fig:lb_hbha}\emph{c} we plot the ratio of the
observed \hb{} luminosity, \lhbobs, corrected for stellar absorption,
to \sfrha, versus \lb.  Below $\lb\lesssim3\times10^{9}$~\lbsun, the
median ratio is nearly constant at $\sim-0.4$~dex, and the scatter
ranges from $0.06$ to $0.13$~dex, or $15-35\%$.  Toward higher
luminosity, the median ratio decreases progressively (reaching a
minimum of $\sim-0.9$~dex), while the scatter increases systematically
from $0.13$~dex near $\sim3\times10^{9}$~\lbsun, to $0.20$~dex
($\pm60\%$) at $\sim10^{11}$~\lbsun.  In \S\ref{sec:hb_results} we
parameterize the observed non-linear dependence of \lhbobs/\sfrha{} on
\lb{} and discuss it in more detail.  Finally, we note that if we do
not correct \hb{} for stellar absorption in
Figure~\ref{fig:lb_hbha}\emph{c}, the trend with luminosity steepens
(since the correction is more important for luminous galaxies), the
median \hb{} SFR conversion factor is $\sim70\%$ lower, and the
scatter increases to $0.36$~dex, or $\pm130\%$.  We find that even
applying a simple statistical correction of $4\pm1$~\AA{} reduces the
scatter in \lhbobs/\sfrha{} to $\sim0.24$~dex, although the systematic
trend with blue luminosity remains.




The SDSS sample provides the opportunity to investigate the same
empirical correlations explored above using more than two
orders-of-magnitude more galaxies.  In Figure~\ref{fig:sdss_lb_hbha}
we plot the \hbha{} ratio as a function of \lb{} for the SDSS.  In
order to accomodate the large number of data points, each panel
displays the logarithm of the number of galaxies in a $100\times100$
square grid.  The trends exhibited by the SDSS galaxies are very
similar to those observed in Figure~\ref{fig:lb_hbha} using the
integrated sample.  Once again, we find that correcting \hb{} for
stellar absorption reduces both the luminosity dependence and the
scatter in the observed \hahb{} ratio (compare
Figs.~\ref{fig:sdss_lb_hbha}\emph{a} and
\ref{fig:sdss_lb_hbha}\emph{b}).  For reference, the mean \hb{}
absorption correction for the SDSS sample is $2.4\pm0.6$~\AA.  The
median $(\hbha)_{\rm obs}$ ratios in
Figures~\ref{fig:sdss_lb_hbha}\emph{a} and
\ref{fig:sdss_lb_hbha}\emph{b} are systematically lower by $0.07$~dex
than the corresponding panels in Figure~\ref{fig:lb_hbha} due to the
higher mean extinction and near-absence of any galaxies less luminous
than $\mb\simeq-18$~mag in the SDSS sample
(Fig.~\ref{fig:sdssproperties}).  Finally,
Figure~\ref{fig:sdss_lb_hbha}\emph{c} shows the transformation between
\lhbobs{} and \sfrha{} as a function of \lb.  The median conversion
factor varies from $-0.65$~dex around $\sim3\times10^{9}$~\lbsun, to
$-0.95$~dex near $\sim10^{11}$~\lbsun, while the dispersion at fixed
luminosity is fairly constant at $\sim0.23$~dex, or $\pm70\%$.

In conclusion, we find that both variations in dust reddening and
stellar absorption limit the precision of \lhbobs{} as a quantitative
SFR indicator.  Correcting \hb{} for underlying stellar absorption
either statistically or using population synthesis modeling reduces
the uncertainty in \sfrhb{} substantially and should not be neglected.
Finally, we find that dust reddening introduces an average error of
$0.1-0.25$~dex in \sfrhb{}, depending on the luminosity distribution
of the sample.  In \S\ref{sec:hb_results} we attempt to improve the
statistical precision of \lhbobs{} as a SFR indicator by
parameterizing $\lhbobs/\sfrha$ in terms of \lb.

\subsection{[O~{\sc ii}]~$\lambda3727$ Star-Formation
  Rates}\label{sec:oii_sfr}  

As we discuss in \S\ref{sec:intro}, the \oiilam{} nebular emission
line is frequently used as a qualitative and quantitative tracer of
star formation in galaxies \citep[e.g.,][]{songaila94, hammer97,
hogg98, hippelein03, teplitz03b}.  SFRs based on \oii, however, are
subject to considerable systematic uncertainties due to variations in
dust reddening, chemical abundance, and ionization among star-forming
galaxies.  In this section we explore these systematic effects using
empirical correlations in order to understand the limitations of
\oii{} as a SFR indicator.

\subsubsection{Variations in Dust Reddening}\label{sec:oii_dust}

We begin our analysis by studying the effects of dust reddening on the
\oiiha{} flux ratio.  In Figure~\ref{fig:lb_oiiha} we plot \oiiha{}
versus \lb{} for the integrated spectroscopic sample.
Figure~\ref{fig:lb_oiiha}\emph{a} shows the observed \oiiha{} ratio,
$(\oiiha)_{\rm obs}$, uncorrected for dust reddening.  As shown by
\citet{jansen01}, the observed ratio varies systematically with
luminosity in the sense that more luminous galaxies have, on average,
lower \oiiha{} ratios.  We find a median (mean) logarithmic ratio of
$-0.17$~dex ($-0.21\pm0.22$~dex), corresponding to a linear ratio
$(\oii/\ha)_{\rm obs}\simeq0.68$ and a scatter of $\pm65\%$.  For
comparison, \citet{kenn92b} obtains a median line-ratio of
$0.45\pm0.26$ based on a smaller sample of normal luminous galaxies.
Figure~\ref{fig:lb_oiiha}\emph{b} shows the reddening-corrected
\oiiha{} ratio, $(\oiiha)_{\rm cor}$, versus \lb.  Correcting both
\oii{} and \ha{} for dust reddening removes almost all the luminosity
dependence seen in panel (\emph{a}), increases the median (mean) ratio
to $0.00$~dex ($-0.01$~dex), and reduces the scatter to $0.12$~dex.
Consequently, we find that even with an optimal reddening correction
(i.e., using \hahb), the scatter in \oiiha{} is $\sim32\%$, which, in
the absence of any other corrections, places a lower limit on the
uncertainty in \oii{} SFRs.  However, if \ha{} were available to make
this dust correction then \ha{} should be used in place of \oii{} to
derive the SFR.  A Spearman rank correlation test on the data in
Figure~\ref{fig:lb_oiiha}\emph{b} yields a correlation coefficient of
$-0.18$; the probability of obtaining this coefficient by chance is
$<1\%$.  In \S\ref{sec:oii_metallicity} we show that metallicity,
which also correlates with luminosity (see \S\ref{sec:calibrations}),
drives this residual correlation.  Finally, in
Figure~\ref{fig:lb_oiiha}\emph{c} we plot the \loiiobs/\sfrha{} ratio
versus \lb.  We find that the median \oii{} SFR transformation varies
systematically with blue luminosity, from $0.02-0.05$~dex below
$3\times10^{9}$~\lbsun, to $-0.67$~dex near $\sim10^{11}$~\lbsun.
This luminosity dependence is driven by the combined effects of
reddening and metallicity, which we discuss in detail below, and
parameterize in \S\ref{sec:oii_results}.  In the absence of any other
information, however, SFRs based on the observed \oii{} luminosity are
susceptible to a $\sim0.39$~dex uncertainty, or a factor of $\sim2.5$.



Figure~\ref{fig:sdss_lb_oiiha} plots the correlation between \oiiha{}
and \lb{} for the SDSS sample.  The mean and median ratios of the SDSS
galaxies are systematically lower than the corresponding ratios for
the integrated sample (Fig.~\ref{fig:lb_oiiha}).  These differences
are partly due to the narrower optical luminosity distribution in the
SDSS sample (Fig.~\ref{fig:sdssproperties}\emph{a}): the
low-luminosity galaxies in the integrated sample act to elevate the
median ratio.  However, even at fixed luminosity the distribution of
\oiiha{} in the SDSS is shifted to lower values.  For example, the
median (mean) $(\oiiha)_{\rm cor}$ ratio around
$(2-5)\times10^{10}~\lbsun$ is $-0.22$~dex ($-0.22\pm0.15$~dex),
compared to $-0.04$~dex ($-0.06\pm0.12$~dex) in the integrated sample.
Finally, comparing Figures~\ref{fig:lb_oiiha}\emph{b} and
\ref{fig:sdss_lb_oiiha}\emph{b}, we find that the reddening correction
applied to the SDSS galaxies is much less effective at removing the
luminosity dependence of the \oiiha{} ratio than for the galaxies with
integrated spectroscopy.  We explore this point next.

In Figure~\ref{fig:ehbha_oiiha} we show the observed \oiiha{} flux
ratio as a function of \ehbha{} for the integrated sample and the
SDSS.  Confirming \citet{jansen01}, we find that $(\oiiha)_{\rm obs}$
correlates tightly with reddening \citep[see
also][]{salamanca02,kewley04}.  In the integrated sample
$(\oiiha)_{\rm obs}$ decreases by an order-of-magnitude over a factor
of $\sim2.5$ change in \ehbha.  Following \citet{jansen01}, we
overplot the predicted reddening at \oii{} and \ha{} for the
\citet{odonnell94} Galactic extinction curve (\emph{solid} line), and
the \citet{charlot00} attenuation curve (\emph{dashed} line).  Other
reddening curves in this wavelength regime \cite[e.g., for the Small
Magellanic Cloud;][]{gordon03} are very similar to the two curves
plotted.  We define the intercept of these curves to be
$\log\,(\oiiha)_{\rm obs}=0.0$~dex for the integrated sample,
corresponding approximately to the intrinsic (de-reddened) flux ratio
of the data (see Fig.~\ref{fig:lb_oiiha}\emph{b}).  The slope of the
relation between $(\oiiha)_{\rm obs}$ and \ehbha{} is remarkably
consistent with the expected reduction in flux between $3727$~\AA{}
and $6563$~\AA{} due to a simple foreground extinction curve.  The
SDSS galaxies, by comparison, exhibit a more complex relationship
between reddening and observed \oiiha{}] ratio
(Fig.~\ref{fig:ehbha_oiiha}, \emph{right}).  The \emph{solid} and
\emph{dashed} lines show the same two extinction curves described
above, this time normalized to $\log\,(\oiiha)_{\rm obs}=-0.16$~dex
(see Fig.~\ref{fig:sdss_lb_oiiha}\emph{c}).  The SDSS sample exhibits
a much larger range in \oiiha{} ratio at fixed reddening.  Within
$\ehbha=0.4\pm0.05$~mag, the total range in $(\oiiha)_{\rm obs}$ spans
a factor of $\sim11$, compared to a factor of $\sim3.5$ in the
integrated sample.  In \S\ref{sec:oii_metallicity} we show that these
differences arise because the reddening-corrected \oiiha{} ratio
correlates with oxygen abundance, which affects the SDSS sample to a
greater extent. 

Before investigating the metallicity sensitivity of \oii, we turn our
attention to an interesting set of outliers in
Figure~\ref{fig:ehbha_oiiha} (\emph{left}).  In the integrated sample
we find galaxies with widely varying \oiiha{} ratios,
$-0.8$~dex~$\lesssim\log\,(\oiiha)_{\rm obs}\lesssim0.1$~dex, but very
little dust reddening, $\ehbha\lesssim0.2$~mag.  These objects are the
low-luminosity, low-metallicity galaxies in our sample, which exhibit
low \oiiha{} ratios because they are deficient in heavy elements such
as oxygen.  Although obtaining high-quality spectroscopy of the
high-redshift counterparts to these low-luminosity objects is very
difficult (see, e.g., \S\ref{sec:applications}), their existence
indicates that any \oii-based SFR calibration must be applied
carefully to chemically unevolved objects, which may be much more
common at high redshift \citep[e.g.,][]{maier05}.  We do not identify
a corresponding branch of extremely low-metallicity galaxies in the
SDSS sample, which is consistent with its luminosity distribution and
the study by \citet{tremonti04}.

\subsubsection{Variations in Oxygen Abundance}\label{sec:oii_metallicity}

In the preceding analysis we have shown that the observed \oiiha{}
ratio correlates very tightly with reddening in the integrated sample
and less well in the SDSS sample.  To investigate this difference and
the origin of the residual scatter, we study how variations in
metallicity affect the \oiiha{} ratio.  One widely used abundance
indicator is the \pagel{} parameter \citep{pagel78}, given by

\begin{equation}
\pagel \equiv \frac{\oiilam+\oiiidoublet}{\hblam},
\label{eq:r23}
\end{equation}

\noindent where each emission line must be corrected for dust
reddening \citep[but see][]{kobulnicky03a}.  Observations of \hii{}
regions and photoionization modeling reveal that \pagel{} depends on
the oxygen abundance, although the relationship is not monotonic
\citep{edmunds84, skillman89b, mcgaugh91, kobulnicky99a, pilyugin00,
pilyugin01, kewley02a}.  In addition, \pagel{} depends on the
ionization parameter, particularly at low metallicity
\citep[e.g.,][]{kobulnicky99a}, and possibly even at high metallicity
\citep{pilyugin01}.  Recent observations have also shown that the
theoretical calibration of \pagel{} is discrepant by $0.2-0.5$~dex
relative to electron-temperature abundance measurements, and may
require revision to lower abundances \citep[J.~Moustakas et~al., 2006,
in preparation;][]{pilyugin01, kenn03, garnett04, bresolin04}.
Unfortunately, we cannot use \pagel{} to probe the metallicity
sensitivity of \oii{} because it depends explicitly on the \oii{}
intensity.  To emphasize this point, we adopt $\hahb=2.86$ as the
de-reddened Balmer decrement and rewrite equation~(\ref{eq:r23}) as

\begin{eqnarray}
\begin{array}{lcl}
\log\,\left(\frac{\oiilam}{\ha}\right) & = & \log\,(\pagel) - \\
 & & \log\,\left(1+\frac{\oiiidoublet}{\oiilam}\right) - \log\,(2.86).
\label{eq:r23rewrite}
\end{array}
\end{eqnarray}

\noindent This equation shows that $\log\,(\oiiha)$ and
$\log\,(\pagel)$ are linearly proportional with a slope of unity and
an intercept that depends weakly on the \oiiioii{} ratio.  Due to this
covariance, therefore, we argue that \pagel{} should not be used to
quantify the \oii{} metallicity dependence.

Fortunately, there are several other strong-line abundance diagnostics
that do not rely on the \oii{} emission-line flux.  One such indicator
is the \onpettini{} ratio \citep{alloin79, pagel79, dutil99,
pettini04}.  A recent empirical calibration between \onpettini{} and
the oxygen abundance is given by \citet{pettini04}:

\begin{equation}
\logoh = 8.73 - 0.32\,\log\,\left(
\frac{\oiiilam/\hblam}{\niilam/\halam} \right).
\label{eq:O3N2OH}
\end{equation}

\noindent Equation~(\ref{eq:O3N2OH}) is only appropriate in the range
$-1.0\lesssim\log\,\{\onpettini\}\lesssim1.9$~dex, or
$8.12\lesssim\logoh<9.05$~dex.  Below $\logoh\simeq8.12$~dex,
therefore, we adopt an empirical calibration based on the
\niilam/\ha{} ratio \citep{pettini04}:

\begin{equation}
\logoh = 8.9 + 0.59 \log\,(\niilam/\ha).
\label{eq:N2OH}
\end{equation}

\noindent Note that equation~(\ref{eq:N2OH}) should not be applied at
high metallicity because \niiha{} converges to an approximately
constant value in star-forming galaxies (see, e.g.,
Fig.~\ref{fig:bpt}).  Due to their small wavelength separation, both
\onpettini{} and \niiha{} are insensitive to dust reddening, and
therefore equations~(\ref{eq:O3N2OH}) and (\ref{eq:N2OH}) can be
applied to the observed emission-line fluxes.

In Figure~\ref{fig:oh12_oiiha} we plot the reddening-corrected
\oiiha{} ratio as a function of oxygen abundance for the integrated
sample, \hii{} regions, and the SDSS sample.  The \hii{} region
comparison is valuable because \hii{} regions span a broader range of
\oiiha{} flux ratios and abundances relative to both our galaxy
samples.  Not unexpectedly, we find that $(\oiiha)_{\rm cor}$ varies
with metallicity.  Heavy elements, particularly oxygen, are an
important source of radiative cooling in \hii{} regions because they
have collisionally excited energy levels that are well-populated at
$\sim10,000$~K \citep{osterbrock89}.  Figure~\ref{fig:oh12_oiiha},
therefore, reflects the behavior of oxygen cooling with metallicity.

In order to characterize the dependence of \oiiha{} on oxygen
abundance we divide Figure~\ref{fig:oh12_oiiha} into a metal-poor
regime, $\logoh<8.15$~dex, an intermediate-metallicity regime,
$8.15<\logoh<8.7$~dex, and a metal-rich regime, $\logoh>8.7$~dex.  In
\S\ref{sec:oii_dust} we identified the most metal-poor galaxies in the
integrated spectroscopic sample as outliers in the correlation between
$(\oiiha)_{\rm obs}$ and dust reddening (Fig.~\ref{fig:ehbha_oiiha},
\emph{left}).  In Figure~\ref{fig:oh12_oiiha} (\emph{left}) we see
that their $(\oiiha)_{\rm cor}$ ratios obey an almost linear
dependence on oxygen abundance, rising by a factor of $\sim5$ over a
factor of $\sim2.5$ increase in metallicity.  In the SDSS there are
only $57$ galaxies in this part of the diagram, all with metallicities
around $\sim8.1$~dex.  In the intermediate-metallicity regime
$(\oiiha)_{\rm cor}$ plateaus to a median (mean) value of $0.04$~dex
($0.03\pm0.07$~dex) in the integrated sample, and to $0.01$~dex
($0.00\pm0.11$~dex) in the SDSS.  In this regime we find $\sim75\%$ of
the integrated spectroscopic sample and $\sim50\%$ of the SDSS
galaxies.  The weak metallicity dependence and small scatter
($\pm15-35\%$) of the reddening-corrected \oiiha{} ratio in the
metallicity interval $\logoh=8.15-8.7$~dex constitutes one of the
principal results of this paper.  Finally, above $\logoh=8.7$~dex we
find that $(\oiiha)_{\rm cor}$ decreases by a factor of $\sim1.6$ in
the integrated sample and by more than a factor of $\sim3$ in the SDSS
over a very short interval in metallicity, $\sim0.2$~dex.  In this
regime the median (mean) ratio of the integrated sample is $-0.13$~dex
($-0.12\pm0.11$~dex) compared with $-0.24$~dex ($-0.24\pm0.13$~dex) in
the SDSS.  We conclude, therefore, that correcting \oiiha{} for
extinction in the SDSS sample (Fig.~\ref{fig:sdss_lb_oiiha}\emph{b})
marginally decreases the scatter because of the steep metallicity
dependence of $(\oiiha)_{\rm cor}$ for $\logoh>8.7$~dex.

To better understand Figure~\ref{fig:oh12_oiiha} we turn to
photoionization models.  Recently, \citet{kewley01b} have studied the
variation of optical emission-line ratios as a function of
metallicity, ionization parameter, ionizing spectral energy
distribution, and star-formation history using a large grid of
photoionization model calculations.  We use these models to plot in
Figure~\ref{fig:oh12_oiiha} the theoretical \oiiha{} ratio as a
function of oxygen abundance for six values of the ionization
parameter, $U\propto (n_{\rm e}f^2 Q)^{1/3}$, where $n_{\rm e}$ is the
electron density, $f$ is the filling fraction, and $Q$ is the rate of
photoionizing photons injected into the gas by massive stars.  These
theoretical curves have been computed using input spectral energy
distributions for evolved star-clusters from {\tt STARBURST~99}
\citep{leitherer99}, which are themselves based on the
metallicity-dependent stellar atmospheres from \citet{lejeune97} and
the
Geneva\footnote{\url{http://obswww.unige.ch/\~{}mowlavi/evol/stev\_database.html}}
stellar evolutionary tracks.  For this comparison we have adopted the
$8$~Myr continuous star-formation history models from
\citet{kewley01b} based on the \citet{salpeter55} IMF, evaluated
between $0.1$ and $120$~\msun.

We find very good correspondence between the observed and the
theoretical metallicity dependence of \oiiha.  In the metal-poor
regime, the models show that both a reduction in the heavy-element
abundance and a hardening of the ionizing radiation field, toward
higher values of $U$, drive the rapid decline in \oiiha.  The plateau
in the intermediate-metallicity regime, and the subsequent decline in
\oiiha{} at high metallicity, is due to the increasing importance of
nebular cooling via the infrared fine-structure lines and a consequent
weakening of the \oii{} line-flux.  Finally, according to the models,
the scatter in \oiiha{} at fixed metallicity arises from differences
in ionization parameter.  For example, around $\logoh=8.4\pm0.1$~dex
the total variation in ionization parameter in the integrated sample
is a factor of $\sim10$ ($-3.8\lesssim \logu \lesssim-2.9$~dex), which
results in the measured $\sim15\%$ scatter in the observed \oiiha{}
ratio at fixed metallicity.

\subsubsection{Summary of Physical Sources of Scatter in
  \oiiha{}}\label{sec:oii_summary}

The empirical correlations shown in
Figures~\ref{fig:lb_oiiha}-\ref{fig:oh12_oiiha} enable us to evaluate
the importance of dust extinction, metallicity, and ionization on the
\loiiobs/\sfr{} ratio.  Figure~\ref{fig:ehbha_oiiha} shows that dust
reddening in the integrated sample accounts for most of the variation
in the observed \oiiha{} ratio, in accord with the findings of
\citet{jansen01}. However, the same is not true for the SDSS.  We
reconcile this discrepancy in Figure~\ref{fig:oh12_oiiha}, where we
plot the reddening-corrected \oiiha{} ratio as a function of
metallicity.  Both the integrated and SDSS samples map out similar
sequences, which agree well with the theoretical models by
\citet{kewley01b}.  We find that most of the integrated sample
galaxies have metallicities in the range $\logoh=8.15-8.7$~dex, where
the metallicity dependence of the \oiiha{} ratio is weak, while the
majority of SDSS galaxies have $\logoh>8.7$~dex, where the correlation
becomes steep.  Thus, for the SDSS sample, metallicity dominates over
reddening as a source of the scatter in the \oiiha{} ratio, whereas it
is only of minor significance for the less metal-rich integrated
sample.  We note that galaxies occupy a fairly narrow range in
ionization parameter, so this contributes minimally to the observed
spread in \oiiha.  From a practical standpoint, we are unlikely to
measure accurately metallicity, ionization, and dust attenuation at
high redshift.  Therefore, in \S\ref{sec:oii_results} we focus on the
empirical correlation between \loiiobs/\sfr{} and \lb.  The well-known
correlations between luminosity, metallicity, and reddening (see
\S\ref{sec:calibrations}), and the wide availability of \lb{}
measurements of high-redshift galaxies, motivate this choice.

\subsection{[O~{\sc iii}]~$\lambda5007$ Star-Formation
  Rates}\label{sec:oiii_sfr}

\citet{kenn92b} investigated the \oiiilam{} nebular emission line as a
quantitative SFR diagnostic and concluded that the large dispersion in
the \oiiiha{} ratio among star-forming galaxies precluded its
suitability for lookback studies.  Nevertheless, the observed \oiii{}
luminosity, \loiiiobs, has been used in the literature to estimate a
crude SFR \citep[e.g.,][]{teplitz00b}.  Therefore, we use our
integrated and SDSS spectra of star-forming galaxies to assess the
systematic uncertainties in \oiii-based SFRs.

In Figure~\ref{fig:lb_oiiiha} we plot the \loiiiobs/\sfrha{} ratio
versus \lb{} for the integrated sample and the SDSS sample.  The
median (mean) $\loiii_{\rm obs}/\sfrha$ ratio in the integrated sample
is $-0.65$~dex ($-0.68\pm0.58$~dex), or $-0.95$~dex
($-0.94\pm0.52$~dex) if we only consider the most luminous galaxies in
our sample ($\mb\lesssim-18.3$~mag).  In the SDSS the median (mean)
ratio is lower, $-1.21$~dex ($-1.16\pm0.42$~dex), which is consistent
with the higher average metallicity of the SDSS galaxies.  Variations
in excitation and chemical abundance dominate over reddening as
physical sources of scatter in the observed $\loiii_{\rm obs}/\sfrha$
ratio.  For example, correcting \oiii{} for reddening reduces the SFR
uncertainty by $\lesssim15\%$.  With no additional corrections,
therefore, converting the observed \oiii{} luminosity into a SFR is
subject to a factor of $3-4$ uncertainty ($1\sigma$), considerably
worse than \hb{} (\S\ref{sec:hb_sfr}), \oii{} (\S\ref{sec:oii_sfr}),
or even the $U$-band luminosity (\S\ref{sec:uband_sfr}).

\subsection{$U$-Band Star-Formation Rates}\label{sec:uband_sfr} 

In the preceding analysis we have focused exclusively on empirically
calibrating nebular emission lines as quantitative SFR diagnostics.
Here, we consider the $U$-band luminosity ($\lambda_{\rm
eff}\simeq3600$~\AA) as a SFR indicator.  Our objective is to provide
a fiducial for evaluating the efficiency of spectroscopy as a means of
deriving galaxy SFRs.  Observationally, the $U$-band continuum is
accessible to ground-based spectroscopy at similar redshifts as
\oiilam{} (Fig.~\ref{fig:zlambda}).  The primary advantage of the
$U$-band over the more commonly used far-UV luminosity
\citep[$\lambda\lambda1250-2500$~\AA;][]{kenn98} as a SFR tracer is
that it suffers less dust attenuation.  However, the $U$-band
luminosity also depends on the recent star-formation history of the
galaxy, and its sensitivity as a SFR indicator may be compromised by
evolved ($\gtrsim100$~Myr) stellar populations \citep{kenn98, cram98,
hopkins03}.

In Figure~\ref{fig:D4000_Uha} we plot the $U$-band-to-\ha{} luminosity
ratio, \lulha, versus \dnbreak, the amplitude of the $4000$-\AA{}
break \citep{balogh99, kauffmann03a} for the integrated spectroscopic
sample.  The $4000$-\AA{} break characterizes the luminosity-weighted
age of the stellar population, is relatively insensitive to dust
reddening, and depends weakly on stellar metallicity only at ages
$\gtrsim1$~Gyr \citep{bruzual83, kauffmann03a}.  Here, we utilize
\dnbreak{} to characterize the effect of intermediate-age and older
stellar populations on $U$-band SFRs.  To ensure that \lu{} and \lha{}
originate from the same physical region, we synthesize the $U$-band
magnitude directly from our spectra using the \citet{bessell90}
$U$-band filter curve and the \citet{lejeune97} theoretical spectrum
of Vega tied to the \citet{hayes85} Vega zero-point.  Due to the blue
wavelength cutoff of the integrated spectra ($\sim3600$~\AA), we use
the best-fitting stellar continuum (\S\ref{sec:sample}) to extrapolate
the data blueward to $\sim3000$~\AA.  The error in this extrapolation
is significantly smaller than physical sources of scatter in the
\lulha{} ratio.  We adopt a fixed $15\%$ uncertainty in our
synthesized $U$-band magnitudes.  Finally, we correct \dnbreak{} for
nebular emission-line contamination from \oiilam, \neonlam, and the
high-order Balmer lines, but not the $U$-band magnitudes.

Figure~\ref{fig:D4000_Uha}\emph{a} plots the observed \lulha{}
luminosity ratio for our integrated sample.  We find that
$[\lulha]_{\rm obs}$ increases systematically by a factor of $\sim2$
between $\dnbreak=1.1$ and $1.5$.  We interpret the observed trend as
a sequence in star-formation history, from continuous star formation
at low values of \lulha{} and \dnbreak, to an increasing fraction of
old-to-young stellar populations at large values of \lulha{} and
\dnbreak.  At fixed \dnbreak, the typical scatter in $[\lulha]_{\rm
obs}$ is $0.18$~dex, which we attribute to differential reddening.  In
addition, observations suggest that the stellar continuum experiences
a fraction of the dust attenuation suffered by star-forming regions
\citep[MK05;][]{calzetti94, kenn94, mayya96, calzetti97b, charlot00,
poggianti00, poggianti01, stasinska01, zaritsky02}.  Therefore,
variations in this fraction among the galaxies in our sample
contributes additional scatter to the observed \lulha{} ratio.
Finally, we note that the median (mean) $[\lulha]_{\rm obs}$ ratio for
the integrated sample is $1.99$~dex ($1.98\pm0.24$~dex).

In Figure~\ref{fig:D4000_Uha}\emph{b} we plot the $\lu_{\rm
obs}/\sfrha$ ratio versus \dnbreak.  Converting \ha{} to an
extinction-corrected SFR flattens the observed trend with \dnbreak{}
and increases the typical dispersion to $0.33$~dex.  A Spearman rank
test yields a correlation coefficient of $0.09$ for these data.  The
probability of obtaining this correlation by chance is $6\%$,
indicating that variations in star-formation history among the
galaxies in our sample contribute systematically to the scatter in
$\lu_{\rm obs}/\sfrha$, particularly for $\dnbreak\gtrsim1.4$.
However, at fixed \dnbreak, the typical scatter is $0.31$~dex, which
accounts for nearly all of the measured dispersion in $\lu_{\rm
obs}/\sfrha$.  We conclude, therefore, that differences in reddening
among the galaxies in our sample dominate the factor of $\sim2.1$
uncertainty in converting the observed $U$-band luminosity into a
SFR.  


Next, in Figure~\ref{fig:sdss_D4000_Uha} we plot the \lulha{}
luminosity ratio as a function of \dnbreak{} for the SDSS sample.  As
before, we correct \dnbreak{} for emission-line contamination, but not
the $U$-band magnitude.  We estimate the $U$-band magnitude (on the
Vega system) from the SDSS $u_{\rm AB}$-band fiber magnitude
\citep{abazajian04} and $(u-g)_{\rm AB}$ color according to the
following relation (M.~R.~Blanton et~al. 2006, in preparation):

\begin{equation}
U_{\rm Vega} = u_{\rm AB} - 0.0140\,(u-g)_{\rm AB} - 0.841, 
\label{eq:U}
\end{equation}

\noindent where the SDSS magnitudes have been corrected for foreground
Galactic extinction and k-corrected to $z=0$ as in
\S\ref{sec:sdsssample}.  In this equation the standard deviation of
the color term is $0.25$~mag, which we add in quadrature to the formal
photometric uncertainties.  

Using our SDSS sample we confirm in Figure~\ref{fig:sdss_D4000_Uha}
the general trends in \lulha{} versus \dnbreak{} seen in the
integrated galaxy sample.  Figure~\ref{fig:sdss_D4000_Uha}\emph{b}
demonstrates that variations in dust reddening effectively scramble
the observed correlation between $[\lulha]_{\rm obs}$ and \dnbreak{}
(Fig.~\ref{fig:sdss_D4000_Uha}\emph{a}), except at large values of
\dnbreak{} where an increasing contribution from evolved stellar
populations begins to dominate the scatter arising from reddening
variations.  Neglecting this effect, however, we find that the
conversion from \luobs{} to a SFR in the SDSS sample is approximately
constant, and susceptible to an uncertainty of $\sim0.3$~dex, or a
factor of $\sim2$, comparable to the SFR uncertainty based on the
integrated sample.


\section{RESULTS}\label{sec:results}

\subsection{Empirical Star-Formation Rate
  Calibrations}\label{sec:calibrations} 

In the preceeding analysis we have compared the \hblam, \oiilam,
\oiiilam, and $U$-band luminosities against the extinction-corrected
\ha{} luminosity to evaluate their suitability as quantitative SFR
diagnostics.  We find that dust reddening affects all four diagnostics
at some level; it is the dominant source of uncertainty for \hb{}
(corrected for stellar absorption) and the $U$-band.  We find that
variations in reddening, oxygen abundance, and ionization parameter
generate significant scatter in \oii{} as a SFR indicator, although
reddening is the dominant effect, at least in the integrated sample.
Finally, we find that SFRs based on the \oiiilam{} line are
susceptible to uncertainties that are factors of $2-4$ higher than any
of the other diagnostics considered due to the extreme sensitivity of
the \oiiilam/\ha{} ratio on oxygen abundance and excitation.
Consequently we do not discuss \oiiilam{} as a quantitative SFR
indicator any further.

In our analysis we have relied on two highly complementary samples: an
integrated spectroscopic sample of $412$ galaxies, and $120,846$
galaxies drawn from the SDSS.  The integrated sample covers the
broadest possible range of galaxy luminosity, metallicity, and dust
extinction to provide the greatest lever-arm on systematic trends in
SFR indicators with these properties. The SDSS galaxies provide a
near-complete magnitude-limited sample which is ideal for evaluating
the intrinsic dispersion in SFR indicators as a function of galaxy
properties. Figures~\ref{fig:lb_hbha}-\ref{fig:sdss_D4000_Uha} show
qualitative agreement between the two samples in the systematic trends
in SFR diagnostics with luminosity, extinction, and metallicity.
However a quantitative comparison reveals differences consistent with
the spatial undersampling of the SDSS galaxies.  As highlighted in
\S\ref{sec:sdsssample}, the SDSS fiber-spectra sample the central
regions of galaxies ($\sim25\%$ of the total light) which are dustier
and more metal rich.  The effects of aperture bias are most pronounced
on the \oii{} SFR calibration because of its sensitivity to both
metallicity and dust. For $\mb^{*}$ galaxies the
\loiiobs/\sfrha{} ratio is $\sim2$ times lower in the SDSS than in our
spatially integrated spectra.  The net effect is that calibrations
derived from the SDSS and applied to spatially unresolved spectra of
distant galaxies could be in error by a factor of two.  Consequently,
we elect to use only the integrated sample to develop quantitative SFR
calibrations for high-redshift applications.

To first order, Figures~\ref{fig:lb_hbha}\emph{c},
\ref{fig:lb_oiiha}\emph{c}, and \ref{fig:D4000_Uha}\emph{c} provide
the median $L_{\rm obs}/\sfr$ conversion factor needed to transform,
respectively, the observed \hb, \oii, and $U$-band luminosity into an
estimate of the SFR.  The scatter in the observed ratios is
$70-150\%$.  We have demonstrated that each of these conversions
depends on some combination of metallicity, dust attenuation, and
recent star-formation history.  To create precision SFR calibrations,
it is desirable to account directly for this dependence on physical
properties.  However, from a practical standpoint, measuring these
properties for high-redshift galaxies is very difficult.  To
circumvent this shortcoming as much as possible, we choose to
parameterize our SFR calibrations as a function of $B$-band
luminosity, \lb.  Mounting evidence shows that many galaxy attributes,
including metallicity, dust attenuation, and star formation history,
correlate strongly with stellar mass \citep[e.g.,][]{tremonti04,
kauffmann03b, brinchmann04}.  Blue luminosity is an imperfect
surrogate for stellar mass, because star-forming galaxies can exhibit
a wide range of mass-to-light ratios \citep[e.g.,][]{bell01b}.
However, \lb{} is a direct observable which is available for our local
sample and virtually all intermediate- and high-redshift samples.  In
Figure~\ref{fig:lb_correlations} we plot the observed correlations
between luminosity, dust extinction, and metallicity using the
integrated sample, and in \S\ref{sec:applications} we discuss how
evolutionary changes in these underlying correlations might impact our
locally derived empirical SFR calibrations.

Finally, we consider two other effects on our empirical SFR
calibrations.  At intermediate redshift, the complete set of
emission-line diagnostic diagrams used in \S\ref{sec:sample} to remove
type~2 AGN from our sample generally are unavailable.  Therefore,
since AGN are likely to contaminate optical/UV selected samples of
distant star-forming galaxies, we consider their distribution and
effect on each SFR diagnostic.  Furthermore, observations show that
the characteristic infrared luminosity of star-forming galaxies
increases as a function of redshift \citep[e.g.,][]{flores99, chary01,
hammer05, bell05}.  Consequently, we divide our sample into three
broad categories based on infrared luminosity: $\lir>10^{11}~\lsun$;
$\lir<10^{11}~\lsun$; and galaxies undetected at either $60$ or
$100$~\micron{} with IRAS (see \S\ref{sec:ha_sfr}).  The set of
objects without IRAS detections are either bona-fide dust-poor
galaxies below IRAS's sensitivity limits, or individual galaxies in an
interacting system that are unresolved by the IRAS beam
($5\arcmin\times2\arcmin$ at $60$~\micron).

\subsection{\hblam}\label{sec:hb_results}

Figure~\ref{fig:lb_hbha} and \S\ref{sec:hb_sfr} demonstrate the
importance of accounting for stellar absorption underlying the \hb{}
nebular emission line.  Therefore, in the following section we assume
that \hb{} has been suitably absorption-corrected, either
statistically or using population synthesis modeling of the stellar
continuum \citep[e.g.,][]{savaglio05}.  Furthermore, as in
\S\ref{sec:hb_sfr}, we only consider objects having $\ewhb>5$~\AA{} in
emission to ensure a well-measured \hb{} line-flux.

In Figure~\ref{fig:sfrha_lhb} we plot the \sfr/\lhbobs{} ratio versus
\lb, using various symbols to characterize the infrared luminosity of
each galaxy, and the relative positions of AGN and star-forming
galaxies.  The large open circles with error bars give the median,
$25\%$, and $75\%$ quartile of the \sfr/\lhbobs{} ratio in $0.5$~dex
bins of luminosity.  Including or excluding AGN has no significant
effect on these statistics.  In Table~\ref{table:hb_sfr} we list these
quartiles, and the mean, median, and standard deviation of the
distribution in each luminosity bin.  We recommend interpolating
between bins of \lb{} to obtain the relevant conversion factor from
\lhbobs{} to \sfr.  Minus a constant offset,
Figure~\ref{fig:sfrha_lhb} is equivalent to
Figure~\ref{fig:lb_correlations} (\emph{left}) since we use the
observed \hahb{} ratio to estimate the nebular reddening (see
\S\ref{sec:reddening}).  However, because we are interested in SFRs
rather than reddening values, to generate this figure we correct \ha{}
for extinction, convert it into a SFR using
equation~(\ref{eq:ha_sfr}), and divide by \lhbobs{} to obtain the
conversion factor from \lhbobs{} to \sfr{} as a function of \lb.

We find that the logarithmic \sfr/\lhbobs{} ratio increases
non-linearly with increasing luminosity, and that the distribution of
\sfr/\lhbobs{} at fixed luminosity is highly asymmetric, particularly
above $\sim3\times10^{8}~\lbsun$.  This behavior reflects the average
increase in the dust content of galaxies of increasing
luminosity/mass, and the large variation in dust extinction at fixed
luminosity \citep[e.g.,][]{buat96, wang96}.  Infrared-luminous galaxies
constitute a large fraction of the most significant outliers,
deviating by factors of $1.5-3$ from the median relation.  Finally,
the AGN in our sample tend to be optically luminous, and therefore
also exhibit a large variation in \sfr/\lhbobs, although to first
order these AGN exhibit a similar distribution in \sfr/\lhbobs{} as
our star-forming sample.  However, since an unknown fraction of the
Balmer emission in these objects arises from the nucleus, applying the
median SFR conversion factor listed in Table~\ref{table:hb_sfr} to a
sample of AGN may lead to spurious results.

In Figure~\ref{fig:compare_sfrhb} we compare \sfrhb{} estimated using
two different methods against \sfrha.  In panel (\emph{a}) we correct
\lhbobs{} for extinction using the measured \hbhg{} ratio, and we
apply equation~(\ref{eq:ha_sfr}) assuming $\lhb=\lha/2.86$.  As in
\S\ref{sec:reddening}, we restrict the sample plotted in
Figure~\ref{fig:compare_sfrhb}\emph{a} to objects having
$\ewhb>10$~\AA{} in emission and ${\rm S/N}(\hg) > 7$.  We find that
\sfrhb{} computed in this way over-estimates the true SFR by a median
(mean) amount $0.12$~dex ($0.16$~dex).  The scatter is $0.26$~dex, or
$\pm80\%$.  As shown in Figure~\ref{fig:ebv_compare} and discussed in
\S\ref{sec:reddening}, the measured \hbhg{} ratio in our integrated
sample over-estimates systematically the reddening measured from
\hahb, which leads to the observed systematic in
Figure~\ref{fig:compare_sfrhb}\emph{a}.  In
Figure~\ref{fig:compare_sfrhb}\emph{b} we compare \sfrhb{} derived
using Table~\ref{table:hb_sfr} (see also Fig.~\ref{fig:sfrha_lhb})
against \sfrha.  We find a median (mean) residual of $-0.00$~dex
($-0.03\pm0.15$~dex).  Our empirical calibration, therefore, yields
SFRs based on \lhbobs{} with a precision of $\pm40\%$ ($1\sigma$),
although individual star-forming galaxies deviate from the median
relation by up to a factor of $\sim4$.  In \S\ref{sec:applications} we
test whether these results apply at intermediate redshift.


\subsection{[\ion{O}{2}]~$\lambda3727$}\label{sec:oii_results}

In \S\ref{sec:oii_sfr} we show that converting \loiiobs{} into a SFR
is subject to considerable uncertainty.  The total range in
\sfr/\loiiobs{} spans more than a factor of $\sim50$ with a scatter of
a factor of $\sim2.5$.  We have shown that this scatter arises from
variations in dust reddening, metallicity, and ionization among
star-forming galaxies.  Here we attempt to account for these
systematic effects by deriving an \oii{} SFR calibration that uses
\lb{} as the independent variable.  The observed correlations between
\lb, attenuation, and metallicity (Fig.~\ref{fig:lb_correlations})
motivate our empirical \oii{} SFR calibration.  In
\S\ref{sec:applications} we discuss how evolutionary changes in the
luminosity-dust and luminosity-metallicity correlations at
intermediate redshift may affect our results.

In Figure~\ref{fig:sfrha_loii} we plot the \sfr/\loiiobs{} ratio
versus \lb{} for the AGN and star-forming galaxies in our sample.  The
large open circles with error bars indicate the median \sfr/\loiiobs{}
ratio in $0.5$~dex wide luminosity bins and the $25\%$ and $75\%$
quartiles.  In Table~\ref{table:oii_sfr} we give these statistics, as
well as the mean and standard deviation of the data in each bin for
the star-forming galaxies.  Given a $B$-band luminosity and an
observed \oii{} luminosity we recommend interpolating the column which
gives the median conversion factor to estimate the SFR.  Below
$\sim10^{9}$~\lbsun{} ($\mb\gtrsim-17$~mag), where reddening effects
are negligible, the metallicity distribution of these objects
determines the median \sfr/\loiiobs{} ratio, as we discuss in
\S\ref{sec:oii_metallicity}.  With increasing luminosity, the
\sfr/\loiiobs{} ratio increases monotonically because of the combined
effects of extinction of metallicity.  However, as
Figure~\ref{fig:oh12_oiiha} (\emph{left}) demonstrates, metallicity
effects only become important above solar metallicity,
$\logoh\simeq8.7$~dex.  Therefore, variations in dust reddening
primarily drive this trend in luminosity.  Most of the AGN in our
sample overlap the sequence traced by star-forming galaxies, although
several infrared-luminous AGN deviate by factors of $5-15$ from the
median relation.

In Figure~\ref{fig:compare_sfroii} we derive \sfroii{} using four
different techniques and compare the results against \sfrha.  For this
comparison we ensure that the Salpeter IMF defined in
\S\ref{sec:ha_sfr} is used throughout.  In
Figure~\ref{fig:compare_sfroii}\emph{a} we compare \sfrha{} against
the widely used \citet{kenn98} \oii{} SFR calibration, which is an
average of the calibrations published by \citet{gallagher89} and
\citet{kenn92b}.  We find that the \citet{kenn98} calibration
understimates \sfroii{} by a median (mean) $-0.13$~dex ($-0.20$~dex),
and that the scatter in \sfroii/\sfrha{} is $0.39$~dex, or a factor of
$\sim2.5$.  The residual dependence on \sfrha{} occurs because
\citet{kenn98} adopt a mean \oiiha{} ratio measured from the
\citet{kenn92b} sample, and a constant $\aha=1$~mag to derive their
\sfroii{} calibration.  Our analysis has shown, however, that these
are poor approximations given the wide variation in extinction
properties among star-forming galaxies.

Next we attempt to correct \loiiobs{} for dust extinction using the
\hbhg{} decrement.  This panel only includes galaxies satisfying
$\ewhb>10$~\AA{} in emission and ${\rm S/N}(\hg) > 7$ (see
\ref{sec:reddening}).  To obtain an estimate of \sfroii, we assume a
fixed intrinsic flux ratio $\oii/\ha=1$, and apply
equation~(\ref{eq:ha_sfr}).  Figure~\ref{fig:compare_sfroii}\emph{b}
compares \sfroii{} derived this way against \sfrha.  We find that
\sfroii{} over-estimates \sfrha{} by a median (mean) $0.12$~dex
($0.19$~dex), with a scatter of $0.41$~dex, or a factor of $2.5$.  The
systematic offset is a direct consequence of the fact that for our
sample the \hbhg{} ratio tends to over-estimate the reddening derived
using \hahb.  To remove the systematic we would have to adopt an
intrinsic \oii/\ha{} ratio of $1.3$, which is a poor assumption, on
average (e.g., Fig.~\ref{fig:oh12_oiiha}, \emph{left}).


In Figure~\ref{fig:compare_sfroii}\emph{c} we estimate \sfroii{} using
the methodology developed by \citet{kewley04}, and plot the results
against \sfrha.  For this comparison we follow the recommended,
six-step method outlined in \S6.1 of their paper.  First, we derive
the reddening-corrected \oii{} luminosity using their equation~(18),
and apply their equation~(16) to estimate \ebv.  Next, we compute the
reddening-corrected \pagel{} ratio and apply it to their equation~(11)
to determine the oxygen abundance, \logoh.  \citet{kewley04} recommend
the \citet{zaritsky94} \pagel{} abundance calibration.  Finally, armed
with an estimate of the metallicity for each galaxy, we apply their
equation~(10) to obtain the reddening- and abundance-corrected
\sfroii.  The results of our comparison do not change if we instead
utilize their theoretical \sfroii{} calibration [their equation~(15)].
We find a median (mean) residual of $0.10$~dex ($0.08$~dex) and a
scatter of $0.33$~dex between their \sfroii{} and \sfrha.  This
technique removes much of the systematic dependence on \sfrha{} seen
in panel~(\emph{a}) because they use a luminosity-dependent extinction
correction.  However, as these authors emphasize, the applicability of
this empirical reddening correlation has not been tested on
high-redshift samples.  Finally, we note that their methodology relies
on measuring at least three nebular emission lines, \oii, \oiii, and
\hb.  In this case, we recommend using \hb{} to estimate the SFR since
it provides a more reliable measurement of the SFR than \oii{} (see
\S\ref{sec:hb_results}).

Finally, Figure~\ref{fig:compare_sfroii}\emph{d} compares \sfrha{}
against our empirical calibration of \sfroii{} using the median
conversion factors in Table~\ref{table:oii_sfr} (see also
Fig.~\ref{fig:sfrha_loii}).  We find a median (mean) residual
systematic of $0.02$~dex ($-0.02$~dex) and a scatter of $0.28$~dex, or
$\pm90\%$.  Although this method for deriving \sfroii{} should be
applied carefully to individual galaxies, which deviate by up to a
factor $\sim10$ from the mean relation, our calibration should be
useful for computing the statistical star formation properties of
sufficiently large galaxy samples.

\subsection{$U$-band}\label{sec:uband_results}

In \S\ref{sec:uband_sfr} we studied the relative effects of dust
reddening and the presence of evolved stellar populations on the
$U$-band luminosity as a quantitative SFR indicator.  We found that
reddening variations dominate the systematic uncertainty in \luobs{}
as a SFR indicator across a wide range of star-formation histories and
luminosity-weighted ages, as characterized by the $4000$-\AA{} break.
From the median and standard deviation of the data in
Figure~\ref{fig:D4000_Uha}\emph{b} we derive the following $U$-band
SFR calibration:

\begin{equation}
\sfruband = (1.4\pm1.1)\times10^{-43}\ \frac{\luobs}{\lunits}\ \ \
\sfrunits.
\label{eq:uband_sfr}
\end{equation}

\noindent This calibration is difficult to compare with previous
results because for the first time it has been derived empirically
using the extinction-corrected \ha{} luminosity.
Equation~(\ref{eq:uband_sfr}) implicitly includes the effects of dust
attenuation so that it can be applied to high-redshift samples without
spectroscopic information.  \citet{cram98} derive an approximate
$U$-band SFR calibration by bootstrapping the far-UV ($2500$~\AA)
calibration from \citet{cowie97} to $\sim3600$~\AA{} neglecting dust
reddening and assuming extended, continuous star formation.  This
extrapolation is subject to considerable uncertainty given the
potentially large variation in star-formation histories among
galaxies, and the significant effects of dust reddening.
\citet{hopkins03} present an empirical $u$-band SFR calibration for
the SDSS, roughly corrected for aperture bias, using the radio
($1.4$~GHz) luminosity as their fiducial SFR tracer \citep{bell03}.
However, this calibration also requires $u$-band luminosities that
have been corrected for dust attenuation.

Although we experimented with a second-order $U$-band SFR calibration
parameterized in terms of \lb, we conclude that the significant
scatter in \sfr/\luobs{} does not justify such a calibration.
However, for high-redshift surveys that target the bright end of the
optical luminosity function, we provide the median transformation from
\luobs{} to SFR for galaxies brighter than
$\sim3\times10^{9}$~\lbsun{} ($\mb\lesssim-18.3$~mag):

\begin{equation}
\sfruband = (1.8\pm1.0)\times10^{-43}\ \frac{\luobs}{\lunits}\ \ \
\sfrunits.
\label{eq:uband_sfr_bright}
\end{equation}

\section{APPLICATIONS AT INTERMEDIATE
  REDSHIFT}\label{sec:applications}  

In this paper we have developed a set of empirical SFR calibrations
for the \hb{} and \oii{} nebular emission lines, and for the $U$-band
luminosity.  However, these calibrations are built on empirical
correlations obeyed by star-forming galaxies in the local universe
(e.g., Fig.~\ref{fig:lb_correlations}) that must be tested at
intermediate and high redshift.  For example, the
luminosity-metallicity correlation has been shown to evolve in a
mass-dependent way by $z\sim1$ \citep[J.~Moustakas et~al., 2006, in
preparation;][]{lilly03, kobulnicky03b, kobulnicky04, liang04a,
maier04, maier05, savaglio05, shapley05}.  Therefore, it is crucial to
quantify how evolutionary changes in galaxy properties as a function
of redshift might affect our locally derived SFR calibrations.

Testing our local calibrations requires a complete set of optical
emission-line measurements, including \ha, \hb, \oiii, and \oii, of a
statistically significant sample of intermediate-redshfit galaxies
spanning a broad range of luminosity or stellar mass.  In practice,
obtaining all these nebular diagnostics for individual high-redshift
galaxies is extremely challenging.  Nevertheless, several
intermediate-redshift samples recently have become available that
facilitate some simple comparisons.  Where necessary, we convert
absolute magnitudes to our adopted cosmology and onto the Vega system
assuming $B_{\rm AB}-B_{\rm Vega}=-0.09$~mag (M.~R.~Blanton
et~al. 2006, in preparation).

First we assess whether the luminosity-dust correlation holds at
intermediate redshift by plotting the observed \hahb{} ratio versus
\lb{} in Figure~\ref{fig:highz_hahb}.  For this comparison we use
different symbols to distinguish starburst galaxies from AGN in our
local sample.  We compile the relevant intermediate-redshift flux
measurements from \citet{maier05}, \citet{shapley05},
\citet{savaglio05}, and \citet{liang04a} for star-forming galaxies at
a median redshift $0.7$, $1.4$, $0.9$, and $0.7$, respectively.
\citet{maier05} and \citet{shapley05} measure \ha{} and \hb{} in their
respective intermediate-redshift samples, while \citet{savaglio05} and
\citet{liang04a} only measure \hb{} and \hg.  Therefore, to include
the data from the latter two samples in this comparison, we use the
observed \hbhg{} ratio to predict the reddened \hahb{} ratio assuming
case~B recombination and the \citet{odonnell94} extinction curve.  The
\emph{solid} line in Figure~\ref{fig:highz_hahb} indicates the
intrinsic \hahb{} ratio.  The \hahb{} ratio of every object below this
line is unphysical, either due to measurement error or uncertainties
in the stellar absorption correction.  Stellar absorption corrections
to the Balmer lines of the intermediate-redshift galaxies have been
applied either statistically, or are negligible due to large
emission-line equivalent widths.

Overall, we find qualitative agreement between the local and the
intermediate-redshift luminosity-dust relation.  As expected, most of
the intermediate-redshift galaxies overlap the bright end of the local
relation, where galaxies experience the highest dust attenuation, on
average.  Several of the \citet{maier05} galaxies and two of the
objects from \citet{shapley05} have \hahb{} ratios that are consistent
with zero reddening.  We find, however, that nearly half the
\citet{liang04a} sample and several galaxies from the other samples
have $\log\,(\hahb)_{\rm obs}\gtrsim0.8$~dex.  The amount of reddening
implied by this Balmer decrement, $\ebv\gtrsim0.8$~mag, assuming the
optical Milky Way extinction curve applies at intermediate redshift,
is larger than any value measured in our local sample of star-forming
galaxies.  We postulate three possibilities to explain the position of
these galaxies relative to the local sample: (1) they are AGN; (2)
measurement uncertainties; or (3) a substantial fraction of
intermediate-redshift star-forming galaxies are dustier than the most
highly obscured nearby galaxies in our sample.

To explore the possibility that these objects host AGN we note the
position of one highly obscured AGN in the MK05 sample, IC~1623~B.
The reddening implied by this object's \hahb{} ratio,
$\ebv\simeq1.5$~mag, is larger than any of the intermediate-redshift
galaxies.  Although we do not have a measurement of IC~1623~B's
infrared luminosity, from its highly reddened optical continuum (MK05)
and large Balmer decrement we infer that it must be significant.
Therefore, the intermediate-redshift galaxies with the largest \hahb{}
ratios may be highly obscured galaxies hosting an AGN.  On the basis
of their \niiha{} ratios and the \citet{kewley01a} classification
scheme (e.g., Fig.~\ref{fig:bpt}), all the objects studied by
\citet{maier05} and \citet{shapley05} are star-forming galaxies.
However, adopting the \citet{kauffmann03c} empirical boundary we would
re-classify as an AGN one of the \citet{maier05} galaxies with
$\log\,(\hahb)_{\rm obs}>0.8$~dex.  An additional $5$ galaxies from
\citet{maier05} and $2$ from \citet{shapley05} with
$\log\,(\hahb)_{\rm obs}<0.8$~dex are AGN on the basis of the
\citet{kauffmann03c} classification scheme.  Consequently, without
additional \niiha{} measurements for these samples we cannot exclude
the hypothesis that some of these objects are AGN.

Alternatively, measurement errors may explain the position of at least
some of the intermediate-redshift galaxies with Balmer decrements in
excess of what we observe in our $z\sim0$ sample.  The Balmer
decrements of the \citet{savaglio05} and \citet{liang04a} galaxies, in
particular, are consistent (within the errors) with having been drawn
from the distribution of local galaxies.  However, all four
intermediate-redshift samples have at least one galaxy with
$\log\,(\hahb)_{\rm obs}>0.8$~dex.  Therefore, we conclude that
measurement error cannot explain the position of \emph{all} the
intermediate-redshift galaxies on this diagram.

Finally, we consider that intermediate-redshift star-forming galaxies
may experience more dust obscuration than any of the galaxies in our
local galaxy sample.  For example, $75\%$ of the galaxies from
\citet{liang04a} have $\lir>10^{11}~\lsun$ because they were selected
on the basis of their $15$~\micron{} flux.  We note, however, that the
MK05 sample includes $\sim100$ infrared-selected galaxies and a
handful of ultra-luminous infrared galaxies \citep[ULIRGs;
$\lir>10^{12}~L_{\sun}$;][]{sanders96}.  Adopting a standard Milky Way
extinction curve, the Balmer decrements of the most obscured
intermediate-redshift galaxies imply $V$-band optical depths of $2-4$.
For comparison, a $V$-band optical depth of unity corresponds to
$\hahb=4.1$ \citep{odonnell94}.  Different dust geometries other than
the assumed screen model \citep[e.g.,][]{calzetti01} would
underestimate the true extinction since the observed nebular emission
lines are always weighted toward the lines-of-sight with the lowest
obscuration \citep[e.g.,][]{witt92, calzetti94, witt00}.  Finally,
from a multi-wavelength analysis of distant galaxies, \citet{flores99}
estimate that the global extinction at $z<1$ is in the range
$\ebv=0.15-0.3$~mag, and that highly obscured star-forming galaxies
constitute $<1\%$ of galaxies up to $z=1$.  If, however, high-redshift
star-forming galaxies typically suffer $\ebv>0.8$~mag of dust
reddening, then the empirical \hb{} SFR calibration developed in
\S\ref{sec:hb_results} would systematically underestimate the true SFR
by a factor of $1.5-3$.  In summary, some combination of the
possibilities outlined above may result in the displacement of
intermediate-redshift galaxies to larger \hahb{} ratios than our local
sample.  A more detailed comparison awaits \ha, \nii, and \hb{}
measurements for a larger number of intermediate- and high-redshift
galaxies.

Next we test the evolution in the observed \oiiha{} ratio as a
function of \lb{} between $z\sim0$ and $z\sim1$.  In
Figure~\ref{fig:highz_oiiha} we plot measurements presented by
\citet{maier05}, \citet{tresse02}, \citet{hicks02}, and
\citet{glazebrook99} for galaxies at median redshifts $0.7$, $0.7$,
$1.1$, and $0.8$, respectively.  For the \citet{glazebrook99} sample
we obtain \lb{} from \citet{hammer97}.  Based on this comparison we
find that the local \oiiha-\lb{} relation for star-forming galaxies
qualitatively holds at intermediate redshift.  Nearly all of the
intermediate-redshift points overlap with the bright end of the nearby
galaxy \oiiha{} sequence.  Several of the \citet{maier05} galaxies and
one of the objects from \citet{tresse02} have $(\oiiha)_{\rm obs}>1$,
which is higher than any nearby star-forming galaxy at the same
luminosity.  The position of these galaxies contradicts the
expectation that they are either highly extincted, or have extremely
low or high oxygen abundances (Fig.~\ref{fig:oh12_oiiha}).  However,
several AGN in our local sample exhibit similarly enhanced \oiiha{}
ratios, which suggests qualitatively that these intermediate-redshift
galaxies harbor an active nucleus.  At the other extreme, we find a
handful of objects with significantly lower \oiiha{} ratios than we
observe in our local sample.  The most severely discrepant points are
from the study by \citet{hicks02} of \ha-selected galaxies
\citep{mccarthy99} at $z=0.8-1.5$.  \citet{hicks02} discuss in detail
the displacement of their galaxies to lower \oiiha{} ratios than
observed locally, and conclude that they must be experiencing up to
$\ebv=0.6$~mag more reddening.  Alternatively, these objects may be
metal-rich (${\rm Z}>\zsun$) and have ionization parameters that are
an order-of-magnitude higher than nearby star-forming galaxies of
comparable luminosity (e.g., Fig.~\ref{fig:oh12_oiiha}).  If the low
\oiiha{} ratios exhibited by the \citet{hicks02} sample are typical
for intermediate- and high-redshift galaxies then the empirical SFR
calibration presented in \S\ref{sec:oii_results} may systematically
underestimate the true SFR by a factor of $3-10$.

In Figure~\ref{fig:highz_oiiioii} we compare the ionization properties
of nearby and intermediate-redshift galaxies by plotting the observed
\oiiioii{} ratio versus \lb.  We take the observed oxygen
emission-line measurements for galaxies at median redshifts of $0.7$,
$0.8$, $0.7$, and $0.7$ from \citet{maier05}, \citet{savaglio05},
\citet{liang04a}, and \citet{lilly03}, respectively.  We note that
$\ebv=0.3$~mag of reddening increases an object's observed \oiiioii{}
ratio by $0.14$~dex on this plot.  To indicate roughly how \oiiioii{}
translates into a measure of the ionization parameter, \logu, we use
the \citet{kewley01b} photoionization models for starburst galaxies
assuming solar metallicity.  We find that the ionization properties of
the intermediate-redshift galaxies are broadly consistent with local
star-forming galaxies.  Several of the intermediate-redshift points
from \citet{liang04a} and \citet{lilly03} exhibit higher ionization
parameters than the typical galaxy in our local sample, although both
AGN contamination and reddening may enhance the observed \oiiioii{}
ratio.  The average ionization parameter of the intermediate-redshift
galaxies may be slightly higher than locally, although at present we
do not consider this difference significant given the systematic
effects of reddening.

To summarize, we find that the excitation, metal abundance, and dust
attenuation properties of our local sample of star-forming galaxies
are broadly consistent with galaxies at $z\sim1$, although individual
intermediate-redshift galaxies may deviate significantly from the
median spectral sequences defined locally.  A more in-depth comparison
of the physical properties of high-redshift galaxies awaits
measurements of the complete set of rest-frame optical emission-line
diagnostics, from \ha{} to \oii, for larger samples of distant
galaxies.

\section{CONCLUSIONS}\label{sec:conclusions} 

We have used integrated optical spectrophotometry for $412$
star-forming galaxies at $z\sim0$, and fiber-aperture
spectrophotometry for $120,846$ SDSS galaxies at $z\sim0.1$, to
investigate the \halam, \hblam, \oiilam, and \oiiilam{} nebular
emission lines and the $U$-band luminosity as quantitative SFR
indicators.  The integrated sample enables us to study SFR diagnostics
across the broadest possible range in optical/infrared luminosity,
metallicity, dust extinction, and excitation.  A comparison dataset
drawn from the SDSS provides a near-complete magnitude-limited sample
which is ideal for evaluating the intrinsic dispersion in SFR
indicators.

We find that the \ha{} luminosity, when corrected for dust extinction
using the \hahb{} Balmer decrement and the O'Donnell Milky Way
extinction curve, reliably measures the SFR even in highly obscured
infrared galaxies.  The precisions of the observed \hb, \oii, and
$U$-band luminosities as SFR diagnostics are limited to factors of
$\sim1.7$, $\sim2.5$, and $\sim2.1$, respectively, because of
variations in dust reddening among star-forming galaxies.  Correcting
\hb{} for underlying stellar absorption, even statistically, improves
its precision as a SFR indicator significantly.  The
reddening-corrected \oiiha{} ratio depends weakly on changes in
metallicity over a wide range in oxygen abundance,
$\logoh=8.15-8.7$~dex ($\zzsun=0.3-1.0$).  In this metallicity
interval we find that galaxies occupy a narrow range in ionization
parameter ($-3.8\lesssim\logu\lesssim-2.9$~dex).  By contrast, below
$\logoh=8.15$~dex and above $\logoh=8.7$~dex, this ratio depends
steeply on oxygen abundance.  We find that the scatter in \oiiilam{}
as a SFR indicator is a factor of $3-4$ due to its sensitivity to
oxygen abundance and ionization, considerably worse than any of the
other empirical diagnostics considered.  Finally, our analysis reveals
that dust reddening dominates the systematic uncertainty in the $U$-band
luminosity as a SFR tracer over variations in star-formation history
among the galaxies in our sample.

Through a quantitative comparison of optical emission-line diagnostics
from integrated and fiber-aperture spectrophotometry, we conclude that
SFR calibrations derived from SDSS observations and applied to
spatially unresolved spectra of distant galaxies are susceptible to a
factor of $\sim2$ systematic uncertainty due to spatial undersampling
(aperture bias) in the SDSS spectrophotometry.  We develop empirical
SFR calibrations for \hb{} and \oii{} parameterized in terms of the
$B$-band luminosity, motivated by the observed correlations between
luminosity, dust reddening, and oxygen abundance obeyed by local
galaxies.  These calibrations remove the luminosity-dependent
systematic bias in \hb{} and \oii{} SFRs and reduce the scatter to
$\pm40\%$ and $\pm90\%$, respectively.  However, individual galaxies
may deviate by up to a factor of $\sim4$ and $\sim10$ from the median
\hb{} and \oii{} SFR calibrations, respectively.  Finally, we compare
the relations between luminosity and reddening, ionization, and
\oiiha{} ratio for our local, $z\sim0$ galaxies to those of galaxies
observed at $z\sim1$ and find broad agreement, although some
intermediate-redshift galaxies may be more dust-obscured than any of
the star-forming galaxies in our local sample.  We conclude that
optical emission-line measurements for larger samples of intermediate-
and high-redshift galaxies are needed to test the applicability of our
locally derived empirical SFR calibrations to distant galaxies.

\acknowledgements

The authors would like to thank the anonymous referee for comments
that improved the clarity of the paper.  J.~M. would also like to
acknowledge conversations with Daniel Eisenstein, Janice Lee, Amy
Stutz, and Dennis Zaritsky on various aspects of the paper.  The data
analysis for this project relied heavily on IDL routines written by
David Schlegel, Scott Burles, and Doug Finkbeiner, and on the IDL
Astronomy User's Library, which is maintained by Wayne Landsman at the
Goddard Space Flight Center.  Funding for this project has been
provided by NSF grant AST-0307386, NASA grant NAG5-8326, and a SINGS
grant, provided by NASA through JPL contract 1224769.  Funding for the
creation and distribution of the SDSS Archive has been provided by the
Alfred P. Sloan Foundation, the Participating Institutions, the
National Aeronautics and Space Administration, the National Science
Foundation, the U.S. Department of Energy, the Japanese
Monbukagakusho, and the Max Planck Society. The SDSS Web site is
\url{http://www.sdss.org}.  The SDSS is managed by the Astrophysical
Research Consortium (ARC) for the Participating Institutions. The
Participating Institutions are The University of Chicago, Fermilab,
the Institute for Advanced Study, the Japan Participation Group, The
Johns Hopkins University, the Korean Scientist Group, Los Alamos
National Laboratory, the Max-Planck-Institute for Astronomy (MPIA),
the Max-Planck-Institute for Astrophysics (MPA), New Mexico State
University, University of Pittsburgh, University of Portsmouth,
Princeton University, the United States Naval Observatory, and the
University of Washington.



\clearpage

\begin{inlinefigure}
  \begin{center}
    \plotone{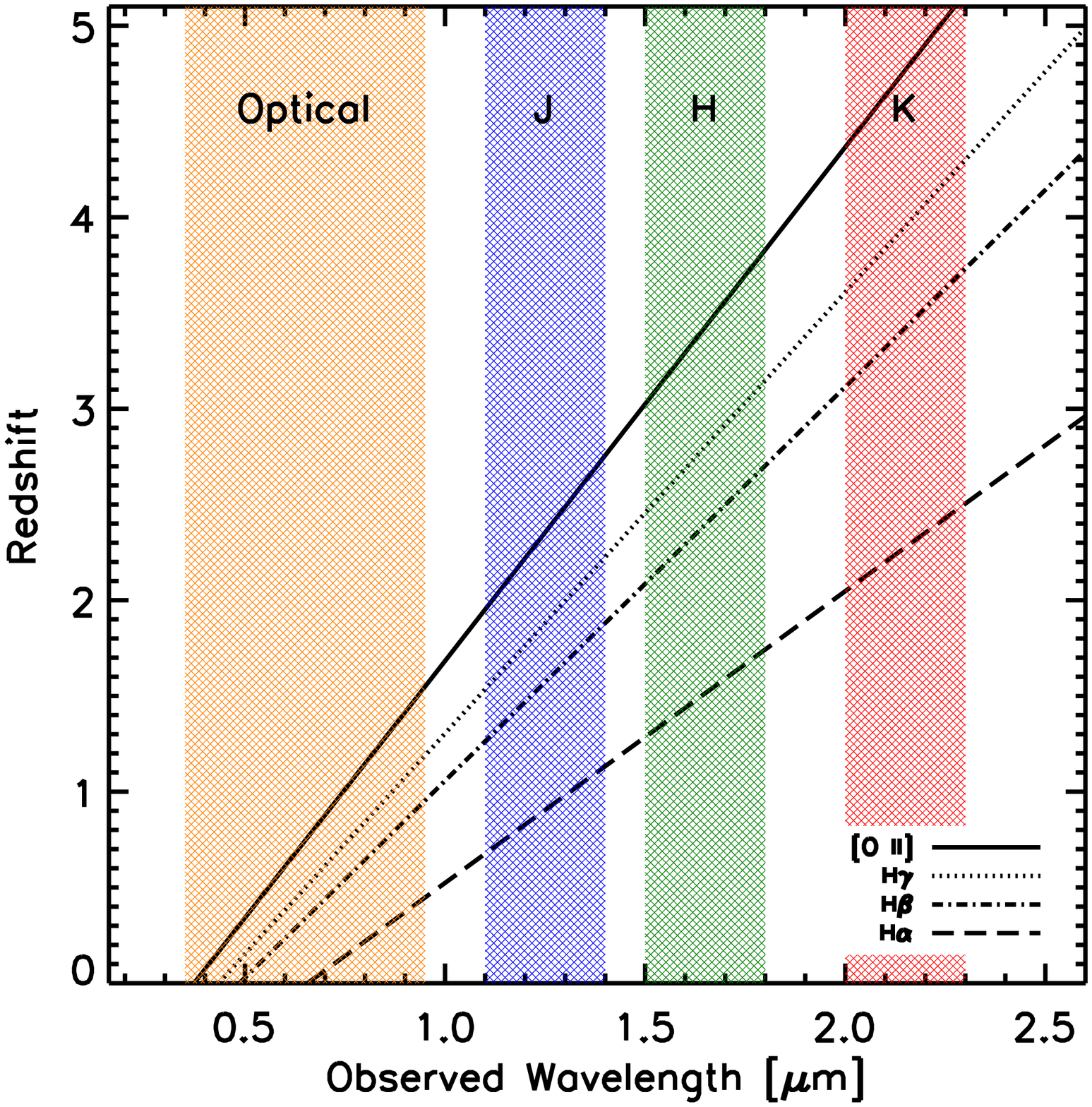}
    \figcaption{Diagram illustrating the redshifts at which \halam, \hblam,
      \hglam, and \oiilam{} are observable from the ground in the optical
      ($3500-9500$~\AA) and through the JHK near-infrared atmospheric
      windows.\label{fig:zlambda}}
  \end{center}
\end{inlinefigure}

\begin{inlinefigure}
  \begin{center}
    \plotone{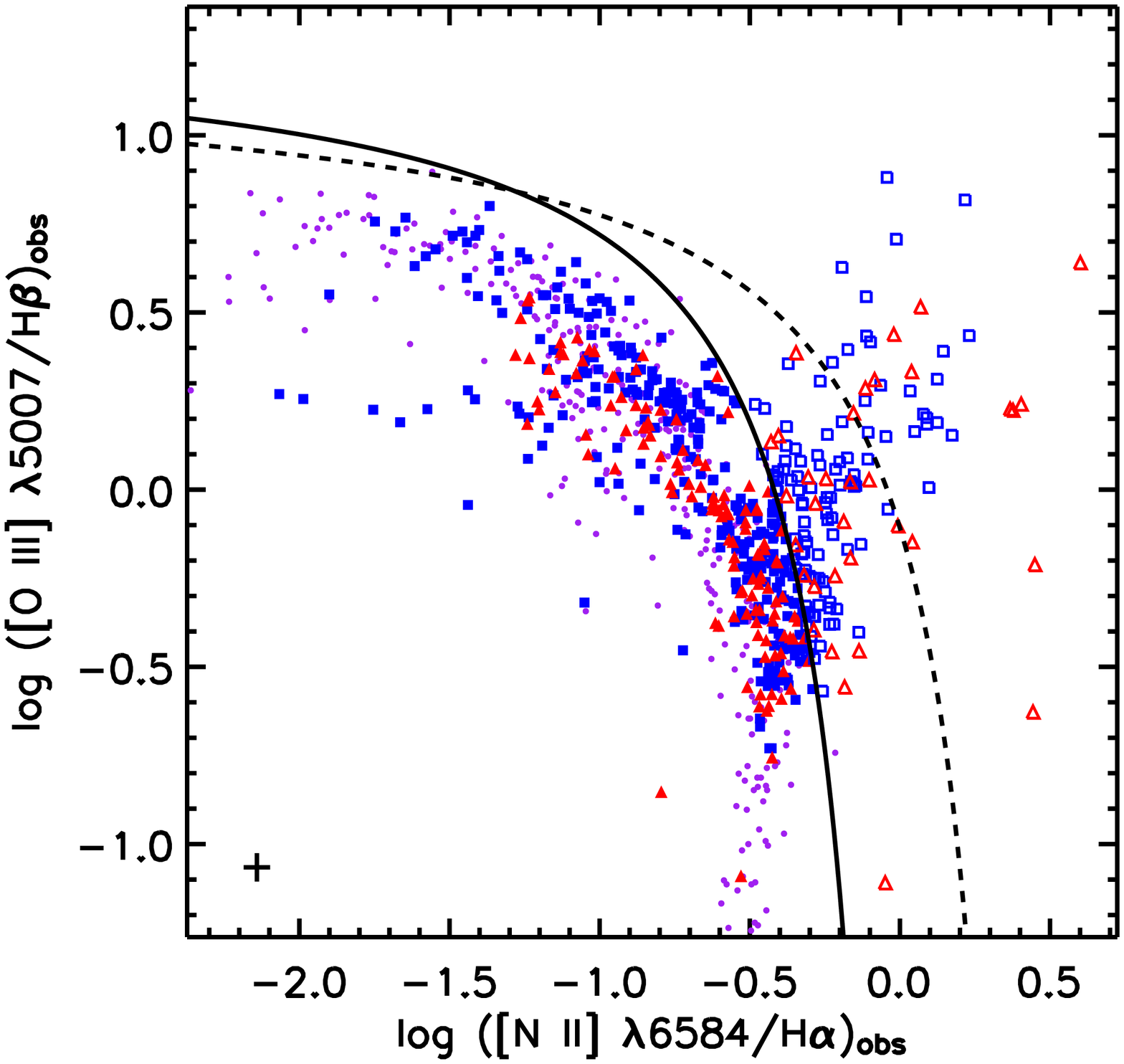}
    \figcaption{Emission-line diagnostic diagram showing the observed
      $(\oiiilam/\hb)_{\rm obs}$ flux ratio versus $(\niilam/\ha)_{\rm obs}$
      for the MK05 survey (squares), the NFGS (triangles), and individual
      \hii{} regions (small points).  (Each sample has been colored blue,
      red, and purple, respectively, in the electronic edition of the
      journal).  We classify objects into star-forming galaxies (filled
      symbols) and AGN (open symbols) using the \emph{solid} curve defined
      empirically by \citet{kauffmann03c}.  For comparison the \emph{dashed}
      line shows the theoretical division between star-forming galaxies and
      galaxies with AGN activity obtained by \citet{kewley01a}.  We plot the
      average uncertainty in the data as a cross in the lower-left part of
      the diagram. \label{fig:bpt}}
  \end{center}
\end{inlinefigure}

\begin{inlinefigure}
  \begin{center}
    \plotone{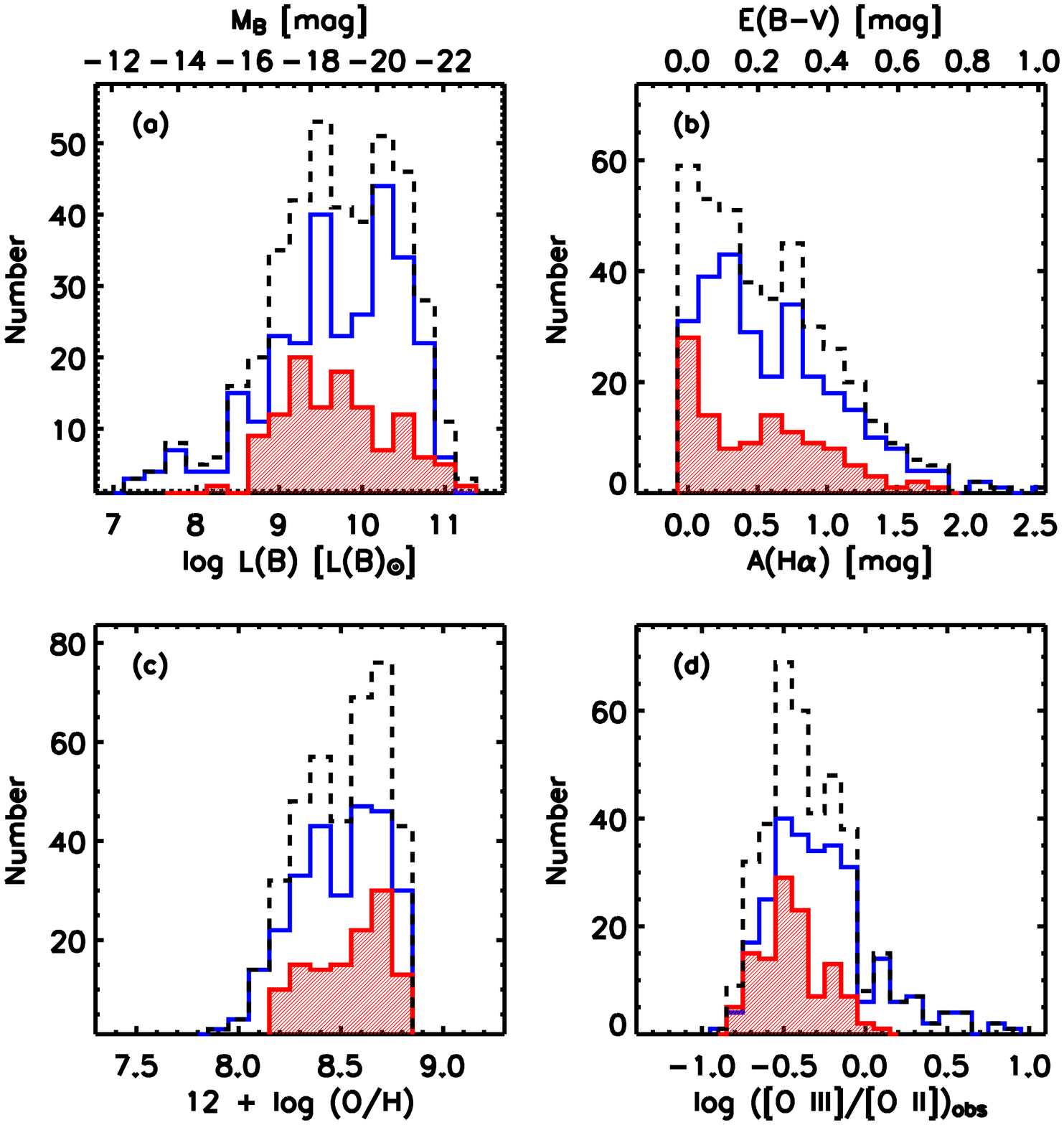}
    \figcaption{Distribution of spectrophotometric properties for the MK05
      survey (unshaded histogram) and the NFGS (shaded histogram) (shown in
      blue and red, respectively, in the electronic edition).  The
      \emph{dashed} histogram shows the distribution for the combined
      integrated galaxy sample.  (\emph{a}) Distribution of $B$-band
      luminosity, \lb, where $M_{\sun, B}=+5.42$~mag can be used to convert
      from \lb{} to \mb.  (\emph{b}) Distribution of \ha{} extinction as
      derived from the \hahb{} Balmer decrement and adopting the
      \citet{odonnell94} Milky Way extinction curve.  (\emph{c})
      Distribution of gas-phase oxygen abundance, \logoh, determined using
      the methodology described in \S\ref{sec:oii_metallicity}.  (\emph{d})
      Distribution of the observed $(\oiiilam/\oiilam)_{\rm obs}$ flux
      ratio, which is indicative of the hardness of the ionizing radiation
      field.  See \S\ref{sec:sample} for more
      details. \label{fig:properties}}
  \end{center}
\end{inlinefigure}

\begin{inlinefigure}
  \begin{center}
    \plotone{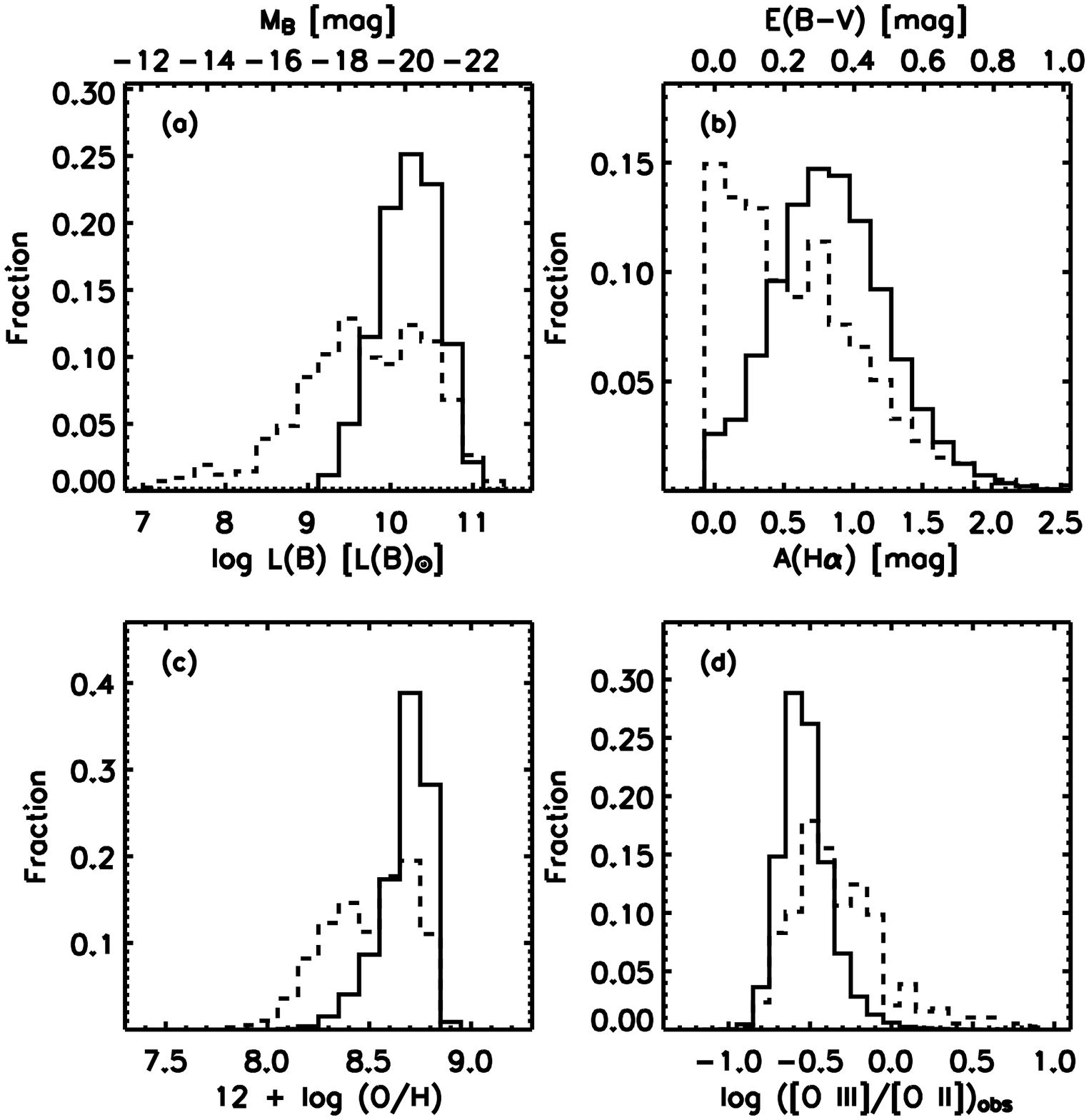}
    \figcaption{Distribution of spectrophotometric properties for the SDSS
      and integrated galaxy samples, plotted as \emph{solid} and
      \emph{dashed} histograms, respectively.  See
      Figure~\ref{fig:properties} and \S\ref{sec:sdsssample} for additional
      details. \label{fig:sdssproperties}}
  \end{center}
\end{inlinefigure}

\begin{inlinefigure}
  \begin{center}
    \plotone{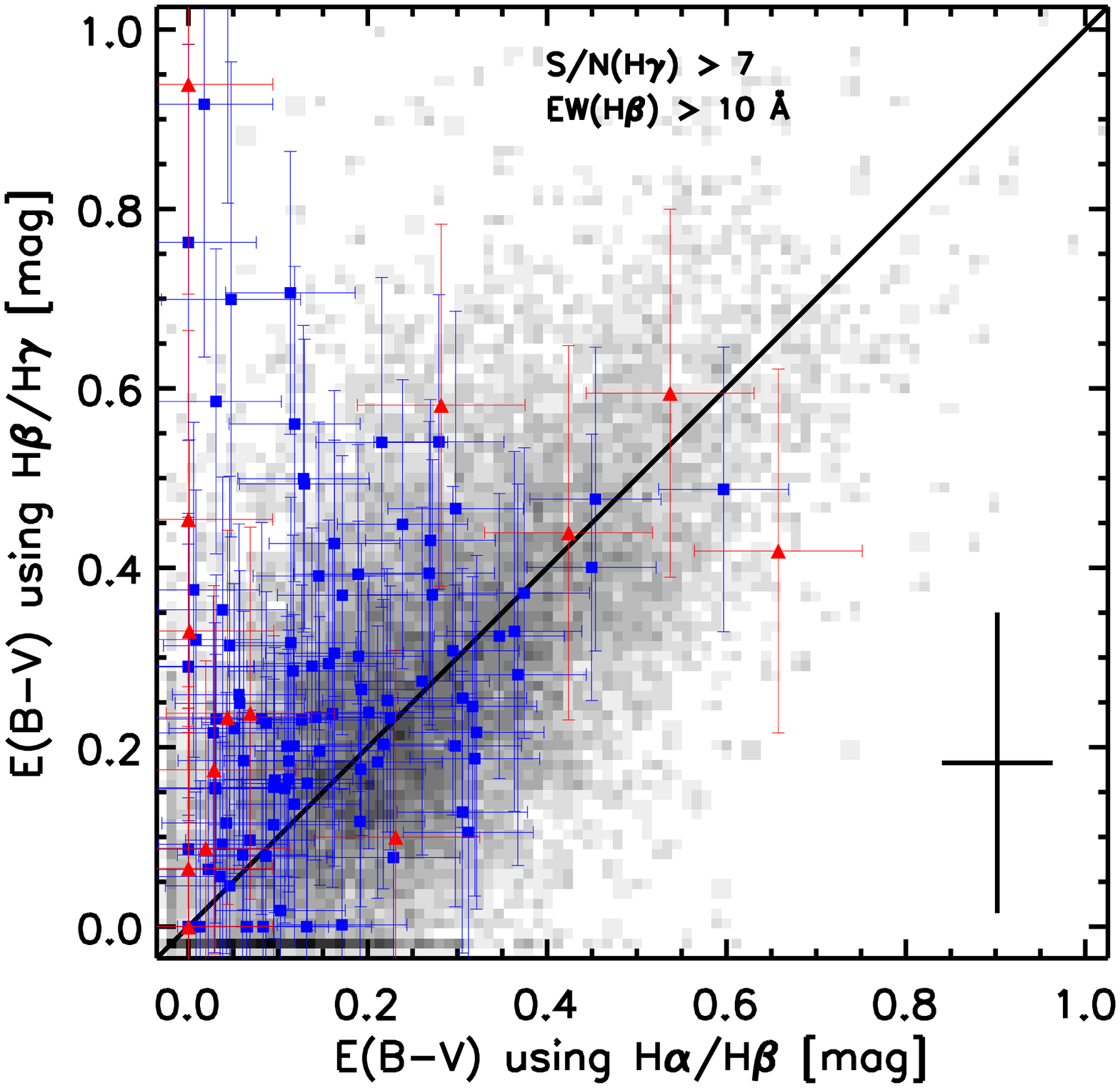}
    \figcaption{Comparison of the reddening values derived using the \hahb{}
      and the \hbhg{} Balmer decrements for the MK05 sample (squares with
      error bars), the NFGS (triangles with error bars), and the SDSS (small
      points without error bars) (plotted in blue, red, and black,
      respectively, in the electronic edition).  The \emph{solid} line
      indicates equality of the two reddening estimates.  We only include
      galaxies with ${\rm S/N(H\gamma)}>7$ and $\ewhb>10$~\AA{} in emission.
      The cross shows the median error of the SDSS
      data. \label{fig:ebv_compare}}
  \end{center}
\end{inlinefigure}

\begin{inlinefigure}
  \begin{center}
    \plotone{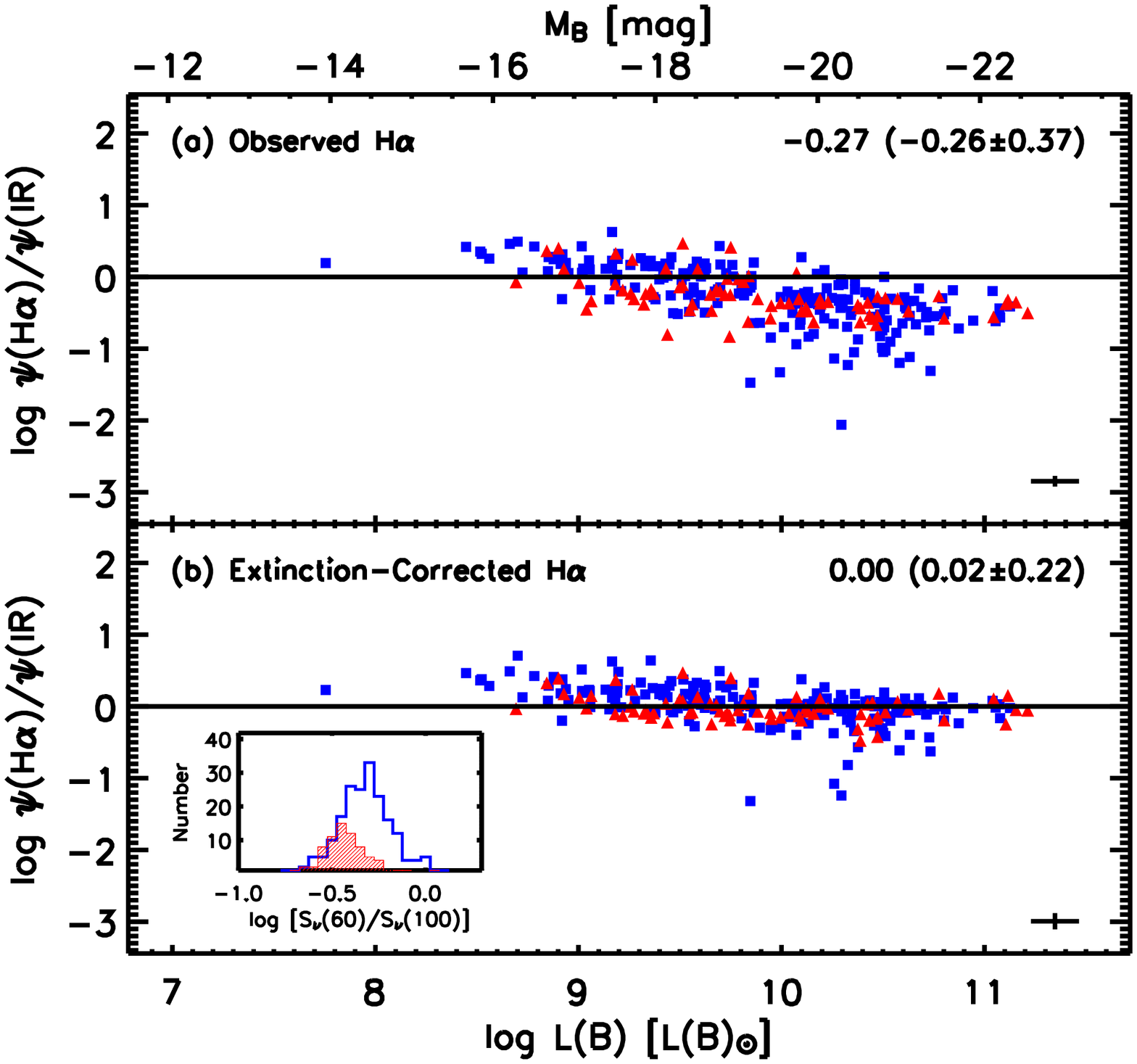}
    \figcaption{Logarithmic ratio of the \ha{} and infrared SFRs,
      $\sfrha/\sfrir$, versus the $B$-band luminosity, \lb{} using (\emph{a})
      the observed \ha{} luminosity and (\emph{b}) the extinction-corrected
      \ha{} luminosity.  The \emph{solid} line indicates equality of the two
      SFR indicators.  The symbols are defined in Figure~\ref{fig:bpt}.  The
      cross in the lower-right of each panel indicates the average
      measurement uncertainty in the data, and the legend gives the median
      logarithmic ratio and, in parenthesis, the mean and standard
      deviation.  The inset in panel (\emph{b}) shows the distribution of
      $S_{\nu}(60~\micron) / S_{\nu}(100~\micron)$ flux ratios for the MK05
      and NFGS samples as \emph{unshaded} and \emph{shaded} histograms,
      respectively (shown in blue and red, respectively, in the electronic
      edition).
      \label{fig:ha_sfr}}
  \end{center}
\end{inlinefigure}

\begin{inlinefigure}
  \begin{center}
    \plotone{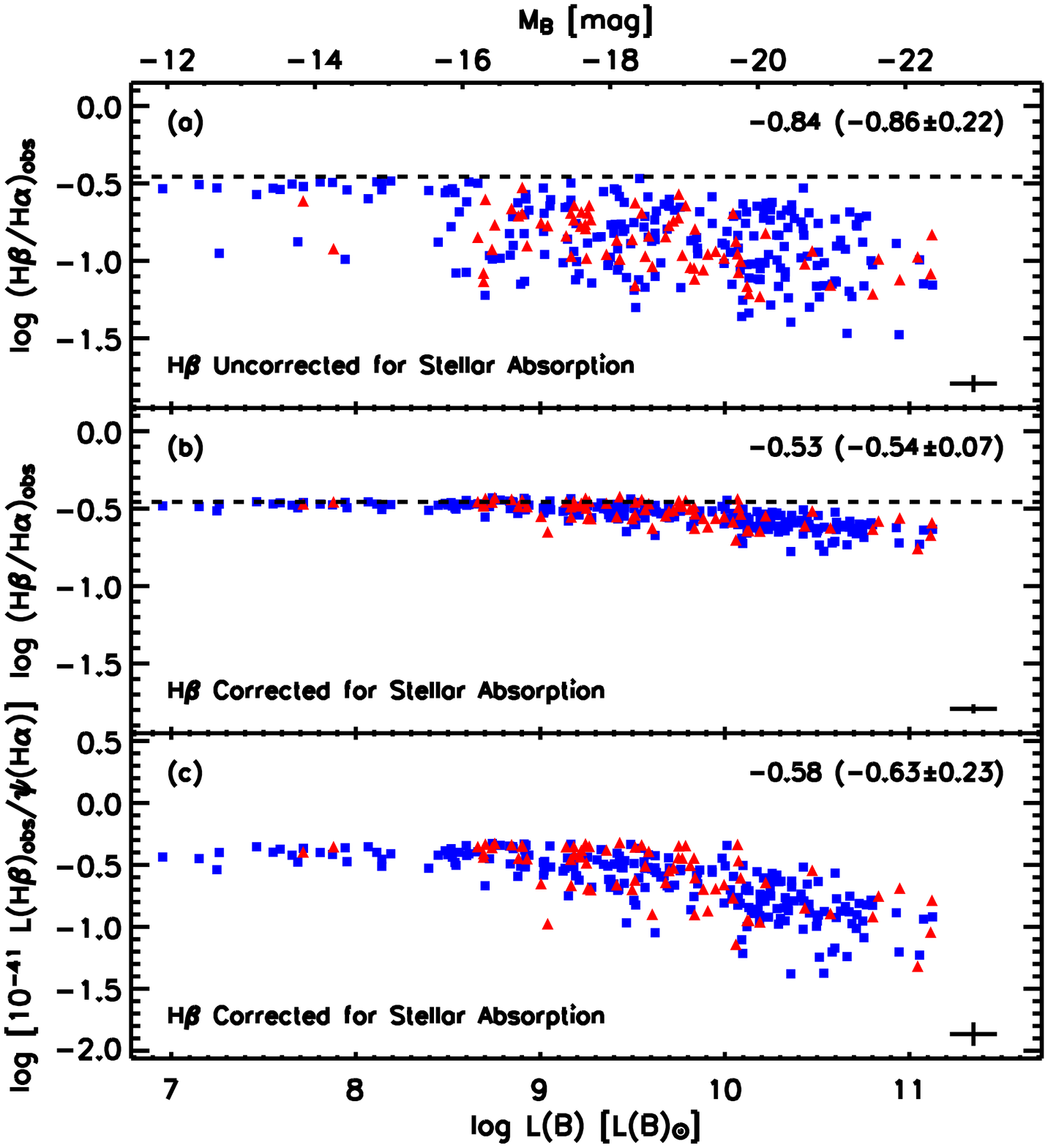}
    \figcaption{Variation in the \hbha{} ratio versus the $B$-band
      luminosity, \lb, for all galaxies in the integrated sample having
      $\ewhb>5$~\AA{} in emission.  The \emph{dashed} line in panels
      (\emph{a}) and (\emph{b}) indicates the intrinsic Balmer decrement,
      $\log\,(\hbha)_{\rm int}\simeq-0.46$~dex (see \S\ref{sec:reddening}).
      The symbols are defined in Figure~\ref{fig:bpt}.  The cross in the
      lower-right of each panel indicates the average measurement
      uncertainty in the data, and the legend gives the median logarithmic
      ratio and, in parenthesis, the mean and standard deviation.
      (\emph{a}) Observed \hbha{} ratio without an \hb{} stellar absorption
      correction.  (\emph{b}) Identical to panel (\emph{a}) except that
      \hb{} is now corrected for underlying stellar absorption.  (\emph{c})
      Ratio of the observed \hb{} luminosity, \lhbobs, to the \ha{}
      star-formation rate, \sfrha, in units of
      $10^{-41}~\sfrunits/(\lunits)$.
      \label{fig:lb_hbha}}
    \end{center}
\end{inlinefigure}

\begin{inlinefigure}
  \begin{center}
  \plotone{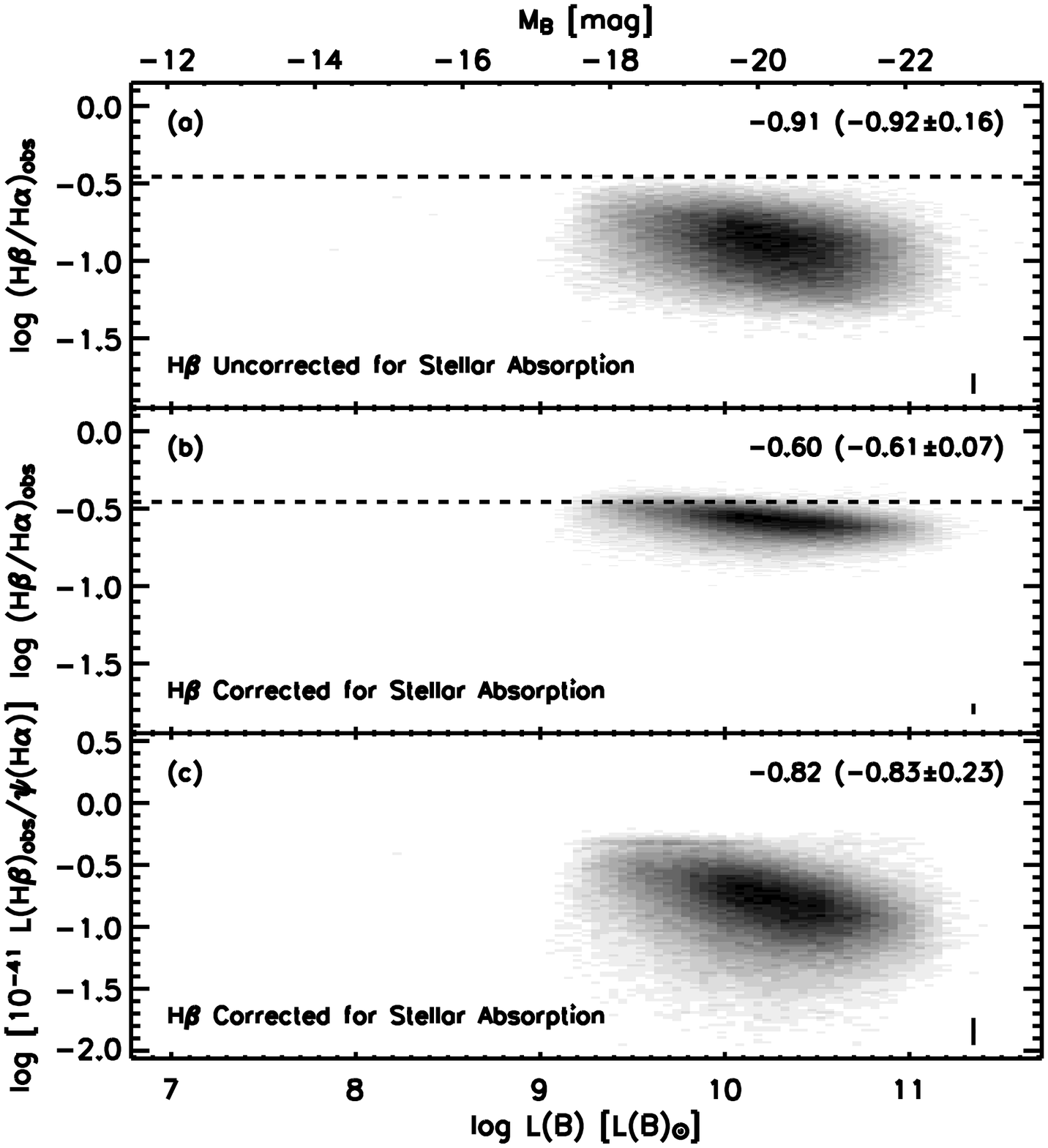}
  \figcaption{Same as Figure~\ref{fig:lb_hbha} but for the SDSS 
    sample.\label{fig:sdss_lb_hbha}}
  \end{center}
\end{inlinefigure}

\begin{inlinefigure}
  \begin{center}
    \plotone{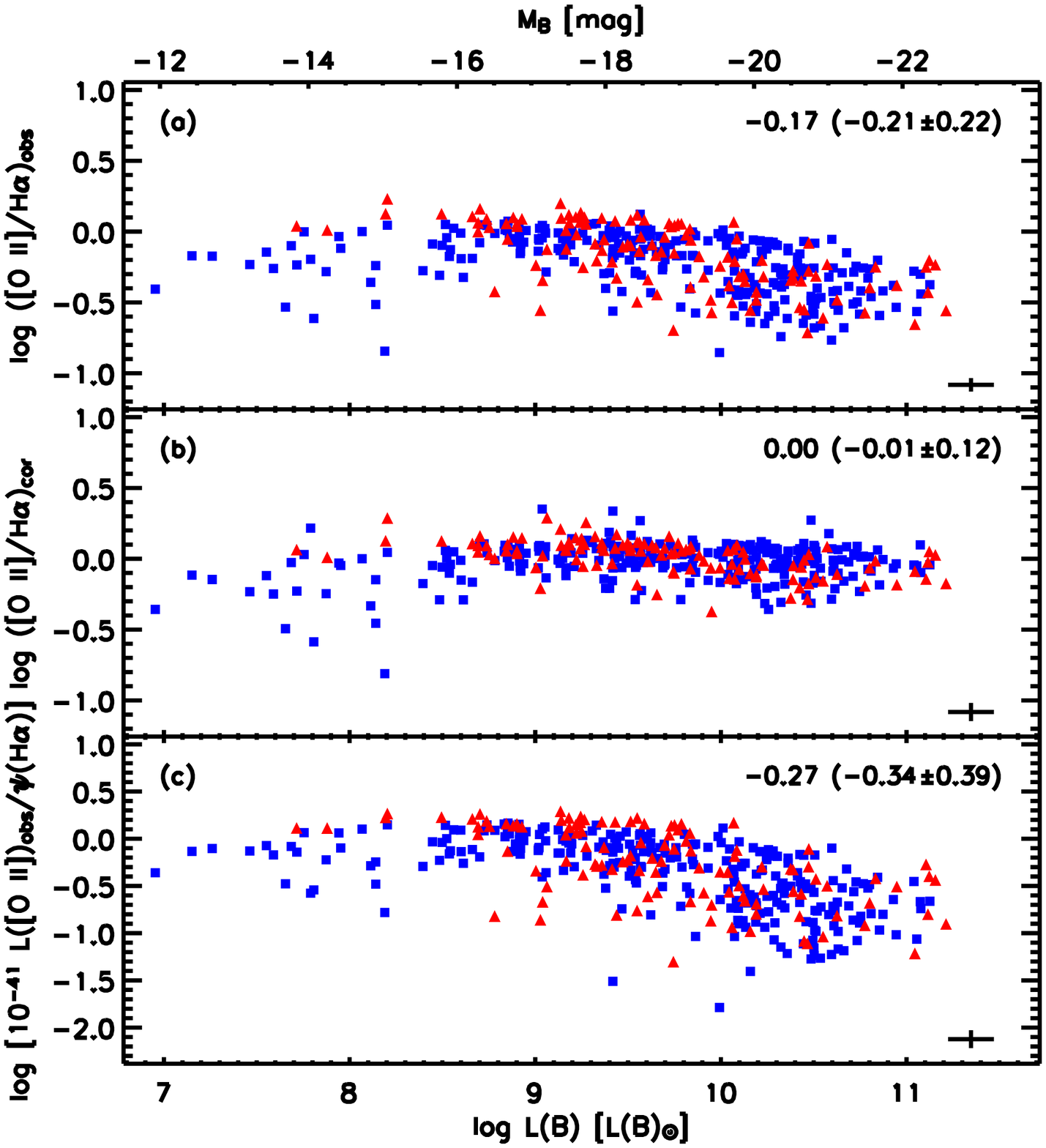}
    \figcaption{Dependence of the \oiiha{} ratio on the $B$-band luminosity,
      \lb, in the integrated sample.  The points have been coded as in
      Figure~\ref{fig:bpt}.  The cross in the lower-right of each panel
      indicates the average measurement uncertainty in the data, and the
      legend gives the median logarithmic ratio and, in parenthesis, the
      mean and standard deviation.  (\emph{a}) Observed \oiiha{} ratio
      uncorrected for dust reddening.  (\emph{b}) Reddening-corrected
      \oiiha{} ratio.  (\emph{c}) Ratio of the observed \oii{} luminosity,
      \loiiobs, to the \ha{} star-formation rate, \sfrha, in units of
      $10^{-41}~\sfrunits/(\lunits)$.
      \label{fig:lb_oiiha}}
  \end{center}
\end{inlinefigure}

\begin{inlinefigure}
  \begin{center}
    \plotone{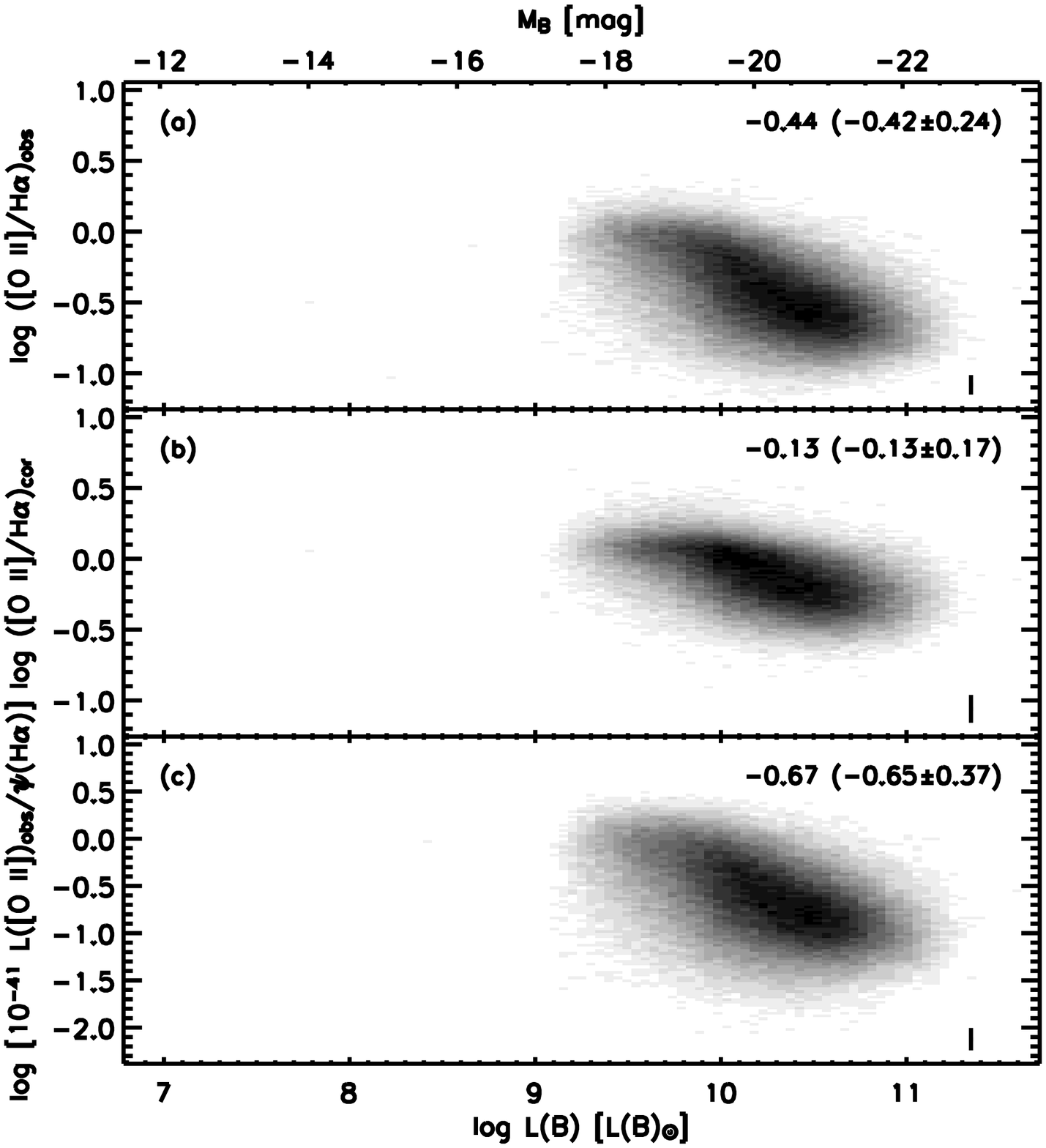}
    \figcaption{Same as Figure~\ref{fig:lb_oiiha} but for the SDSS
      sample.\label{fig:sdss_lb_oiiha}}
  \end{center}
\end{inlinefigure}

\begin{inlinefigure}
  \begin{center}
    \plotone{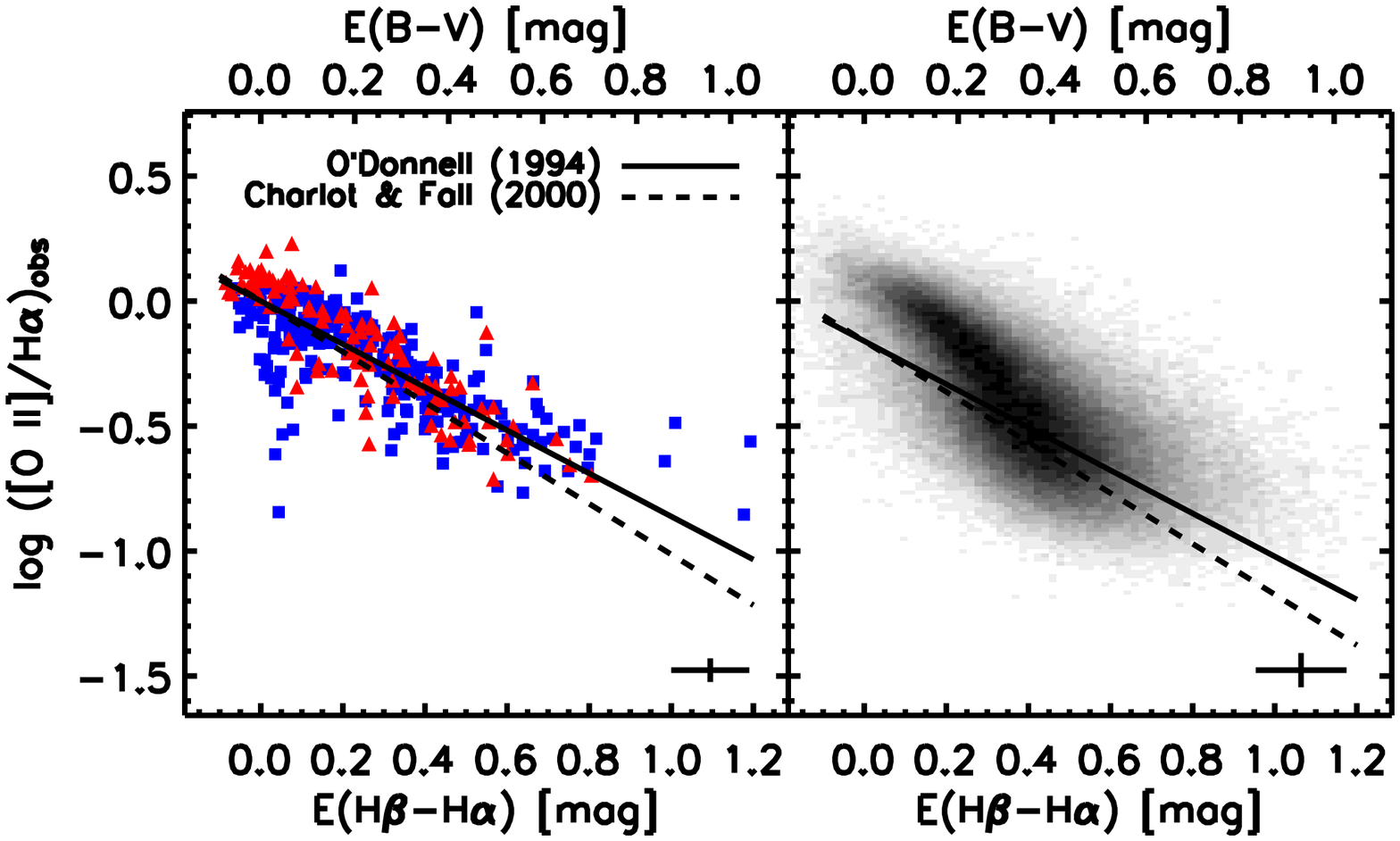}
    \figcaption{Correlation between the observed \oii-to-\ha{} ratio and the
      nebular reddening, \ehbha, for the integrated sample (\emph{left}) and
      the SDSS (\emph{right}).  Refer to Figure~\ref{fig:bpt} for the
      meaning of the symbols.  The cross in the lower-right part of each
      panel indicates the average measurement uncertainty in the data.  We
      plot the \citet{odonnell94} Galactic extinction curve and the
      \citet{charlot00} dust attenuation law as \emph{solid} and
      \emph{dashed} lines, respectively, in both panels.  In the left panel
      we set the normalization of the reddening curves to
      $\log\,(\oiiha)_{\rm obs}=0.00$~dex, and to $\log\,(\oiiha)_{\rm
        obs}=-0.16$~dex in the right panel.
      \label{fig:ehbha_oiiha}}
  \end{center}
\end{inlinefigure}

\begin{inlinefigure}
  \begin{center}
  \plotone{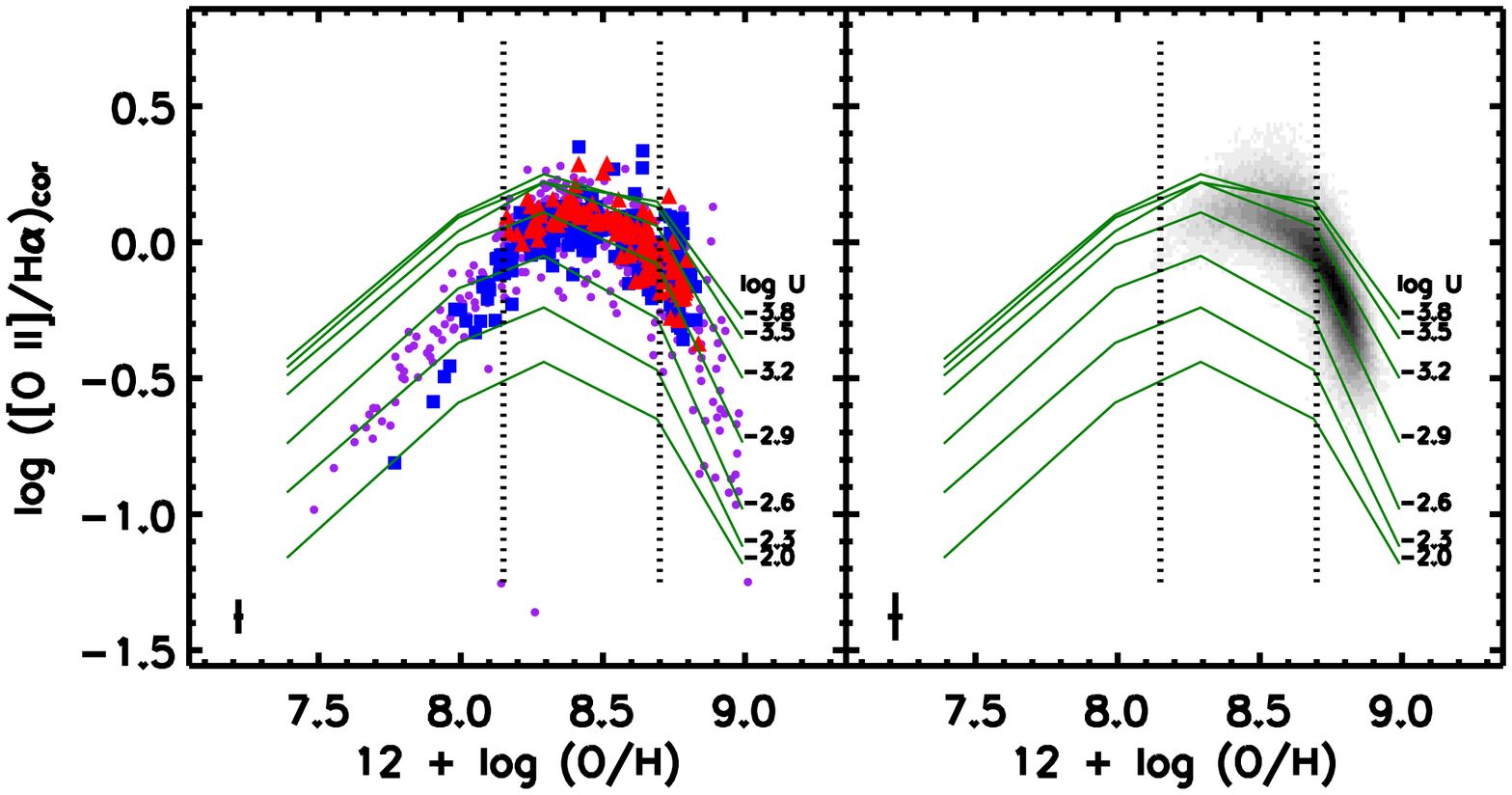}
  \figcaption{Reddening-corrected \oiiha{} ratio versus the nebular oxygen
    abundance, \logoh, for the integrated sample and \hii{} regions
    (\emph{left}), as coded in Figure~\ref{fig:bpt}, and the SDSS sample
    (\emph{right}).  The curves show the results of detailed
    photoionization modeling by \citet{kewley01b} for six values of the
    ionization parameter, $-3.8<\logu<-2.0$.  The vertical dotted lines
    divide each panel into three metallicity regimes: $\logoh<8.15$~dex,
    $8.15<\logoh<8.7$~dex, and $\logoh>8.7$~dex.  The cross in the
    lower-left part of each panel indicates the typical measurement
    uncertainty, excluding any systematic errors in the adopted abundance
    calibration.
    \label{fig:oh12_oiiha}}
  \end{center}
\end{inlinefigure}

\begin{inlinefigure}
  \begin{center}
  \plotone{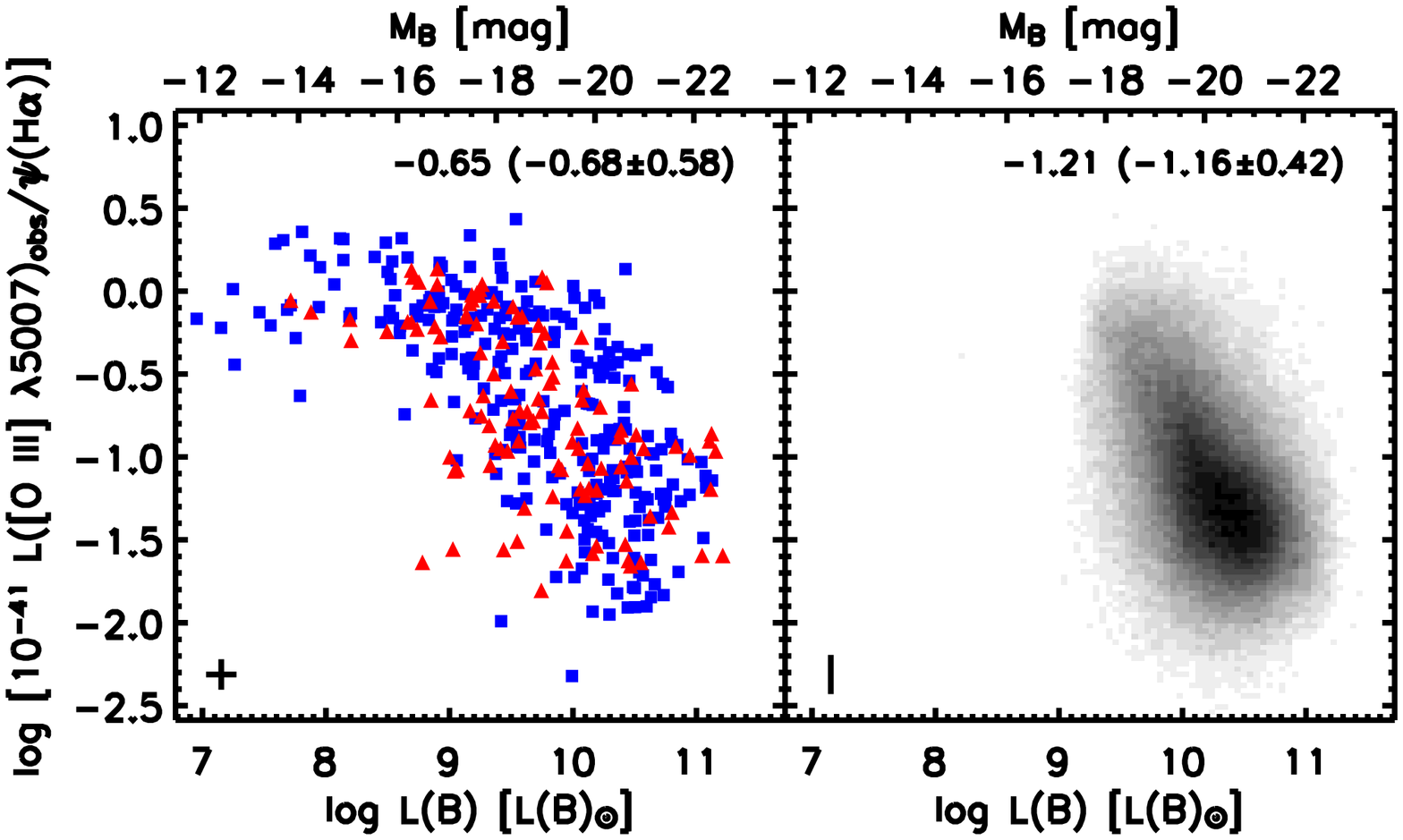}
  \figcaption{Ratio of the observed \oiiilam{} luminosity,
    $(\loiiilam)_{\rm obs}$, to the \ha{} star-formation rate, \sfrha,
    versus the $B$-band luminosity, \lb, for the integrated sample
    (\emph{left}) and the SDSS (\emph{right}).  The symbols in the left
    panel are defined in Figure~\ref{fig:bpt}.  The cross in the
    lower-left corner indicates the average measurement uncertainty in the
    data, and the legend gives the median logarithmic ratio and, in
    parenthesis, the mean and standard deviation.
    \label{fig:lb_oiiiha}}
  \end{center}
\end{inlinefigure}

\begin{inlinefigure}
  \begin{center}
    \plotone{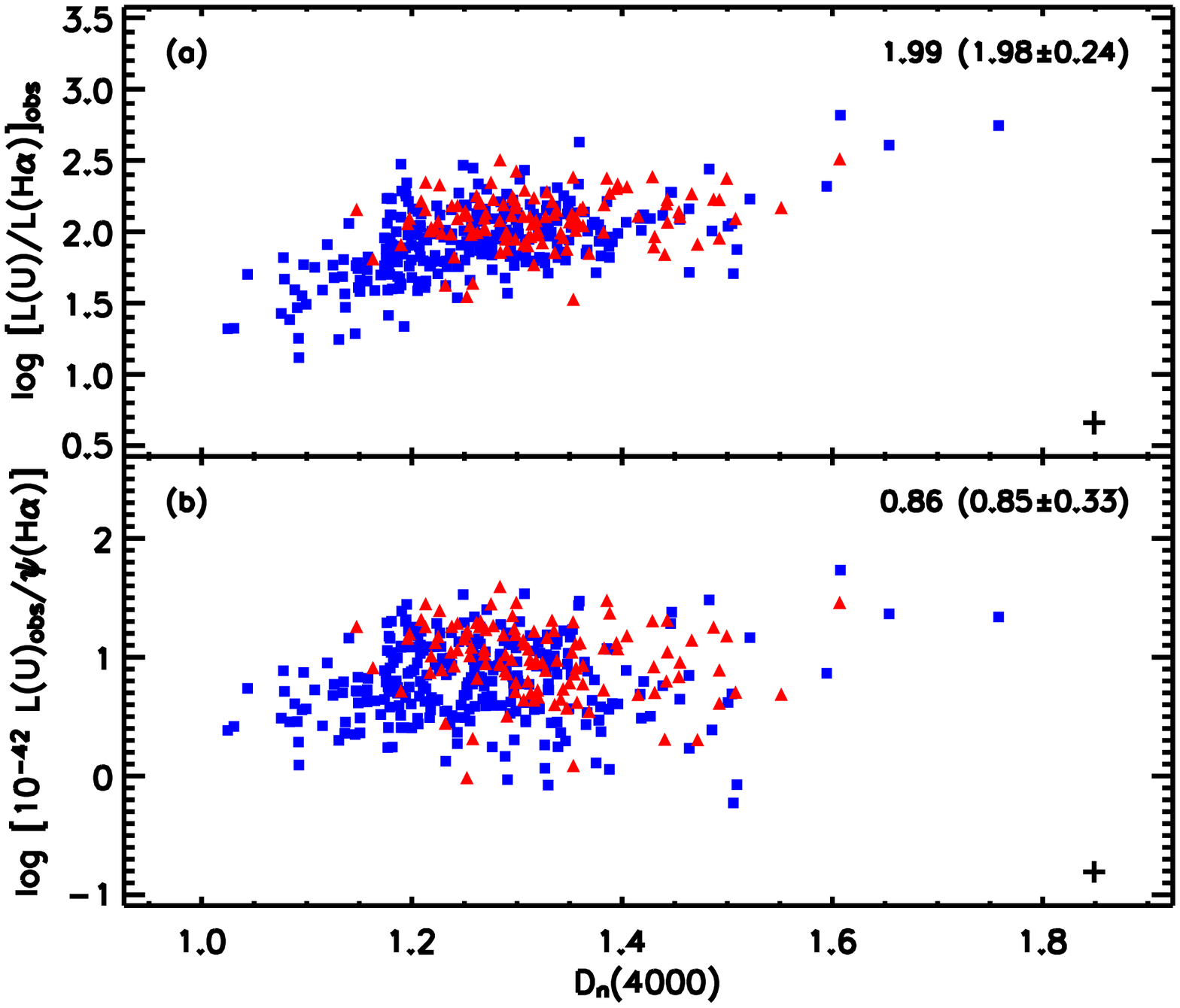}
    \figcaption{Logarithmic $U$-band to \ha{} luminosity ratio versus the
      $4000$-\AA{} break, \dnbreak.  Figure~\ref{fig:bpt} defines the
      symbols used, and the cross in the lower-right indicates the average
      measurement uncertainty in the data.  The legend gives the median
      logarithmic ratio and, in parenthesis, the mean and standard
      deviation.  (\emph{a}) Observed \lulha{} ratio.  (\emph{b}) Ratio of
      the observed $U$-band luminosity, \luobs, to the \ha{} star-formation
      rate, \sfrha, in units of $10^{-42}~\sfrunits/(\lunits)$.
      \label{fig:D4000_Uha}}
    \end{center}
\end{inlinefigure}

\begin{inlinefigure}
  \begin{center}
    \plotone{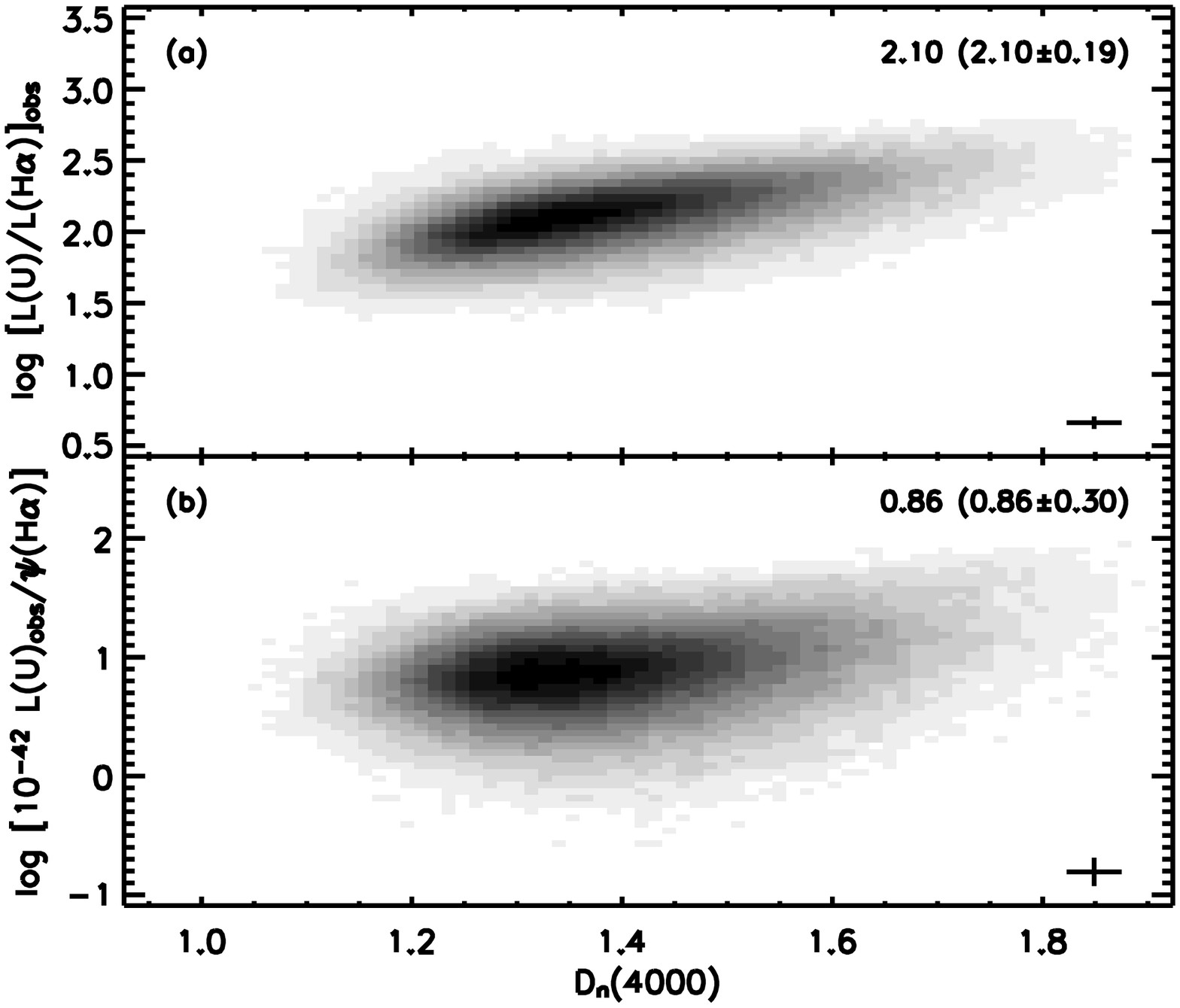}
    \figcaption{Same as Figure~\ref{fig:D4000_Uha} but for the SDSS
      sample. \label{fig:sdss_D4000_Uha}}
    \end{center}
\end{inlinefigure}

\begin{inlinefigure}
  \begin{center}
    \plottwo{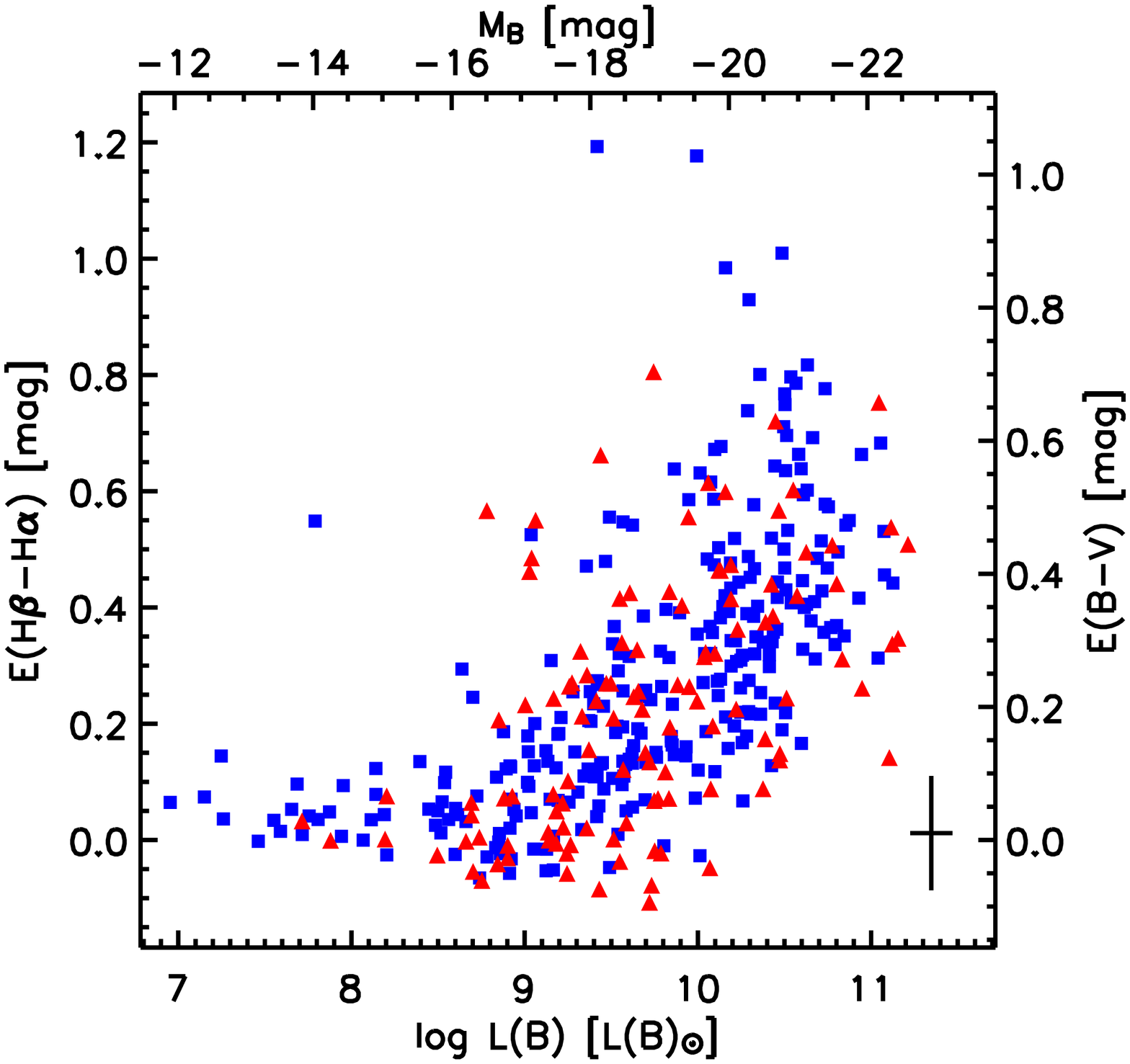}{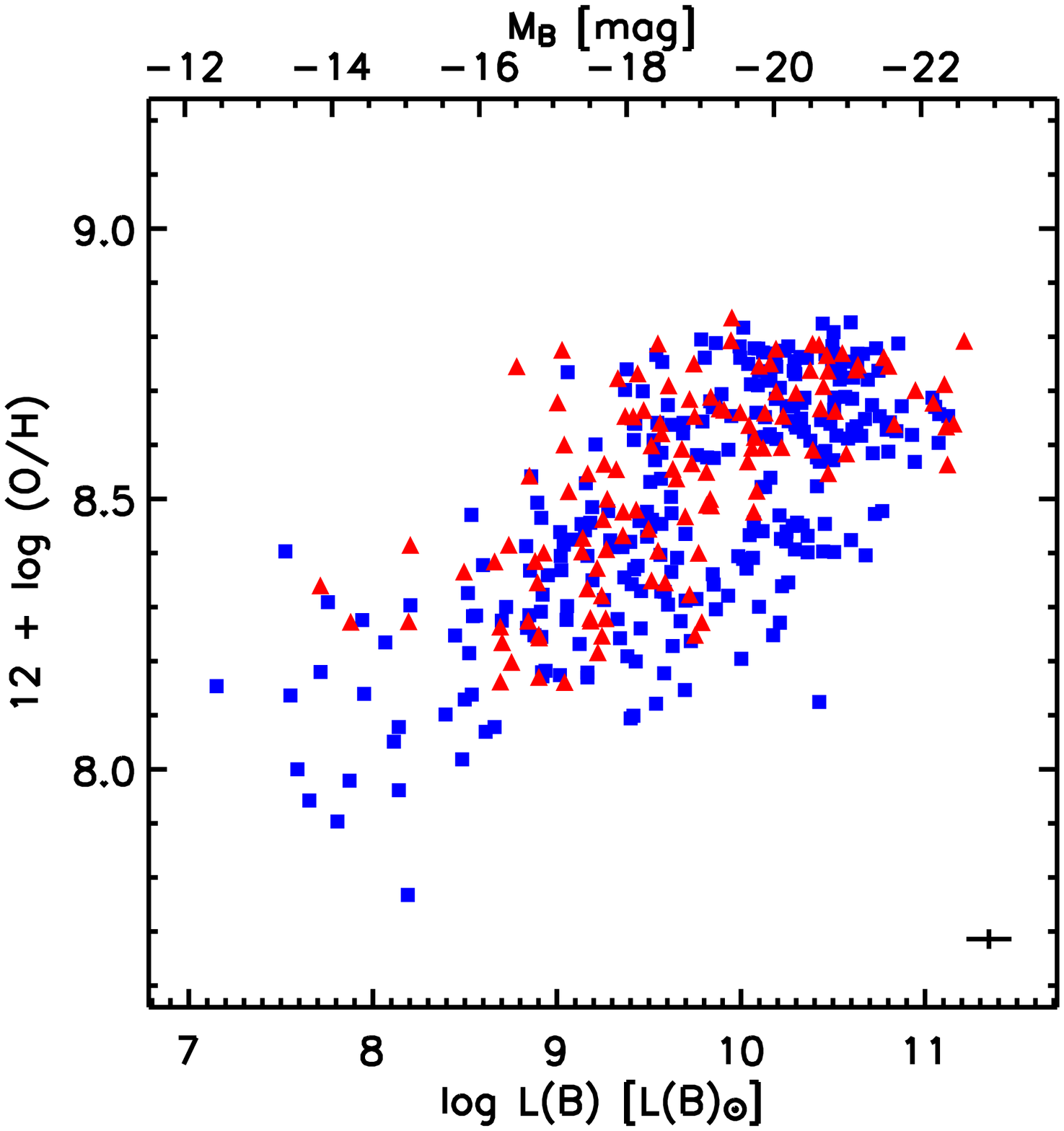}
    \figcaption{Empirical correlations between luminosity and dust extinction
      (\emph{left}), and between luminosity and oxygen abundance
      (\emph{right}).  Figure~\ref{fig:bpt} defines the symbols.  The cross
      in the lower-right corner of each figure shows the average measurement
      uncertainty of the data, excluding any systematic error.
      \label{fig:lb_correlations}}
    \end{center}
\end{inlinefigure}

\begin{inlinefigure}
  \begin{center}
    \plotone{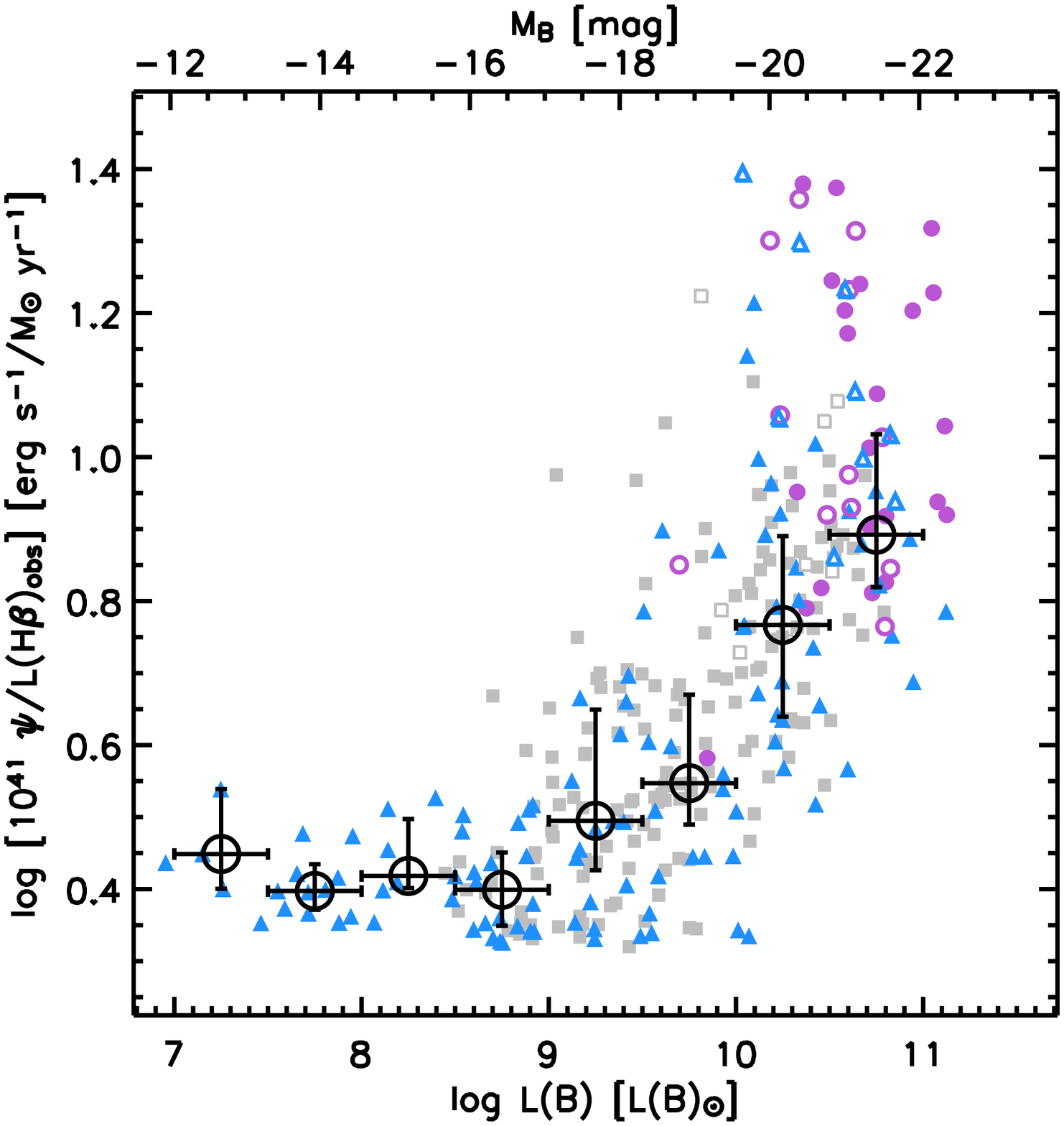}
    \figcaption{Empirical \hblam{} SFR calibration.  The symbols indicate the
      infrared luminosity of each galaxy: circular symbols indicate
      $\lir>10^{11}~\lsun$; squares indicate $\lir<10^{11}~\lsun$; and
      triangles do not have a measured infrared luminosity.  (These symbols
      are colored purple, grey, and blue, respectively, in the electronic
      edition.)  Open and filled symbols indicate the AGN and star-forming
      galaxies in each category, respectively, as classified in
      \S\ref{sec:sample}.  The large open circles give the median
      \sfr/\lhbobs{} ratio in $0.5$~dex luminosity bins, and the lower and
      upper error bars give the $25\%$ and $75\%$ quartile of the
      distribution in each bin, respectively.  One galaxy, IC~1623~B, an AGN
      without a measured infrared luminosity, does not appear on this plot
      because it deviates significantly from the rest of the sample.  Its
      blue luminosity is $\sim10^{10}$~\lbsun{} and its logarithmic
      \sfr/\lhbobs{} ratio is $\sim2.5\times10^{41}$~\lunits/(\sfrunits).
      \label{fig:sfrha_lhb}}
  \end{center}
\end{inlinefigure}

\begin{inlinefigure}
  \begin{center}
    \plotone{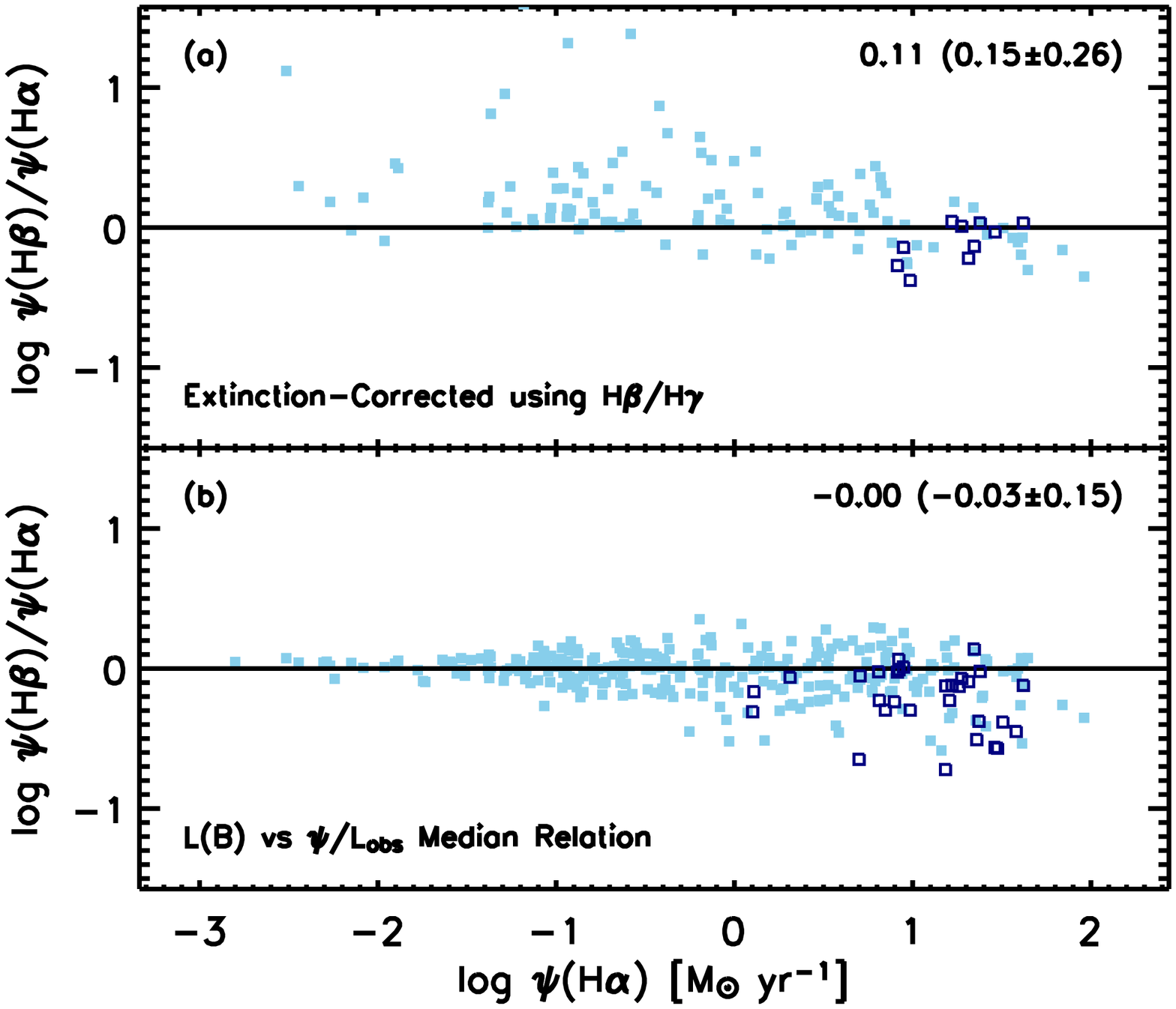}
    \figcaption{Comparison of \sfrhb{} derived using two different methods
      versus \sfrha.  Filled and open symbols indicate star-forming galaxies
      and AGN, respectively, as classified in \S\ref{sec:sample} (colored
      light blue and dark blue, respectively, in the electronic edition).
      (\emph{a}) \sfrhb{} derived using the \hbhg{} ratio to correct for
      dust extinction and appling equation~(\ref{eq:ha_sfr}) assuming
      $\lhb=\lha/2.86$.  As discussed in \S\ref{sec:hb_results}, we only
      include galaxies having $\ewhb>10$~\AA{} in emission and ${\rm
        S/N}(\hg)>7$.  (\emph{b}) \sfrhb{} obtained using the empirical SFR
      calibration developed in \S\ref{sec:hb_results} for all galaxies
      having $\ewhb>5$~\AA{} in emission.
      \label{fig:compare_sfrhb}}
  \end{center}
\end{inlinefigure}

\begin{inlinefigure}
  \begin{center}
    \plotone{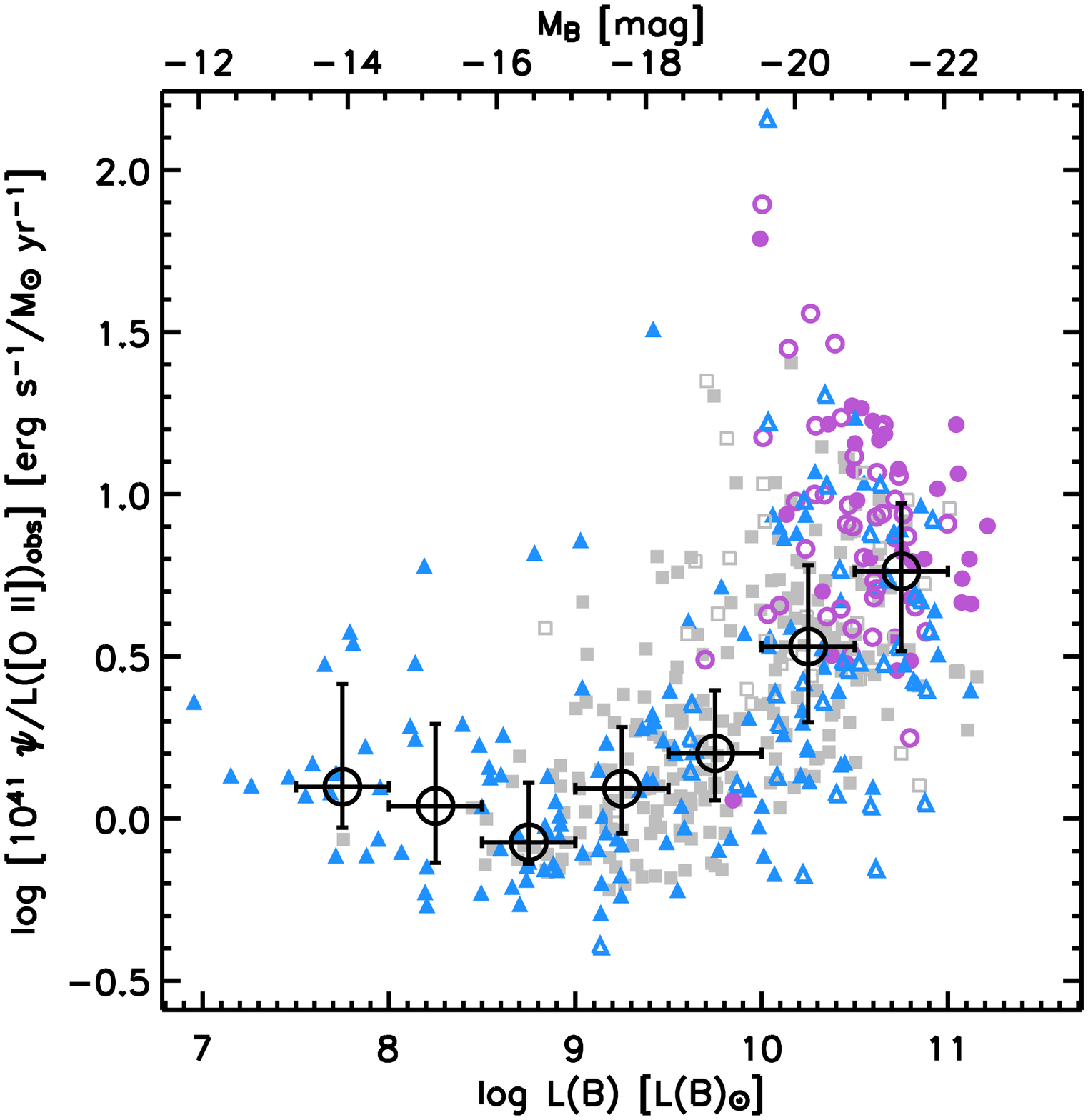}
    \figcaption{Empirical \oiilam{} SFR calibration.  The symbols have been
      defined in Figure~\ref{fig:sfrha_lhb}.  The large open circles give
      the median \sfr/\loiiobs{} ratio in $0.5$~dex luminosity bins, and the
      lower and upper error bars give the $25\%$ and $75\%$ quartile of the
      distribution in each bin, respectively.
      \label{fig:sfrha_loii}}
  \end{center}
\end{inlinefigure}

\begin{inlinefigure}
  \begin{center}
    \plotone{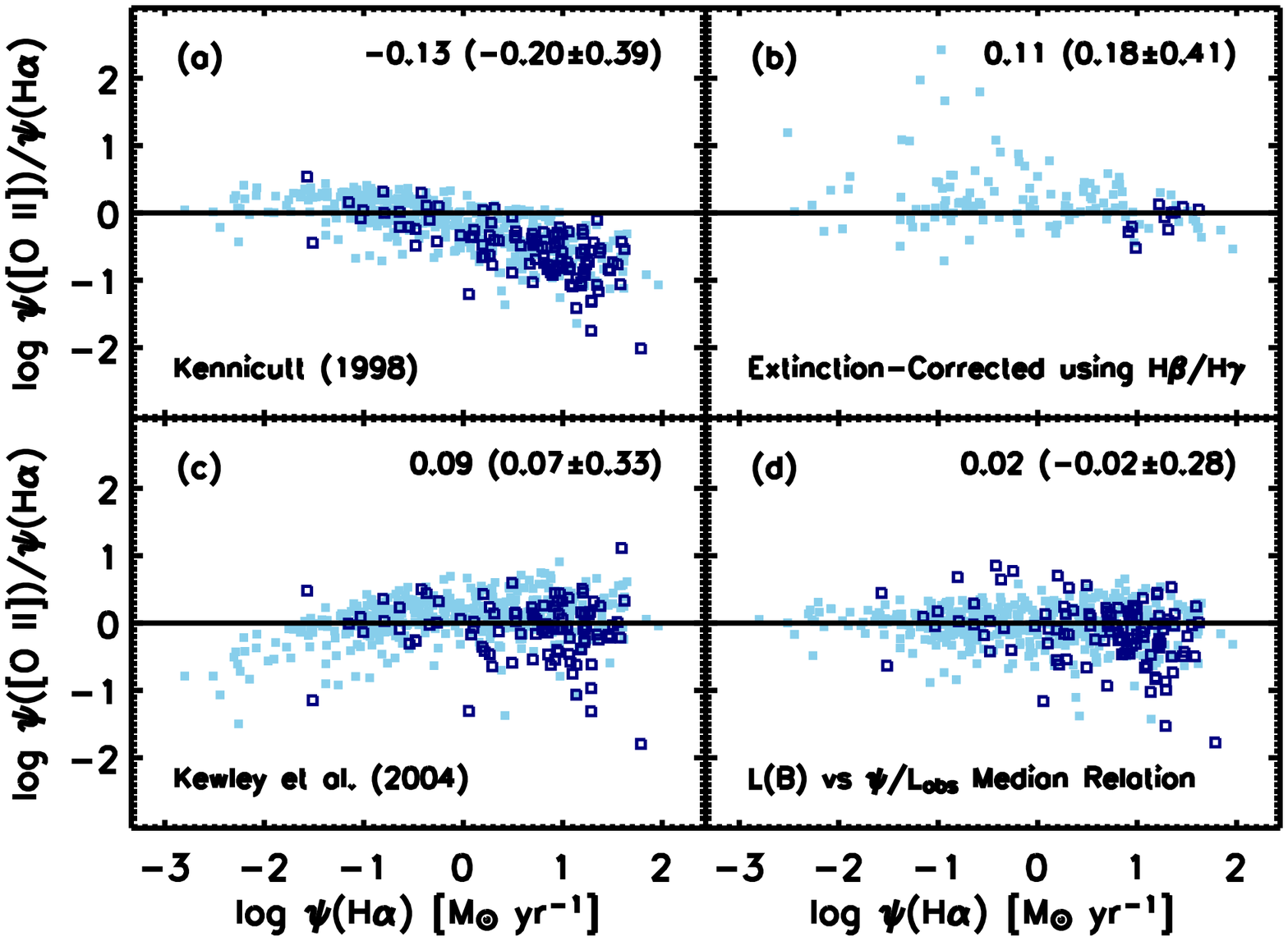}
    \figcaption{Comparison of \sfroii{} using four independent calibrations
      versus \sfrha.  Figure~\ref{fig:compare_sfrhb} defines the different
      symbols.  (\emph{a}) \sfroii{} obtained using the \citet{kenn98}
      calibration. (\emph{b}) Here, we correct \oii{} for extinction using
      the measured \hbhg{} ratio, assume an intrinsic $\oiiha=1$ ratio, and
      apply equation~(\ref{eq:ha_sfr}) to estimate \sfroii.  This panel only
      includes objects having $\ewhb>10$~\AA{} in emission and ${\rm
        S/N}(\hg)>7$.  (\emph{c}) \sfroii{} estimated by applying the
      empirical abundance and extinction corrections advocated by
      \citet{kewley04}.  (\emph{d}) \sfroii{} derived using the calibration
      developed in
      \S\ref{sec:oii_results}. \label{fig:compare_sfroii}}
  \end{center}
\end{inlinefigure}

\begin{inlinefigure}
  \begin{center}
    \plotone{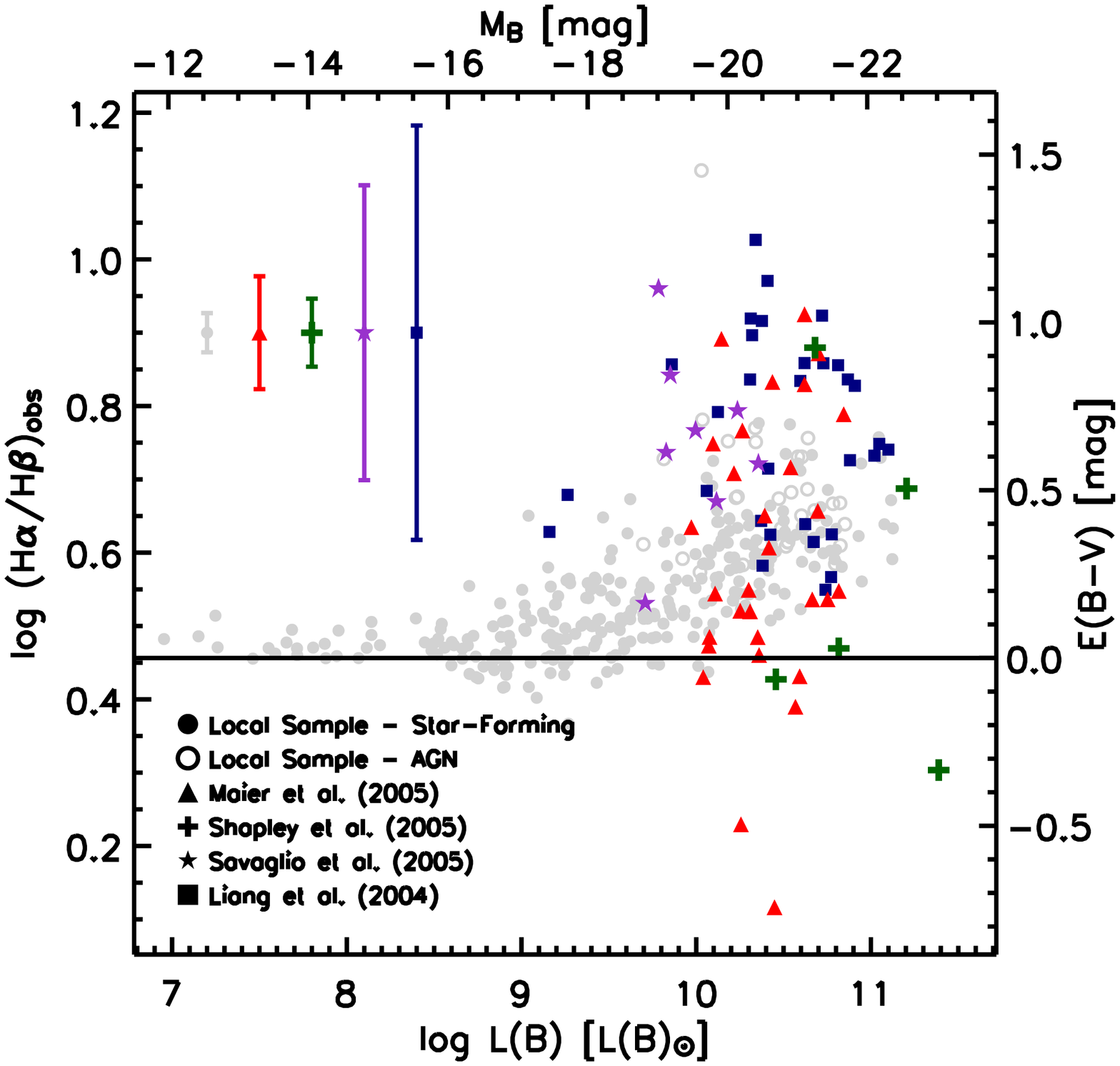}
    \figcaption{Comparison of the local and intermediate-redshift
      ($z=0.7-1.4$) relation between the observed \hahb{} ratio and the
      $B$-band luminosity, \lb.  The \emph{solid} line corresponds to the
      intrinsic \hahb{} Balmer decrement.  The error bars in the upper-left
      part of the figure indicate the median $1\sigma$ uncertainty in the
      data for each sample, as defined in the legend.  Note that the data
      from \citet{savaglio05} and \citet{liang04a} have been derived from
      the \hbhg{} ratio, as described in \S\ref{sec:applications}.
      \label{fig:highz_hahb}}
  \end{center}
\end{inlinefigure}

\begin{inlinefigure}
  \begin{center}
    \plotone{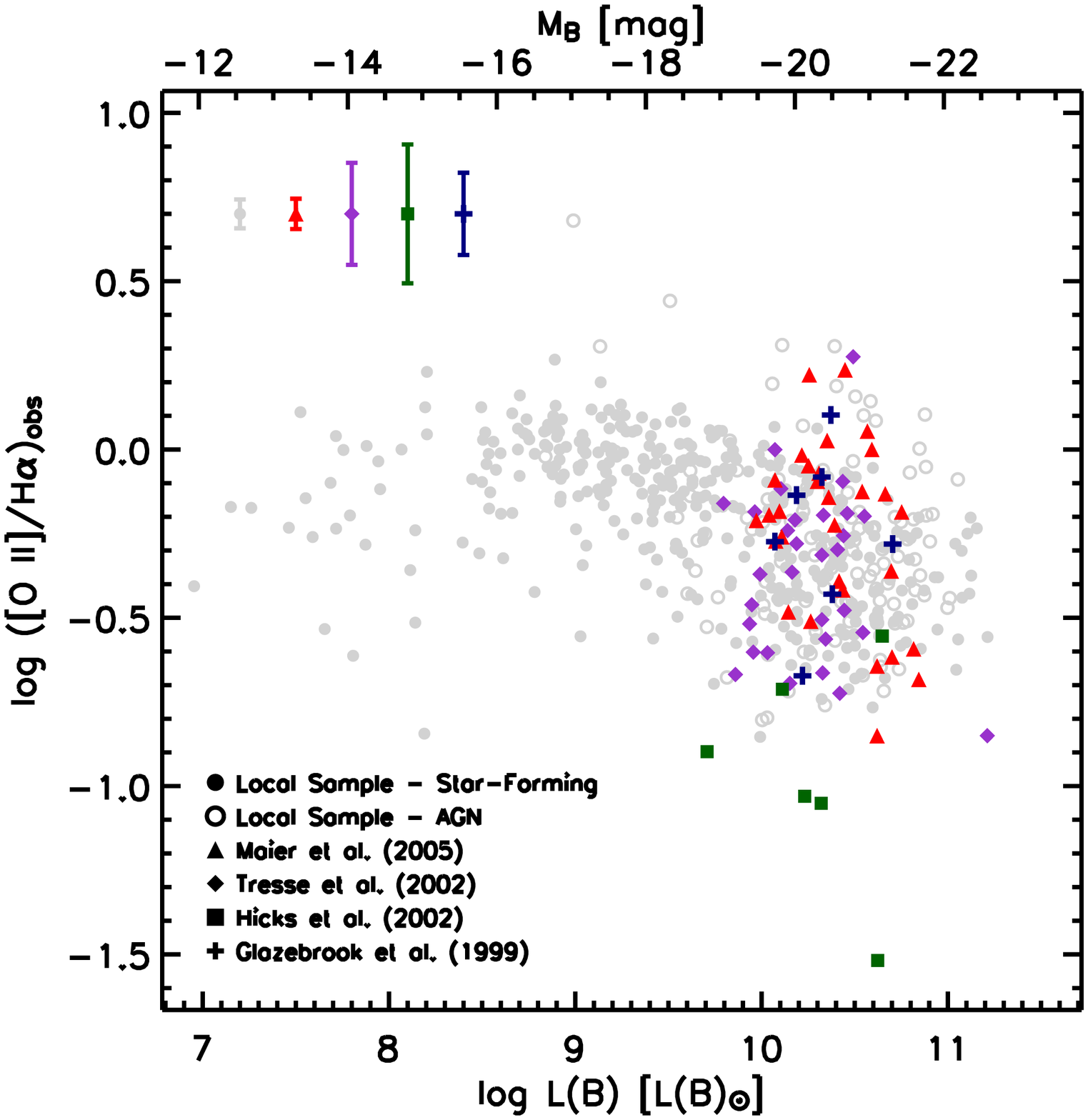}
    \figcaption{Correlation between the observed \oiiha{} ratio and the
      $B$-band luminosity, \lb, for our local spectroscopic sample and
      several intermediate-redshift ($z=0.5-1.5$) galaxy samples.  The error
      bars in the upper-left part of the figure indicate the median
      $1\sigma$ uncertainty in the data for each sample, as defined in the
      legend.
      \label{fig:highz_oiiha}}
  \end{center}
\end{inlinefigure}

\begin{inlinefigure}
  \begin{center}
    \plotone{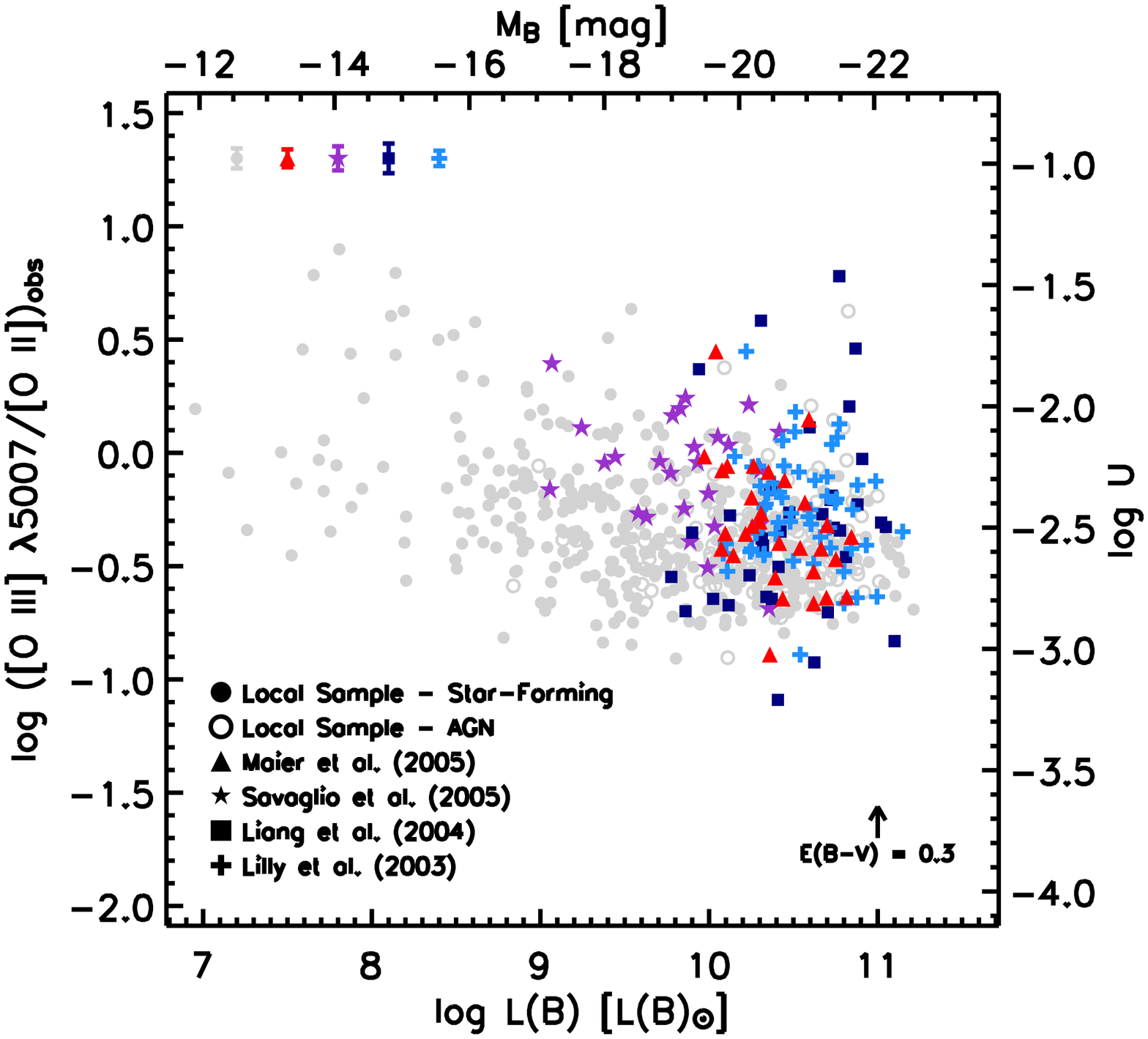}
    \figcaption{Comparison of the local and intermediate-redshift
      ($z=0.4-1.0$) relation between the observed \oiiilam/\oii{} ratio and
      the $B$-band luminosity, \lb.  The error bars in the upper-left part
      of the figure indicate the median $1\sigma$ uncertainty in the data
      for each sample, as defined in the legend.  The arrow indicates the
      effect of $0.3$~mag of reddening on an intrinsic \oiiioii{} ratio of
      $-1.7$~dex, neglecting any reddening effects on \lb.  To estimate the
      sequence in ionization parameter, \logu, from the \oiiioii{} ratio
      (right axis), we utilize the solar-metallicity, continuous star
      formation photoionization models by \citet{kewley01b}.
      \label{fig:highz_oiiioii}}
  \end{center}
\end{inlinefigure}

\clearpage

\begin{deluxetable}{ccccccc}
\tabletypesize{\small}
\tablecolumns{7}
\tablecaption{\hblam{} Star-Formation Rate Conversion Factors \label{table:hb_sfr}}
\tablewidth{0in}
\tablehead{
\colhead{$\log\,\lb$} & 
\colhead{\mb} & 
\multicolumn{5}{c}{$\log\,[\sfr/\lhbobs]$} \\
\colhead{[\lbsun]} & 
\colhead{[mag]} & 
\multicolumn{5}{c}{$[10^{41}\,\lunits/(\sfrunits)]$} \\
\cline{3-7}
\multicolumn{2}{c}{} & 
\colhead{$P_{25}$} & 
\colhead{$P_{50}$} & 
\colhead{$P_{75}$} & 
\colhead{$\langle R\rangle$} & 
\colhead{$\sigma_{R}$}
}
\startdata
 7.25 & -12.68 & 0.400 & 0.449 & 0.539 & 0.435 & 0.079 \\
 7.75 & -13.93 & 0.371 & 0.397 & 0.435 & 0.403 & 0.042 \\
 8.25 & -15.18 & 0.401 & 0.418 & 0.497 & 0.431 & 0.057 \\
 8.75 & -16.43 & 0.349 & 0.399 & 0.451 & 0.414 & 0.080 \\
 9.25 & -17.68 & 0.426 & 0.495 & 0.649 & 0.526 & 0.147 \\
 9.75 & -18.93 & 0.490 & 0.547 & 0.670 & 0.583 & 0.157 \\
10.25 & -20.18 & 0.639 & 0.767 & 0.890 & 0.775 & 0.188 \\
10.75 & -21.43 & 0.819 & 0.892 & 1.032 & 0.927 & 0.190 \\
\enddata
\tablecomments{The columns labeled $P_{25}$, $P_{50}$, and $P_{75}$ give the $25$, $50$ (median), and $75$ percentile of the $\sfr/\lhbobs$ distribution, respectively, in bins of $0.5$~dex in luminosity.  $\langle R\rangle$ and $\sigma_{R}$ give the mean and standard-deviation of the distribution in each bin.}
\end{deluxetable}

\begin{deluxetable}{ccccccc}
\tabletypesize{\small}
\tablecolumns{7}
\tablecaption{\oiilam{} Star-Formation Rate Conversion Factors \label{table:oii_sfr}}
\tablewidth{0in}
\tablehead{
\colhead{$\log\,\lb$} & 
\colhead{\mb} & 
\multicolumn{5}{c}{$\log\,[\sfr/\loiiobs]$} \\
\colhead{[\lbsun]} & 
\colhead{[mag]} & 
\multicolumn{5}{c}{$[10^{41}\,\lunits/(\sfrunits)]$} \\
\cline{3-7}
\multicolumn{2}{c}{} & 
\colhead{$P_{25}$} & 
\colhead{$P_{50}$} & 
\colhead{$P_{75}$} & 
\colhead{$\langle R\rangle$} & 
\colhead{$\sigma_{R}$}
}
\startdata
 7.75 & -13.93 & -0.028 &  0.098 &  0.414 &  0.156 &  0.240 \\
 8.25 & -15.18 & -0.136 &  0.039 &  0.291 &  0.109 &  0.313 \\
 8.75 & -16.43 & -0.140 & -0.074 &  0.111 & -0.017 &  0.185 \\
 9.25 & -17.68 & -0.046 &  0.092 &  0.282 &  0.136 &  0.292 \\
 9.75 & -18.93 &  0.056 &  0.201 &  0.395 &  0.261 &  0.347 \\
10.25 & -20.18 &  0.297 &  0.530 &  0.781 &  0.545 &  0.332 \\
10.75 & -21.43 &  0.517 &  0.762 &  0.972 &  0.749 &  0.285 \\
\enddata
\tablecomments{The columns labeled $P_{25}$, $P_{50}$, and $P_{75}$ give the $25$, $50$ (median), and $75$ percentile of the $\log\,[\sfr/\loiiobs]$ distribution, respectively, in bins of $0.5$~dex in luminosity.  $\langle R\rangle$ and $\sigma_{R}$ give the mean and standard-deviation of the distribution in each bin.}
\end{deluxetable}

\end{document}